\documentclass[aps,prl,reprint,superscriptaddress,showpacs,longbibliography,footnoteinbib]{revtex4-1}
\usepackage[latin9]{inputenc}
\usepackage{xcolor}
\usepackage{float}
\usepackage{bm}
\usepackage{amsmath}
\usepackage{amssymb}
\usepackage{graphicx}
\PassOptionsToPackage{normalem}{ulem}
\usepackage{ulem}
\usepackage[unicode=true,
 bookmarks=false,
 breaklinks=false,pdfborder={0 0 1},backref=false,colorlinks=true]
 {hyperref}
\hypersetup{pdftitle={Title},
 plainpages=false,pdfpagelabels,linkcolor=blue,urlcolor=blue,citecolor=blue,pdfdisplaydoctitle=true,pdfduplex=DuplexFlipLongEdge}

\makeatletter

\newcommand{\lyxdot}{.}

\providecolor{lyxadded}{rgb}{0,0,1}
\providecolor{lyxdeleted}{rgb}{1,0,0}

\DeclareRobustCommand{\lyxdeleted}[3]{{\texorpdfstring{\color{lyxdeleted}\lyxsout{#3}}{}}}
\DeclareRobustCommand{\lyxsout}[1]{\ifx\\#1\else\sout{#1}\fi}

\usepackage{units}\usepackage{wasysym}

\definecolor{orange}{rgb}{0.50, 0.20, 0.0}

\newcommand{\beginsupplement}{%
	\setcounter{page}{1}
	 \renewcommand{\thepage}{SM - \arabic{page}}%
        \setcounter{table}{0}
        \renewcommand{\thetable}{S\arabic{table}}%
        \setcounter{figure}{0}
        \renewcommand{\thefigure}{S\arabic{figure}}%
        \setcounter{section}{0}
        \renewcommand{\thesection}{S\arabic{section}}%
        \setcounter{section}{0}
        \renewcommand{\thesection}{S\arabic{section}}%
        \setcounter{subsection}{0}
        \renewcommand{\thesubsection}{S\arabic{section}.\arabic{subsection}}%
        \setcounter{equation}{0}
        \renewcommand{\theequation}{S\arabic{equation}}%

     }

\usepackage{bm}
\usepackage{braket}

\makeatother

\begin{document}
\noindent\begin{minipage}[t]{1\columnwidth}%
\global\long\def\ket#1{\left| #1\right\rangle }%

\global\long\def\bra#1{\left\langle #1 \right|}%

\global\long\def\kket#1{\left\Vert #1\right\rangle }%

\global\long\def\bbra#1{\left\langle #1\right\Vert }%

\global\long\def\braket#1#2{\left\langle #1\right. \left| #2 \right\rangle }%

\global\long\def\bbrakket#1#2{\left\langle #1\right. \left\Vert #2\right\rangle }%

\global\long\def\av#1{\left\langle #1 \right\rangle }%

\global\long\def\tr{\text{tr}}%

\global\long\def\Tr{\text{Tr}}%

\global\long\def\pd{\partial}%

\global\long\def\im{\text{Im}}%

\global\long\def\re{\text{Re}}%

\global\long\def\sgn{\text{sgn}}%

\global\long\def\Det{\text{Det}}%

\global\long\def\abs#1{\left|#1\right|}%

\global\long\def\up{\uparrow}%

\global\long\def\down{\downarrow}%

\global\long\def\vc#1{\mathbf{#1}}%

\global\long\def\bs#1{\boldsymbol{#1}}%

\global\long\def\t#1{\text{#1}}%
\end{minipage}
\title{Hidden dualities in 1D quasiperiodic lattice models}
\author{Miguel Gonçalves}
\affiliation{CeFEMA, Instituto Superior Técnico, Universidade de Lisboa, Av. Rovisco
Pais, 1049-001 Lisboa, Portugal}
\author{Bruno Amorim}
\affiliation{Centro de Física das Universidades do Minho e Porto, University of
Minho, Campus of Gualtar, 4710-057, Braga, Portugal}
\author{Eduardo V. Castro}
\affiliation{Centro de Física das Universidades do Minho e Porto, Departamento
de Física e Astronomia, Faculdade de Ciências, Universidade do Porto,
4169-007 Porto, Portugal}
\affiliation{Beijing Computational Science Research Center, Beijing 100193, China}
\author{Pedro Ribeiro}
\affiliation{CeFEMA, Instituto Superior Técnico, Universidade de Lisboa, Av. Rovisco
Pais, 1049-001 Lisboa, Portugal}
\affiliation{Beijing Computational Science Research Center, Beijing 100193, China}
\begin{abstract}
We find that quasiperiodicity-induced transitions between extended
and localized phases in generic 1D systems are associated with hidden
dualities that generalize the well-known duality of the Aubry-André
model. These spectral and eigenstate dualities are locally defined
near the transition and can, in many cases, be explicitly constructed
by considering relatively small commensurate approximants. The construction
relies on auxiliary 2D Fermi surfaces obtained as functions of the
phase-twisting boundary conditions and of the phase-shifting real-space
structure. We show that, around the critical point of the limiting
quasiperiodic system, the auxiliary Fermi surface of a high-enough-order
approximant converges to a universal form. This allows us to devise
a highly-accurate method to obtain mobility edges and duality transformations
for generic 1D quasiperiodic systems through their commensurate approximants.
To illustrate the power of this approach, we consider several previously
studied systems, including generalized Aubry-André models and coupled
Moiré chains. Our findings bring a new perspective to examine quasiperiodicity-induced
extended-to-localized transitions in 1D, provide a working criterion
for the appearance of mobility edges, and an explicit way to understand
the properties of eigenstates close to and at the transition.
\end{abstract}
\maketitle
\tableofcontents{}

\newpage
\section{Introduction}

Quasiperiodic systems (QPS) offer a rich playground to study localization
without the presence of disorder. Like disorder, quasiperiodicity
breaks translational invariance, but the nature of both phenomena
is quite distinct - QPS are deterministic, while disordered systems
are random and leads to drastically different localization properties.
For systems with uncorrelated disorder, Anderson localization-delocalization
transitions \citep{Anderson1958} are only possible in 3D. At lower
dimensions, any degree of disorder localizes all eigenstates in the
non-interacting limit \citep{Abrahams1979,MacKinnon1981} as long
as no special symmetries are broken. For QPS, on the other hand, localization-delocalization
transitions are possible even in 1D \citep{AubryAndre}.

QPS host a plethora of exotic phenomena, manifested not only in nature
of the possible localized/de-localized phases they can host , but
also in nontrivial topological properties \citep{Kraus2012,PhysRevLett.109.116404,Verbin2013}.
Interest in QPS has been recently renewed due to their experimental
relevance in optical \citep{PhysRevA.75.063404,Roati2008,Modugno_2009,Schreiber842,Luschen2018,PhysRevLett.123.070405,PhysRevLett.125.060401,PhysRevLett.126.110401}
and photonic lattices \citep{Lahini2009,Kraus2012,Verbin2013,PhysRevB.91.064201,Wang2020},
cavity-polariton devices \citep{Goblot2020Nature} and the promising
recent developments in Moiré systems (see review in Ref.$\,$\citep{Balents2020}
and references therein). For the latter, effects of quasiperiodicity
are often overlooked, but recent studies have shown that in some regimes
these may have important consequences to localization and transport
\citep{PhysRevB.100.144202,Park2018,Carr2020,Fu2020,PhysRevB.101.235121,goncalves2020}.

The first steps to understand quasiperiodic-induced localization were
taken by Aubry and André who considered a nearest-neighbor tight-binding
chain with on-site energies modulated by a sinusoid-the now celebrated
Aubry-André model(AAM) \citep{AubryAndre}. Actually, this model was
previously considered in a different context, to study the peculiar
self similar energy spectrum of Bloch electrons under a uniform magnetic
field \citep{Harper_1955,Hofstadter1976}. Aubry and André showed
that the AAM hosts a localization-delocalization transition arising
simultaneously for all energies. This transition is a consequence
of the real-space/momentum-space duality in the AAM's Hamiltonian.
Generalizations of this model in 1D typically give rise to single-particle
mobility edges \citep{suslov1982localization,PhysRevLett.104.070601,PhysRevB.83.075105,PhysRevLett.113.236403,Liu2015,PhysRevLett.114.146601,Gopalakrishnan2017,Li2017,Szabo2018,PhysRevB.99.054211,PhysRevB.101.064203,Liu2017,PhysRevB.104.064201}
separating regions of the spectrum corresponding to localized and
delocalized states. These may be highly non-trivial to determine and
can only be analytically predicted for few fine-tuned models \citep{PhysRevB.43.13468,PhysRevLett.104.070601,Liu2015,PhysRevLett.114.146601,Wang2020}.
For higher dimensions, self-dual generalizations of the AAM show more
exotic features around the self-dual points: intermediate phases,
for which the wave function is delocalized both in real and momentum
space and transport is diffusive \citep{Devakul2017,PhysRevB.100.144202};
and peculiar partially extended states \citep{Szabo2020}. The interacting
version of the AAM was also recently studied. Here, many-body-localization
for weak interactions and an intermediate regime with slow dynamics
were found both theoretically \citep{Iyer2013,PhysRevA.90.061602,PhysRevLett.121.206601,PhysRevB.100.104203,Xu2019}
and experimentally \citep{Schreiber842,PhysRevLett.119.260401,PhysRevLett.122.170403}.

Even though the study of non-interacting 1D QPS dates back to the
1980s, a simple framework to obtain phase diagrams for generic, non
fine-tuned models and to understand the main ingredients behind localization-delocalization
transitions is still lacking.

In this paper we argue that localization-delocalization \footnote{Note that from this point on, we use the term ``localization-delocalization
transitions'' to refer to transitions between non-critical extended
and localized phases.} transitions in generic 1D QPS occur due to hidden dualities that
are generalizations of the Aubry-André duality. With this insight,
we developed a method to compute mobility edges and duality transformations
through commensurate approximants (CA) of the target 1D QPS. To illustrate
the power of the method, we obtained mobility edges and duality transformations
for a number of models, including generalized Aubry-André models \citep{PhysRevLett.104.070601,PhysRevB.83.075105,PhysRevLett.114.146601}
and coupled Moiré chains, the paradigmatic example of a 1D Moiré system
\citep{Carr2020}. Our results established that dualities between
localized and extended phases, up to now found for only a few fine-tuned
models, are generic for 1D QPS and shows how they can be constructed.

The paper is organized as follows: in Sec.~\ref{sec:WarmUp} we introduce
a minimal set of concepts and give an summarized account of our work
and main findings; in Sec.~\ref{sec:WarmUp} we review in detail
the duality in the Aubry-André model, paving the way for the discussion
of more general hidden dualities. In Sec.$\,$\ref{sec:ModelMethods},
we provide details on the EAAM and LM models used throughout the paper.
Sec.$\,$\ref{sec:SpectralDuality} details how the phase diagram
of a QPS can be known by studying the generalized energy bands of
CA. In Sec.$\,$\ref{sec:DualityTransformation} we provide a complete
definition of our generalized duality transformation for CA and explain
how to extract from it the duality transformation of the limiting
QPS. We show that our duality transformation reduces to the thermodynamic-limit
duality in a model for which the exact transformation is known \citep{PhysRevLett.114.146601}.
We finally apply our duality transformation to models for which no
duality symmetry was shown to exist so far.

\section{Main Results}

\label{sec:main_results}

\textit{Aubry-André duality ---} We start by briefly reformulating
the fundamentals of the Aubry-André duality that will be important
to understand the generalized dualities introduced afterwords. The
Hamiltonian for the AAM is given by \citep{AubryAndre}

\begin{equation}
H=t\sum_{n}\left(c_{n+1}^{\dagger}c_{n}+\text{h.c.}\right)+\sum_{n}V\cos\left(2\pi\tau n+\phi\right)c_{n}^{\dagger}c_{n},\label{eq:AAM_H}
\end{equation}

\noindent where the first term describes the hopping between nearest-neighbors,
the second term is the quasiperiodic potential with strength $V$
and $\tau=a_{1}/a_{2}$ is the ratio between the nearest-neighbor
distance, i.e. the lattice constant, $a_{1}$ and the potential wavelength
$a_{2}$. From this point on, we measure lengths in units of $a_{1}$
by setting $a_{1}=1$, unless otherwise specified.

\noindent We consider a CA of an infinite system with irrational $\tau$
by taking rational approximations, $\tau_{c}=n_{2}/n_{1}\simeq\tau$,
where $n_{1}$ and $n_{2}$ are two co-prime integers \citep{suslov1982localization,Szabo2018}.
The corresponding commensurate system is still infinite but becomes
periodic with a unit cell of length $a_{S}=n_{1}a_{1}=n_{2}a_{2}$
containing $n_{1}$ sites. Being periodic, the Hamiltonian can be
block-diagonalized using Bloch's theorem. The Bloch Hamiltonian depends
on the momenta $k\in\left]-\pi/n_{1},\pi/n_{1}\right]$, and is periodic
under $k\rightarrow k+2\pi/n_{1}$. The block-diagonal sectors are
labeled by the rescaled momentum $\kappa=n_{1}k\in\left]-\pi,\pi\right]$,
measured in units of $a_{S}^{-1}$.\lyxdeleted{Miguel}{Sun Jun  5 14:14:45 2022}{ }

\noindent We also define the rescaled shift variable $\varphi=n_{1}\phi$.
As we will see below, the energy bands are also periodic in $\varphi$
with a period $\Delta\varphi=2\pi$ (or $\Delta\phi=2\pi/n_{1}$).
Importantly, the variables $\varphi$ and $\kappa$ are dual for any
CA of the AAM and the extended and localized phases, as well as the
critical point, can be inferred by studying the dependence of the
energy bands on these variables. To do so, we define the generalized
energy bands $E_{n}(\varphi,\kappa)$, with $n=0,\cdots,n_{1}-1$.
The generalized Fermi surfaces (FS) in the $(\varphi,\kappa)$ plane
are defined by the constant-energy curves $E_{n}(\varphi,\kappa)=E$
.

The Aubry-André duality refers to a mapping between localized and
extended wave functions \citep{AubryAndre}. In the potential-energy,
$V-E$, phase diagram, points $P=(V,E)$ and $P'=(V',E')$ are dual
if $(V',E')=(4/V,2E/V)$, where here and in the following discussion,
we set $t=1$. Exactly at the critical point, $|V|=2$, the wave function
is equal to its Aubry-André dual. This point is therefore commonly
referred to as a self-dual point. For a CA of the AAM, this duality
still holds, but only under a suitable interchange of $\varphi$ and
$\kappa$. Therefore extending AAM duality to commensurate structures
naturally suggests a duality between the phases $\varphi$ and $\kappa$.
We will prove all these claims in detail bellow.

\begin{figure}[h]
\centering{}\includegraphics[width=1\columnwidth]{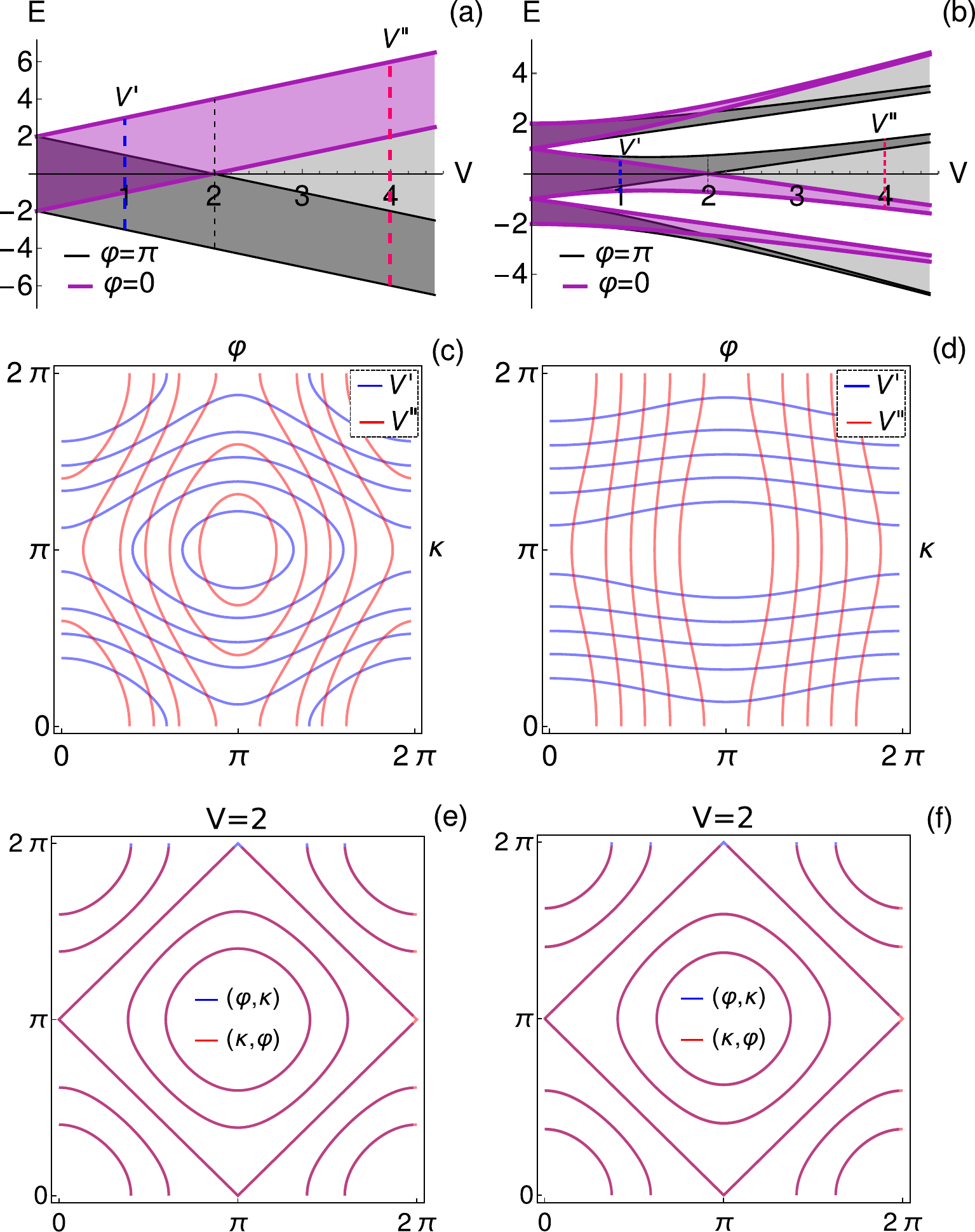}\caption{(a,b) Energy bands for commensurate approximants of the AAM, Eq.$\,$\ref{eq:AAM_H}.
Results for $\varphi=0$ (purple shaded) and $\varphi=\pi$ (dark
grey shaded) and integrated over $\kappa$ for $\tau_{c}=1$ (a) and
$\tau_{c}=2/3$ (b), corresponding respectively to one and three sites
per unit cell. The light grey shading at $V>2$ denotes the energies
that are swept when $\varphi$ is varied between $\varphi=0$ and
$\varphi=\pi$. (c,d) Generalized Fermi surfaces in the $(\varphi,\kappa)$
plane for the dual parameters $V'=1$ and $V''=4$, for the energy
bands indicated in (a,b) respectively. The results in Figs.$\,$(c),(d)
are for $\tau_{c}=1$ and $\tau_{c}=2/3$, respectively. Note that
the contours obtained for $V=V'$ are the same as the ones obtained
for $V=V''$ if $\varphi$ and $\kappa$ are interchanged. (e,f) FS
obtained at the self-dual point $V=2$ for $\varphi$ ($\kappa$)
in the horizontal axis, in blue (red). The results in Figs.$\,$(e),(f)
are for $\tau_{c}=1$ and $\tau_{c}=2/3$, respectively. Note that
the contours are self-dual under switching $\varphi$ and $\kappa$.\label{fig:0}}
\end{figure}

In Fig.$\,$\ref{fig:0} we illustrate the duality of the FS in the
$(\varphi,\kappa)$ plane for very simple CA of the AAM (with one-
and three-site unit cells). At dual points the FS are the same upon
interchanging $\varphi$ and $\kappa$. Moreover, as the unit cell
of the CA is increased, the dependence of the FS on $\varphi$ ($\kappa$)
decreases with respect to the dependence on $\kappa$ ($\varphi$)
in the extended (localized) phase of the limiting QPS {[}compare Figs.$\,$\ref{fig:0}(c,d){]}.
Exactly at the self-dual point, the FS are invariant under interchanging
$\varphi$ and $\kappa$ {[}Figs.$\,$\ref{fig:0}(e,f){]}. The $\varphi\leftrightarrow\kappa$
duality holds irrespectively of the chosen CA.

\begin{figure}[H]
\centering{}\includegraphics[width=1\columnwidth]{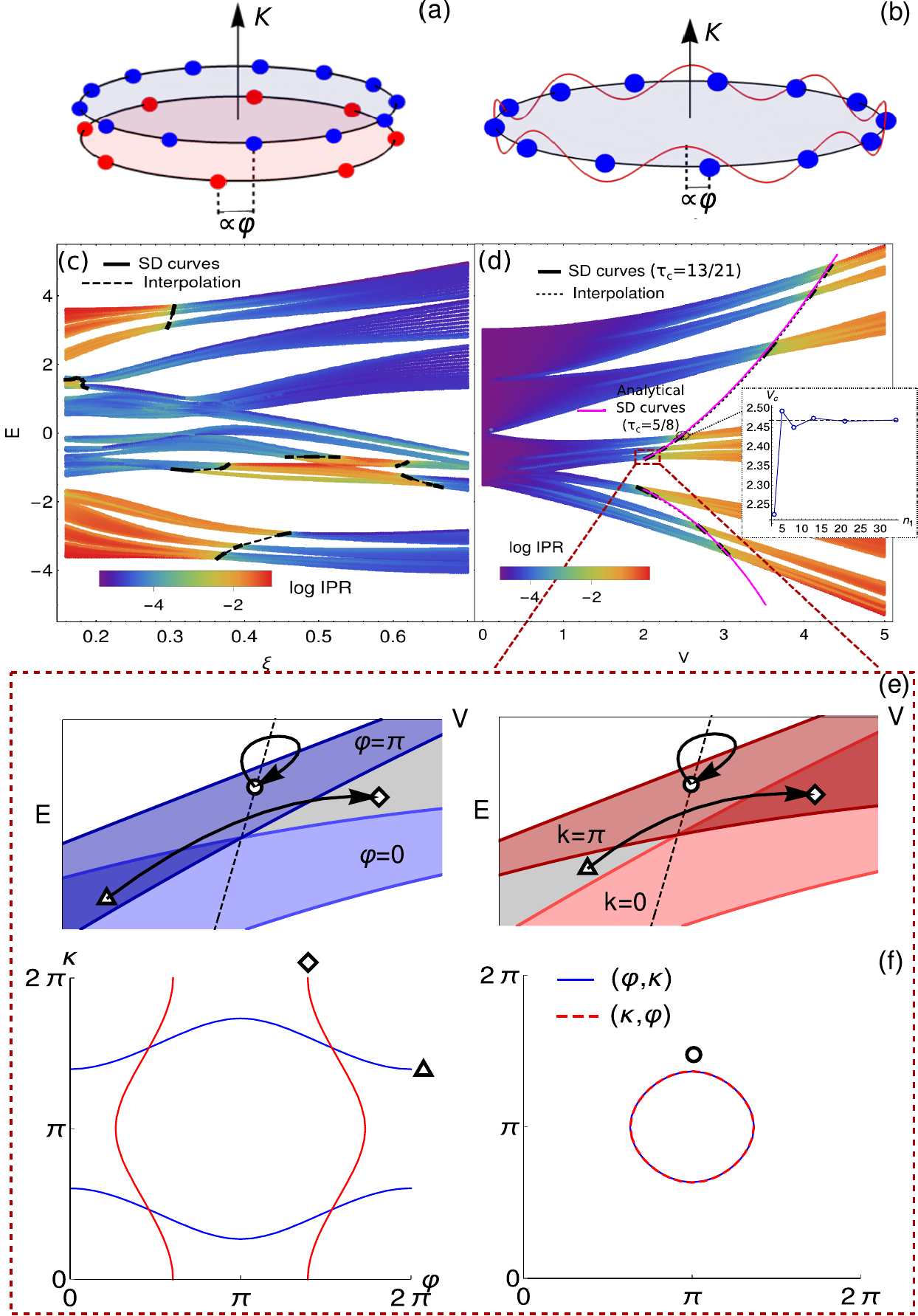}\caption{In this figure we approach a limiting QPS with $\tau=(\sqrt{5}-1)/2$.
(a,b) Illustration of the dual parameters $k$ and $\phi$ for a periodic
ladder (a) and a periodic chain (b). These examples are rings containing
unit cells of CA with $\tau_{c}=8/13$. (c,d) Numerical results for
the energy spectrum and ${\rm IPR}$ of an EAAM (LM with gaussian
decaying interlayer hoppings) as a function of potential's strength
$V$ (interlayer hopping's decay length $\xi$), for $\tau_{i}=89/144$.
EAAM: Aubry-André model with additional next-nearest-neighbor hopping
$t_{2}=0.5$, Eq.$\,$\ref{eq:EAAM_H}; LM: Ladder model with interlayer
hopping strength $V=3$ and cutoff $\Lambda=2.5a$, Eq.$\,$\ref{eq:ILM_H}.
In black, we plot curves of SD points for the CA with $\tau_{c}=13/21$
(obtained numerically) and in magenta, approximate analytical SD curves
using the CA with $\tau_{c}=5/8$. The inset contains the comparison
between the SD points computed for CA with $\tau_{c}=F_{n}/F_{n+1}$
and with unit cell of size $n_{1}=F_{n+1}$, with $F_{n}$ the Fibonnacci
number of order $n$, and the estimation of the critical point through
the IPR results for a system with $\tau=4181/6765$ and $L=6765$
sites. The definitions for the parameters here mentioned, including
the ${\rm IPR}$, $\tau_{i}$ and model parameters, are given in Sec.$\,$\ref{sec:ModelMethods}.
(e) Close-up of an energy band for $\tau_{c}=13/21$ {[}the range
of parameters shown is contained in the red dashed box in (d){]}.
The energy bands obtained for $\varphi=0,\pi$ ($\kappa=0,\pi$) and
integrated over $\kappa$ ($\varphi$) are shown in blue (red) to
illustrate the duality encoded in $\kappa,\varphi$. A generalized
duality transformation for a CA maps points in the extended and localized
phases of the limiting QPS (dual points, e.g. triangle and diamond
markers), switching the roles of phases $\kappa$ and $\varphi$.
The fixed points of this transformation (SD points, e.g. circle marker)
are the manifestation of the mobility edge of the limiting QPS. (f)
FS in the $(\varphi,\kappa)$ plane for the points indicated in (e).
For the SD point (circle) the FS is shown with the standard, $(\varphi,\kappa)$,
and switched, $(\kappa,\varphi)$, axis in order to show their perfect
agreement. \label{fig:main}}
\end{figure}

\textit{General dualities} --- A brief overview of our findings and
their application is summarized in Fig.$\,$\ref{fig:main}. We consider
two simple systems, an extended Aubry-André model (EAAM) and a ladder
model (LM) consisting of two coupled incommensurate chains. For these
two classes of models, $\tau$ is either the ratio between the lattice
constant, $a_{1}$, and the QP potential's wavelength for EAAM, $a_{2}$;
or the ratio between the two lattice constants for LM, $a_{1}$ and
$a_{2}$: $\tau=a_{1}/a_{2}$. Despite the simplicity of these examples,
we argue that the same basic mechanism extends to more complex 1D
systems, including the so-called mosaic-models \citep{Wang2020a,Duthie2020}
for which self-duality is believed to be absent.\lyxdeleted{Miguel}{Sun Jun  5 14:14:45 2022}{ }

Examples of CA for EAAM and LM are shown pictorially as rings containing
a single unit cell in Figs.$\,$\ref{fig:main}(a,b). The spectrum
for a large enough unit cell can be seen in Fig.$\,$\ref{fig:main}(c)
and \ref{fig:main}(d), respectively for LM and EAAM. We also compute
the inverse participation ratio (IPR, see definition in Sec.$\,$\ref{sec:ModelMethods}),
that distinguishes the extended (bluish) and localized (reddish) phases.

For a given CA we can again define the rescaled momentum $\kappa=n_{1}k\in\left]-\pi,\pi\right]$.
The Bloch Hamiltonian is equivalent to a single unit cell of the CA
forming a ring with periodic boundary conditions that is threaded
by a flux $\kappa$, as depicted in Figs.$\,$\ref{fig:main}(a,b).
$\kappa$ is also commonly mentioned as a phase twist. Besides the
periodicity in Bloch momentum, the quasiperiodic models possess an
additional periodicity related to the phase of the QP potential in
EAAM and the stacking between the two chains in the LM. Both cases
can be seen as a slide, or shift, degree of freedom between the potential
and the lattice (in EAAM) of the two lattices (in LM) as depicted
respectively in panels b) and a) of Fig.~\ref{fig:main} . The crucial
point is that, for a given CA, for certain shifts smaller than the
CA unit cell, the system is left invariant (up to a relabeling of
the sites). As we did for the AAM, we encode this slide periodicity
into a rescaled shift $\varphi\in\left]-\pi,\pi\right]$, such that
the system is invariant under $\varphi\rightarrow\varphi+2\pi$.

We found that hidden dualities between the variables $\kappa$ and
$\varphi$ of CA's are at the root of localization-delocalization
transitions in 1D QPS. The identification of such dualities allows
to determine mobility edges. As the size of the CA unit cell increases
the mobility edges approach the ones of the true QPS. Surprisingly,
CA with relatively small unit cells already provide an excellent approximation
to the mobility edges of the limiting QPS.

As for the AAM, the hidden dualities can be identified through the
generalized energy bands $E(\varphi,\kappa)$. Generalized duality
transformations can be obtained for each band $E(\varphi,\kappa)$,
as depicted in Fig.$\,$\ref{fig:main}(e). There, we see that the
$\kappa$-dependent energy dispersion in the extended phase of the
limiting QPS is dual of the $\varphi$-dependent energy dispersion
in the localized phase. This becomes even more clear in Fig.$\,$\ref{fig:main}(f),
where we show the generalized FS. This feature is a consequence of
the duality transformation that maps wave functions at points in the
extended phase to wave functions at their dual points in the localized
phase, exchanging the roles of phases $\varphi$ and $\kappa$, just
like we have seen for the AAM. The generalized FS of $E(\varphi,\kappa)$
at these dual points are related by a suitable interchange of $\varphi$
and $\kappa$. At the fixed points of the duality transformation,
the FS are invariant under this interchange. Such self-dual (SD) points
are the manifestation of the critical points of the limiting QPS in
its CA. In the limit that $\tau_{c}\rightarrow\tau$ (infinite-size
unit cell), the duality transformation of the CA reduces to the duality
transformation of the limiting QPS.

The lines of SD points for CA provide remarkably accurate descriptions
of the mobility edges of limiting QPS, in many cases even for CA with
relatively small unit cells. Examples are shown in Fig.$\,$\ref{fig:main}(c,d)
for an EAAM and a LM. The SD curves obtained for a CA with $\tau_{c}=13/21$
match very accurately the mobility edge separating extended (blue)
and localized (red) phases obtained numerically via IPR. In the inset
of Fig.$\,$\ref{fig:main}(d) we can also see that the predictions
of SD points converge to the critical point computed through the IPR
using large system sizes for CA with relatively small unit cell size.
Moreover, it is possible in some regimes to compute very accurate
approximate analytical SD curves for CA with smaller unit cells, as
shown in Fig.$\,$\ref{fig:main}(d) for $\tau_{c}=5/8$.

For the models with exact mobility edges in Refs.$\,$\citep{PhysRevLett.114.146601,Wang2020a},
the SD curves are exact for any CA and CA-independent, perfectly matching
the mobility edges of the limiting QPS. In these cases, the mobility
edge can be obtained through the simplest possible CA. Additionally,
we show in Sec.$\,$\ref{sec:DualityTransformation} that it is possible
to use CA to define a generalized duality transformation that maps
eigenstates of QPS at dual or SD points.

Our findings establish a strong connection between the localization-delocalization
transitions in widely different 1D QPS in terms of dualities that
are believed to be absent away from fine-tuned models like the AAM.
They not only provide a simple way to characterize the phase diagrams
of these systems in terms of CA, but also to understand whether a
simple duality transformation can be defined for a given model.

\section{Aubry-André Model for Commensurate Structures}

\label{sec:WarmUp}

In this section we review and prove the existence of the Aubry-André
duality for commensurate structures, to justify our claims in the
previous section. We again take the commensurate Aubry-André Hamiltonian
in Eq.$\,$\ref{eq:AAM_H} by setting $\tau=\tau_{c}=n_{2}/n_{1}$,
but now consider a periodic system formed by $N$ supercells. Expressing
the label of each site as $n=m+rn_{1}$, where $m=0,...,n_{1}-1$
runs over sites within a unit cell and $r=0,...,N-1$ runs over supercells,
we can write the electron destruction operator as
\begin{equation}
c_{n}\equiv c_{r,m}=\frac{1}{\sqrt{N}}\sum_{k}e^{ik\left(n_{1}r+m\right)}\tilde{c}_{m}(k)\label{eq:c_tilde}
\end{equation}
where $k=2\pi j/(n_{1}N)$, with $j=0,...,N-1$, is the Bloch momentum.
With this change, the Hamiltonian becomes block diagonal, $H=\sum_{k}H(\phi,k)$,
where $H(\phi,k)$ depends parametrically on $k$ and $\phi$:
\begin{multline}
\tilde{H}(\phi,k)=t\sum_{m=0}^{n_{1}-1}\left(e^{-ik}\tilde{c}_{m+1}^{\dagger}(k)\tilde{c}_{m}(k)+e^{ik}\tilde{c}_{m}^{\dagger}(k)\tilde{c}_{m+1}(k)\right)\\
+\sum_{m=0}^{n_{1}-1}V\cos\left(2\pi\tau_{c}m+\phi\right)\tilde{c}_{m}^{\dagger}(k)\tilde{c}_{m}(k),
\end{multline}
where we used the fact that for $\tau_{c}=n_{2}/n_{1}$, $\cos\left(2\pi\tau_{c}\left(m+rn_{1}\right)+\phi\right)=\cos\left(2\pi\tau_{c}m+\phi\right)$.
It can be seen that $H(\phi,k)$ is left invariant (up to a relabeling
of the sites) under a change $\phi\rightarrow\phi+2\pi/n_{1}$, and
(up to a gauge transformation) under a change $k\rightarrow k+2\pi/n_{1}$.
This allows us to defined the rescaled momenta and shift as $\kappa=n_{1}k$
and $\varphi=n_{1}\phi$, respectively. Writing the eigenstates of
$H(\phi,k)$ as 
\begin{eqnarray}
\left|\psi\right\rangle  & = & \sum_{m}\psi_{m}^{\text{r}}(\phi,k)\tilde{c}_{m}^{\dagger}(k)\left|0\right\rangle ,\label{eq:psi's}
\end{eqnarray}
the amplitudes $\psi_{m}^{\text{r}}(\phi,k)$, where $\t r$ stands
for real space, satisfy the equation
\begin{multline}
te^{-ik}\psi_{m-1}^{\text{r}}(\phi,k)+te^{ik}\psi_{m+1}^{\text{r}}(\phi,k)+\\
+V\cos\left(2\pi\tau_{c}m+\phi\right)\psi_{m}^{\text{r}}(\phi,k)=E\psi_{m}^{\text{r}}(\phi,k),\label{eq:AAM_real_space}
\end{multline}
where the energies depend parametrically on $\varphi$ and $\kappa$,
$E=E(\varphi,\kappa)$. It is easy to see that the hopping term would
be diagonalized with a further Fourier expansion in the $m$ indices.
However, by writing the Aubry-André potential as $V\cos\left(2\pi\tau_{c}m+\phi\right)=V\cos\left(gx_{m}+\phi\right)$,
where $x_{m}=m$ is the position of site $m$ (in units of $a_{1}$)
and $g=2\pi\tau_{c}$ (in units of $a_{1}^{-1}$) is the wavenumber
of the potential, we can see that it will couple different momentum
states by increments of $\pm g$. This motivates the following transformation
\begin{align}
\psi_{m}^{\text{r}}(\phi,k) & =\frac{1}{\sqrt{n_{1}}}\sum_{q=0}^{n_{1}-1}e^{i2\pi\tau_{c}qm}\psi_{q}^{\t d}(\phi,k),\label{eq:AA_transformation}
\end{align}
where $\t d$ stands for dual. The equation for the amplitudes $\psi_{q}^{\t d}(\phi,k)$
thus becomes 
\begin{multline}
2t\cos\left(2\pi\tau_{c}q+k\right)\psi_{q}^{\t d}(\phi,k)+\\
+\frac{V}{2}e^{i\phi}\psi_{q-1}^{\t d}(\phi,k)+\frac{V}{2}e^{-i\phi}\psi_{q+1}^{\t d}(\phi,k)=E\psi_{q}^{\t d}(\phi,k).\label{eq:AAM_momentum_space}
\end{multline}
Comparing Eqs.~\eqref{eq:AAM_real_space} with \eqref{eq:AAM_momentum_space},
we see that under the transformation Eq.~\eqref{eq:AA_transformation},
the roles played by $k$ and $\phi$ (or by $\kappa$ and $\varphi$)
are exchanged. Furthermore, if $V=2t$ the model becomes self-dual.
It is this duality that is at the heart of the localization-delocalization
transition in the Aubry-André model \citep{AubryAndre}. In this work,
we will see how generalized hidden dualities based on the dual roles
played by rescaled shift and momentum, $\varphi$ and $\kappa$, can
be used to determine localization-delocalization transition in general
one-dimensional QPS.

\section{Models and Methods}

\label{sec:ModelMethods}

\begin{figure}[h]
\centering{}\includegraphics[width=1\columnwidth]{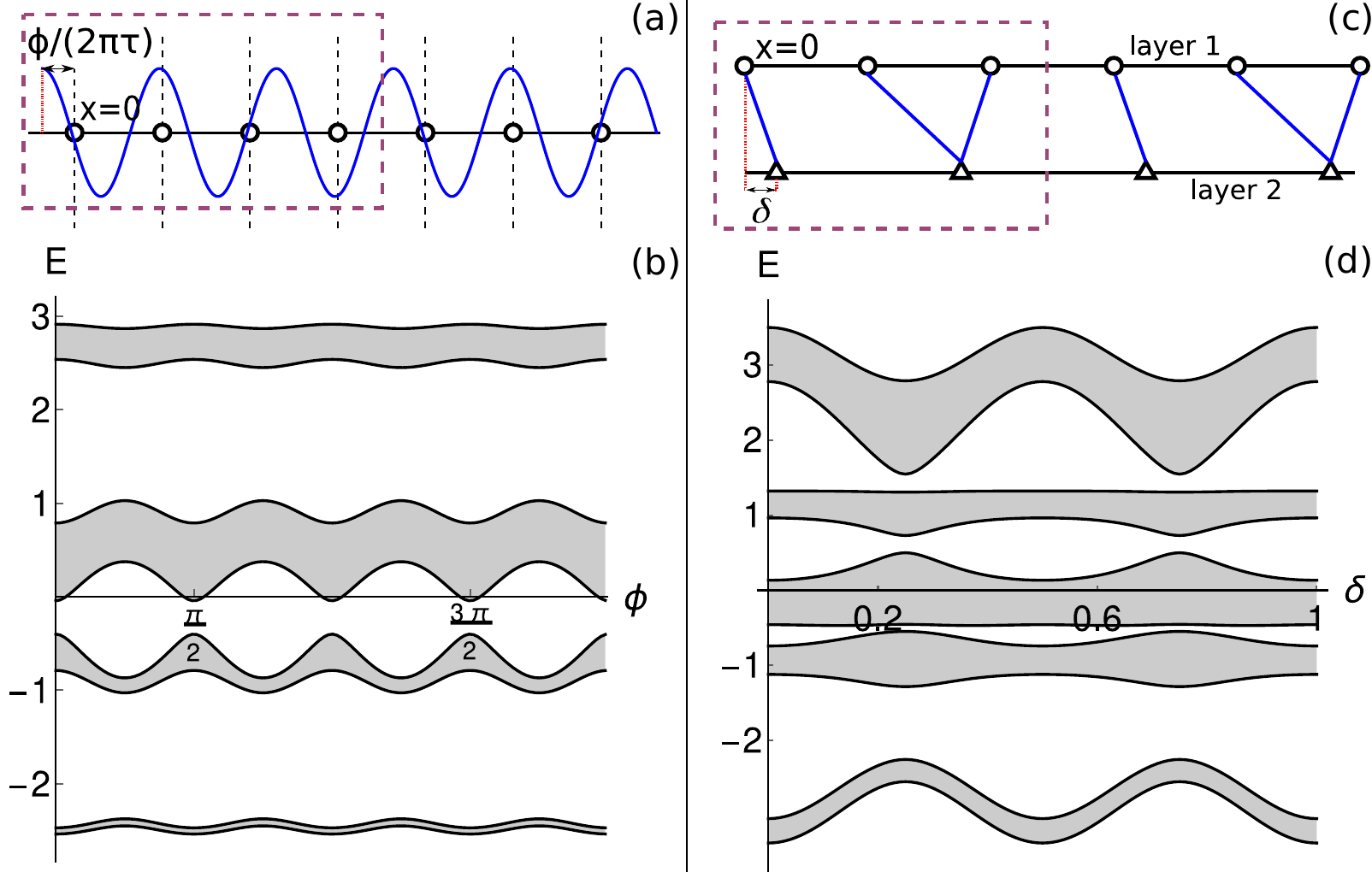}\caption{(a) Sketch of a commensurate EAAM, Eq.$\,$\ref{eq:EAAM_H}. In blue
we represent the potential $\mathcal{V}(2\pi\tau n+\phi)$ for $\tau_{c}=n_{2}/n_{1}=3/4$.
In this case, the unit cell depicted by the purple dashed box is composed
by 4 sites. (b) Band edges for the commensurate system in (a) as a
function of phase $\phi$. We considered the AA-NNN, with next-nearest-neighbor
hoppings of strength $t_{2}=0.2$ and potential of strength $V=1.75$.
(c) Sketch of the LM, Eq.$\,$\ref{eq:ILM_H}, for $\tau_{c}=n_{2}/n_{1}=2/3$.
In this case, the unit cell, depicted by the purple dashed box, has
$5$ sites. (d) Band edges for the commensurate system in (c) as a
function of the displacement $\delta$ defined in there. The parameters
used for the LM were $V=2.5$ (interlayer hopping's strength) and
$\xi=0.25$ (interlayer hopping's decay length). \label{fig:sketch_and_periodic}}
\end{figure}

For the remainder of this work, we will focus on EAAM and LM introduced
previously. Explicitly, the Hamiltonian for the family of EAAM is

\begin{equation}
H=\sum_{n}\mathcal{V}(2\pi\tau n+\phi)c_{n}^{\dagger}c_{n}+\sum_{n}\Big(c_{n}^{\dagger}c_{n+1}+t_{2}c_{n}^{\dagger}c_{n+2}+\textrm{h.c.}\Big)\label{eq:EAAM_H}
\end{equation}

\noindent where the first term contains the quasiperiodic potential,
with $\mathcal{V}(x)$ a $2\pi$-periodic function, and the second
term contains nearest-neighbor and next-nearest-neighbor hoppings.
The ratio between the lattice constant and the wave length of the
quasiperiodic potential is given by $\tau$, and the phase $\phi$
fixes the shift with respect to the first lattice site {[}see Fig.$\,$\ref{fig:sketch_and_periodic}(a){]}.
For the standard AAM, $\mathcal{V}(x)=V\cos(x)$ and $t_{2}=0$. We
also consider a modified AAM (MAAM) \citep{PhysRevLett.114.146601},
where $\mathcal{V}(x)=2V\cos(x)/[1-\alpha\cos(x)]$ and $t_{2}=0$.
In both models, $V$ is the strength of the quasiperiodic potential.
The latter model has an exact mobility edge, in contrast with the
AAM, for which the localization-delocalization transition occurs simultaneously
for all energies at $|V|=2$. Both models are self-dual under different
duality transformations \citep{AubryAndre,PhysRevLett.114.146601},
being associated with $\tau$-independent localization-delocalization
transitions.

If we consider $t_{2}\neq0$, the special self-duality of the previous
models is broken \citep{PhysRevB.83.075105}. In this case, the exact
mobility edge is not known. Therefore, we also consider a next-nearest-neighbor
AA model (AA-NNN), with $\mathcal{V}(x)=V\cos(x)$ and $t_{2}\neq0$.
The EAAM results shown in Fig.$\,$\ref{fig:main}(d) and \ref{fig:main}(f)
are for the AA-NNN with $t_{2}=0.5$.

In the LM, two chains with commensurate or incommensurate lattice
constants are coupled by hopping terms {[}see Fig.$\,$\ref{fig:sketch_and_periodic}(c){]}.
The Hamiltonian can be written as

\begin{equation}
H=\sum_{l,n}c_{ln}^{\dagger}c_{ln}+\sum_{|x_{1n}-x_{2m}|<\Lambda}t_{\perp}(|x_{1n}-x_{2m}|)c_{1n}^{\dagger}c_{2m}+{\rm h.c.}\,,\label{eq:ILM_H}
\end{equation}

\noindent where $c_{ln}^{\dagger}$ creates an electron at site $n$
of chain $l$ and $x_{1n}=a_{1}n$ and $x_{2}=a_{2}n+a_{1}\delta$
is the position of this site in an axis parallel to the layer, with
$\delta$ a shift of layer 2 with respect to layer 1 {[}see Fig.$\,$\ref{fig:sketch_and_periodic}(b){]}.
We consider $t_{\perp}(x)=V\exp(-x^{2}/\xi^{2})$, where $V$ and
$\xi$ are respectively the interlayer hopping strength and decay
length, and again $\tau=a_{1}/a_{2}$. This model is associated with
non-trivial mobility edges for some choices of the parameters. In
what follows, we set $\Lambda=2.5a_{1}$ and $a_{1}=1$.

In order to capture the phase diagram of 1D QPS we analyse its CA,
characterized by the rational $\tau_{c}=n_{2}/n_{1}$. The set of
possible $\tau_{c}$ are the so-called convergents of the irrational
number $\tau$, that can be computed through its continued fraction
expansion. Convergents with larger $n_{1}$ approximate $\tau$ more
accurately. Therefore, we label such convergents as \textit{higher
order approximants}. The unit cell of CA contains $n_{1}$ and $n_{1}+n_{2}$
sites, respectively for the EAAM and the LM, as exemplified in Figs.$\,$\ref{fig:sketch_and_periodic}(a,c).

As discussed in Sec.~\ref{sec:WarmUp}, for a given CA, changing
$\phi\rightarrow\phi+2\pi/n_{1}$ in Eq.$\,$\eqref{eq:EAAM_H} corresponds
only to a redefinition of the unit cell. This follows from the coprimality
between $n_{1}$ and $n_{2}$. Therefore, the Hamiltonian is $\phi$-periodic
with period $2\pi/n_{1}$ {[}see Fig.$\,$\ref{fig:sketch_and_periodic}(b){]}.
With this in mind, we define the rescaled shift as $\varphi=n_{1}\phi$.
A single period of the energy dispersion is covered for $\varphi\in[0,2\pi[$.
In a similar way, we can note that for the LM, the unit cell is redefined
under $\delta\rightarrow\delta+1/n_{2}$ {[}see Fig.$\,$\ref{fig:sketch_and_periodic}(d){]}.
In this case, we define the rescaled shift as $\varphi=2\pi\tau_{c}\delta$.
Note that for more generic models, the definition of $\varphi$ can
change - it should be always bounded by the $\phi(\delta)$-dependent
periodicity of the energy bands (see Sec.$\,$\ref{sec:Discussion}
and Appendix$\,$\ref{sec:Emergence-of-duality}). Notice that the
phase $\phi$ in an EAAM can also be interpreted as a shift $\delta$,
by writing the potential as $\mathcal{V}\left(2\pi\tau_{c}n+\phi\right)=\mathcal{V}\left(2\pi\tau_{c}\left(n-\delta\right)\right)$,
where $\delta=-\phi/(2\pi\tau_{c}).$

In order to characterize the localization properties of eigenstates
of QPS, we also compute the inverse participation ratio (IPR) for
large systems. This quantity is defined for each eigenstate $\ket{\psi(E)}=\sum_{n}\psi_{n}(E)\ket n$,
where $\{\ket n\}$ is a basis localized at each site, as

\begin{equation}
{\rm IPR}(E)=\frac{\sum_{n}|\psi_{n}(E)|^{4}}{(\sum_{n}|\psi_{n}(E)|^{2})^{2}}
\end{equation}
For extended states, we expect ${\rm IPR}(E)\sim L^{-1}$, where $L$
is the number of sites in the system. For localized states, ${\rm IPR}(E)\sim{\rm constant}$.
For a large enough system, the value of the ${\rm IPR}$ in the localized
phase is significantly larger than in the extended phase. In practice,
in order to simulate numerically the IPR of incommensurate systems,
one must consider finite lattices with periodic boundary conditions
in order to avoid defects. Thus, we approximate the incommensurate
system by simulating a single unit cell with periodic boundary conditions
of a CA with $\tau_{i}=n_{2}^{i}/n_{1}^{i}$, for very large $n_{1}^{i}$.
Hereinafter, and even though we will always be considering CA's, we
will denote the commensurability parameter for \textit{high order
approximants} as $\tau_{i}$, which will be considered when studying
IPR; and reserve $\tau_{c}$ for the parameter of a CA with moderate
size, which will be used to construct generalized hidden dualities.

\section{Spectral duality}

In this section, we study the energy bands of CA as a function of
wave vector $\kappa$ and phase $\varphi$. The aim is ultimately
to obtain the phase diagram of the limiting QPS, including possible
mobility edges.

\label{sec:SpectralDuality}

\paragraph{Far from critical point.---}

\begin{figure}[h]
\centering{}\includegraphics[width=1\columnwidth]{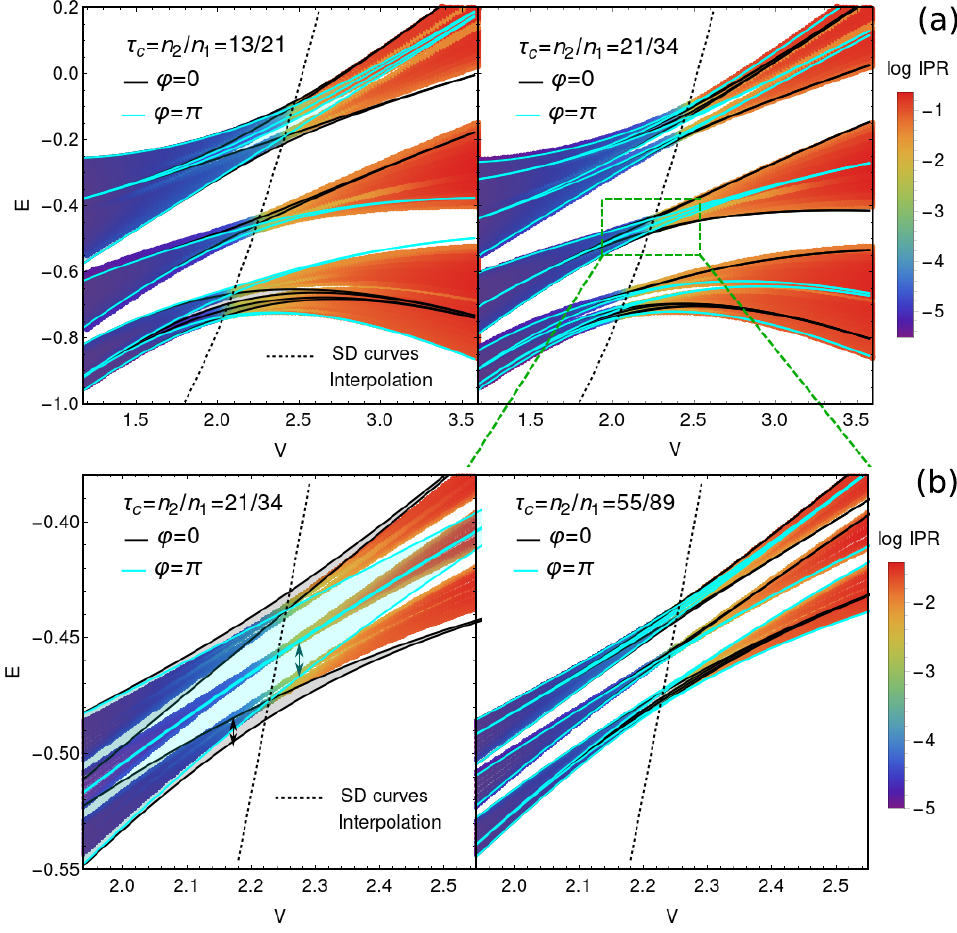}\caption{(a) Band edges for $\varphi=0$ and $\varphi=\pi$, for CA with $\tau_{c}=13/21$
(left) and$\tau_{c}=21/34$ (right), along with the IPR for $\tau_{i}=233/377$.
The light gray and light cyan shading denotes the range of energies
delimited by the band edges of $\varphi=0$ and $\varphi=\pi$ respectively.
Results are for the AA-NNN defined in Eq.$\,$\ref{eq:EAAM_H} and
below it, with next-nearest-neighbor hopping $t_{2}=0.5$ (close-up
of Fig.$\,$\ref{fig:main}(d) for a selected region of parameters
around the critical point). (b) Same as (a), but for a narrower region
of parameters around criticality encompassed by the green dashed box
in (a)-right, and for $\tau_{c}=21/34$ (left) and $\tau_{c}=55/89$
(right). The arrows exemplify band edges that poorly approximate the
region of finite spectral weight. \label{fig:BandEdges_vs_IS}}
\end{figure}

We start by plotting in Fig.$\,$\ref{fig:BandEdges_vs_IS} the band
edges of some CA on top of the phase diagram obtained for the AA-NNN
{[}shown in Fig.$\,$\ref{fig:main}(d){]}, for a selected region
of parameters around the critical point. The band edges are obtained
for the energy bands $E_{n}(\varphi,k)$ that appear in this region
of parameters, for fixed $\varphi=0,\pi$. In the shown examples,
the energy bands for all $\varphi\in[0,2\pi[$ are bounded by the
lowest and highest energy band edges of $\varphi=0$ and $\varphi=\pi$.
In Fig.$\,$\ref{fig:BandEdges_vs_IS}(a), we can see that for a CA
with $\tau_{c}=21/34$, these band edges bound very well the energy
intervals over which a finite density of states (DOS) is observed
for the QPS. However, if we zoom in to a narrower region of parameters
around the critical point, we see in Fig.$\,$\ref{fig:BandEdges_vs_IS}(b)-left
that the bounds are not accurate for $\tau_{c}=21/34$ (some band
edges do not bound any spectral weight, as exemplified with vertical
double arrows). A higher-order approximant is needed for an accurate
bounding, as seen in Fig.$\,$\ref{fig:BandEdges_vs_IS}(b)-right
for $\tau_{c}=55/89$. In fact, the closer we are to the critical
point, the larger the CA's order needed for an accurate bounding.

\begin{figure}[h]
\centering{}\includegraphics[width=1\columnwidth]{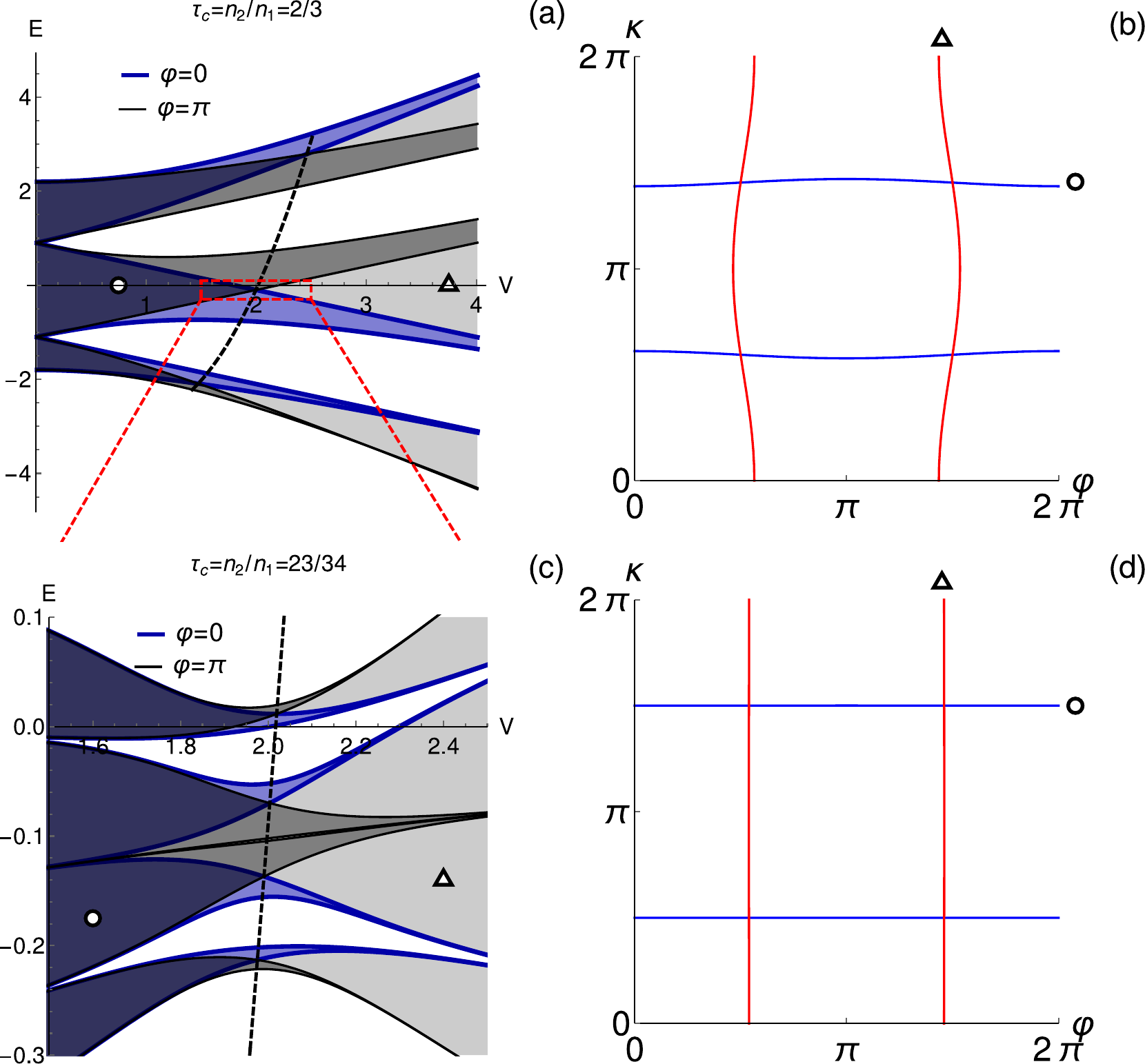}\caption{Energy bands for CA of the irrational $\tau=0.418(\sqrt{5}+1)/2=0.676338\cdots$,
for the AA-NNN defined in Eq.$\,$\ref{eq:EAAM_H} and below it, with
next-nearest-neighbor hopping $t_{2}=0.1$. (a,c) Band edges for the
commensurate approximants $\tau_{c}=n_{2}/n_{1}=2/3$ (a) and $\tau_{c}=n_{2}/n_{1}=23/34$
(c), for $\varphi=0,\pi$. In addition, we plot the SD curves in dashed
black (we interpolate these curves inbetween bands). Note that the
range of energies and $V$ depicted in (c) is contained in the dashed
red box shown in (a). (b,d) FS in the $(\varphi,\kappa)$ plane of
generalized energy bands $E(\varphi,\kappa)$, for the points marked
in (a,c). \label{fig:Ecuts_ext_loc}}
\end{figure}

\begin{figure*}[t]
\centering{}\includegraphics[width=0.75\paperwidth]{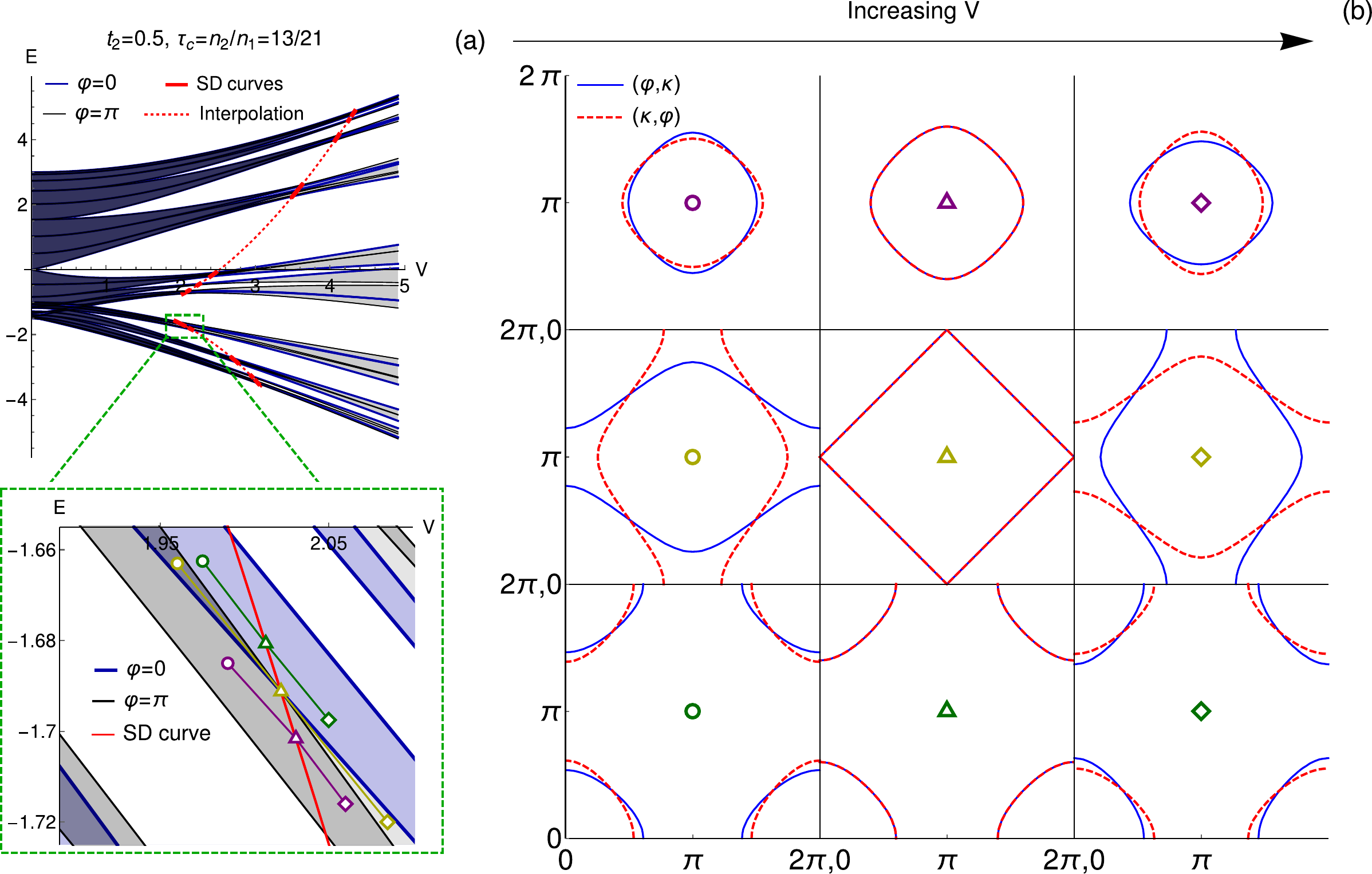}\caption{Example of spectral duality for the AA-NNN, defined in Eq.$\,$\ref{eq:EAAM_H}
and below it. (a) Top: Band edges for the commensurate approximant
$\tau_{c}=n_{2}/n_{1}=13/21$, for $\varphi=0,\pi$ and next-nearest-neighbor
hoppings $t_{2}=0.5$, along with the SD curves. Bottom: Close-up
of randomly chosen bands inside the green dashed box in the upper
figure. The qualitative behaviour of the $\varphi=0,\pi$ bands is
always the same as depicted here: the bands split at a point within
the SD curve. (b) FS for the points depicted at the bottom of Fig.$\,$(a).
The point corresponding to each FS is shown at the center of each
figure. We consider $\varphi$ ($\kappa$) in the $x$($y$)-axis
for the full blue (dashed red) curves. In the SD points, the FS are
invariant under a $\pi/2$-rotation $\mathcal{R}_{0}$ around $(\varphi_{0},\kappa_{0})=(0,0)$
{[}or $(\varphi_{0},\kappa_{0})=(\pi,\pi)${]}, which corresponds
to interchanging $\kappa\leftrightarrow\varphi$. \label{fig:phi_k_curves}}
\end{figure*}

We illustrate how the energy bands of CA behave in the extended and
localized phases of the limiting QPS using the AA-NNN. Fig.$\,$\ref{fig:Ecuts_ext_loc}
depict the energy bands $E_{n}(\varphi,\kappa)$ for CA of the QPS
defined by $\tau=0.418(\sqrt{5}+1)/2=0.676338\cdots$. Figs.$\,$\ref{fig:Ecuts_ext_loc}(a,c),
show the band edges when $\varphi=0$ (blue lines) and $\varphi=\pi$
(black lines) for two CA systems. Filled regions in blue $\left(\varphi=0\right)$
or dark grey $\left(\varphi=\pi\right)$ correspond to states inside
the respective band, while light grey regions indicate all the states
that would appear for $0<\varphi<\pi$. The following observations
can be made:
\begin{enumerate}
\item The energy bands of CA weakly depend on $\varphi$ ($\kappa$) in
the extended (localized) phases of the limiting QPS. On the other
hand, they strongly depend on $\kappa$ ($\varphi$) in the extended
(localized) phases.
\item The energy bands depend equally on $\varphi$ and $\kappa$ at the
SD points;
\item For higher-order approximants, the $\varphi$-dependence ($\kappa$-dependence)
of the energy dispersion decreases abruptly in the extended (localized)
phase. In fact, this decay is exponential in the size of the unit
cell, with a characteristic length scale that correspond to a possible
definition of the correlation length in extended (localized) phase
(see Appendix$\,$\ref{sec:scaling_anal} for details).
\end{enumerate}
Observations 1-to-3 can also be made looking at constant-energy cuts.
Examples are shown in Figs.$\,$\ref{fig:Ecuts_ext_loc}(b,d), where
the FS is observed in points for which $E_{n}(\varphi,\kappa)$ depends
weakly either on $\varphi$ or $\kappa$. The FS corresponds to almost
straight lines for constant $\varphi$ ($\kappa$) and is weakly dependent
on $\kappa$ ($\varphi$).

The region of parameters for which $E_{n}(\varphi,\kappa)$ depends
significantly both on $\kappa$ and $\varphi$ shrinks around the
SD points as we increase the order of the CA. In the limit of high-order
approximants, there is essentially no $\varphi$-dependence ($\kappa$-dependence)
in the extended (localized) phase except for this narrowing region,
that ultimately reduces to the SD point in the limit $\tau_{c}\rightarrow\tau$.

\paragraph{Close to critical point.---}

We now explore the regions in parameter space near the critical SD
points.

Fig.$\,$\ref{fig:phi_k_curves}(a) show the band edges for the CA
defined by $\tau_{c}=13/21$, for $\varphi=0,\pi$. Together with
the band edges, we plot a set of SD curves in Fig.$\,$\ref{fig:phi_k_curves}(a).
Along these curves, the FS is invariant under a suitable interchange
of $\varphi$ and $\kappa$, to an almost perfect approximation. In
fact, the perfect invariance only arises when $\tau_{c}\rightarrow\tau$
(see Appendices$\,$\ref{subsec:estimate_VR_tR_ER} and \ref{sec:Emergence-of-duality}),
but in practice it can be seen even for low-order CA. The transformation
that interchanges $\varphi\leftrightarrow\kappa$ , and that we denote
$\mathcal{R}_{0}$, corresponds to a $\pi/2$-rotation in the $(\varphi,\kappa)$
plane around some point $(\varphi_{0},\kappa_{0})$. To better illustrate
this point, we zoomed into a generalized band $E(\varphi,\kappa)$
as shown in the bottom panel of Fig.$\,$\ref{fig:phi_k_curves}(a).
Note that the $E(\varphi=0,\kappa)$ and $E(\varphi=\pi,\kappa)$
bands split at a point contained in the SD curve. This feature is
generic. Fig.$\,$\ref{fig:phi_k_curves}(b) depicts the FS for different
parameters specified by points in the bottom panel of Fig.$\,$\ref{fig:phi_k_curves}(a).
At the SD points (middle panel), the FS is invariant under $\mathcal{R}_{0}$.
Figure$\,$\ref{fig:phi_k_curves}(b) also suggests that FS for $E>E_{c}(V)$
can be mapped into FS for $E<E_{c}(V)$ upon this rotation, with $E_{c}(V)$
the energy at the SD point: the FS at points with $E<E_{c}(V)$ (blue
curves in leftmost sub-figures of Fig.$\,$\ref{fig:phi_k_curves}(b)
) are identical to the rotated FS at the corresponding dual points
at $E>E_{c}(V)$ (red dashed curves in rightmost sub-figures of Fig.$\,$\ref{fig:phi_k_curves}(b)
). Such observation hints at the existence of a generalized Aubry-André
duality between the extended and localized phases that switches the
roles of phases $\phi$ and $k$, as in the AAM \citep{AubryAndre}.

Remarkably, the SD curves for a fixed CA, obtained by requiring invariance
under $\mathcal{R}_{0}$, approximate unexpectedly well the mobility
edge of the QPS, as seen in Fig.$\,$\ref{fig:main}(d). This approximation
becomes increasingly better as the order of the order of the approximant
increases. Convergence can be controlled by comparing the results
for two consecutive CA (see Fig.$\,$\ref{fig:intersection_points}
in Appendix$\,$\ref{sec:Emergence-of-duality} for examples).\lyxdeleted{Miguel}{Sun Jun  5 14:14:45 2022}{ }

For all the systems that we tested, the SD curves always approximate
very well the mobility edge, even in regimes where a numerical analysis
based on the IPR can fail \footnote{When the ${\rm IPR}$ is computed for QPS with incommensurate ratio
$\tau$ very close to commensurate ratios $\tau_{c}$ of low-order
CA, it can give wrong results. In particular, we can have ${\rm IPR}\sim{\rm constant}$
in the extended phase. This problem occurs for very weakly dispersive
states, in particular when the energy dispersion both in $\varphi$
and $\kappa$ is below machine precision. It can occur for any model,
including the AAM.}. A generic example for the LM can be seen in Fig.$\,$\ref{fig:main}(c),
for which the phase diagram is highly non-trivial. Remarkably, the
FS close-enough to criticality are always of the form observed in
Fig.$\,$\ref{fig:phi_k_curves}(b), for CA of high enough order and
irrespectively of the studied model. In Appendices$\,$\ref{subsec:estimate_VR_tR_ER}
and \ref{sec:Emergence-of-duality} we show examples of additional
models and provide details on the computation of SD points.

\begin{figure}[h]
\begin{centering}
\includegraphics[width=1\columnwidth]{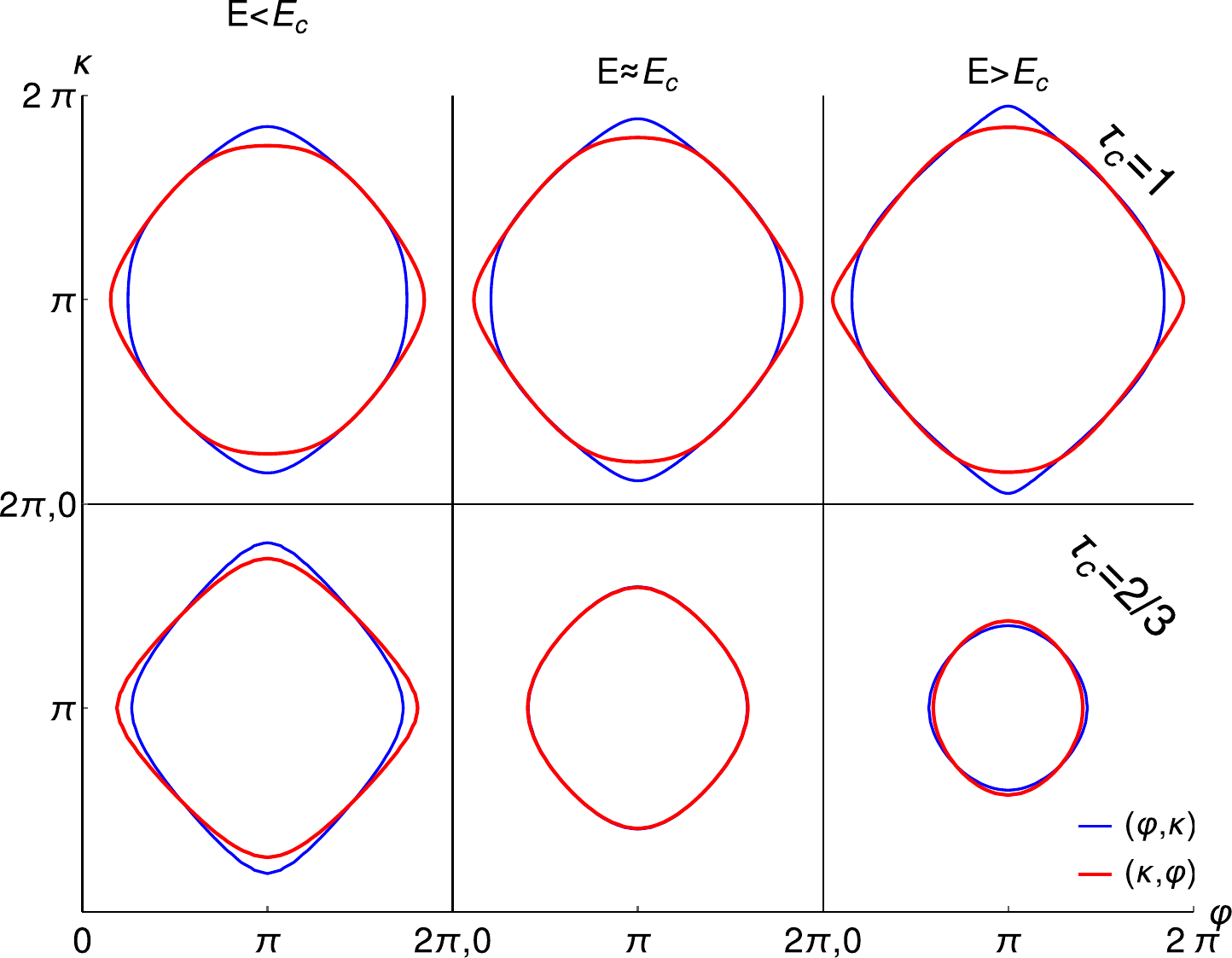}
\par\end{centering}
\caption{Constant energy cuts of the energy bands $E_{n}(\varphi,\kappa)$
for the AA-NNN, defined in Eq.$\,$\ref{eq:EAAM_H} and below it,
with fixed $V=2.1,t_{2}=0.2$ and variable $E$. The results are for
the CA defined by $\tau_{c}=1$ (top row) and $\tau_{c}=2/3$ (bottom
row).\label{fig:AA_NNN_tac1_tac2o3}}
\end{figure}

The results we presented up to now naturally raise the question of
whether a hidden duality transformation generalizing that found by
AA exist for a generic localization-delocalization transitions in
1D QPS. We make the case for its existence in Sec.$\,$\ref{sec:DualityTransformation}
. However, it is worth noting that in general such duality may not
be well defined for low-order approximants. As an example, consider
the FS for CA of the AA-NNN at fixed $V=2.1$ and $t_{2}=0.2$, and
variable $E$ {[}see Fig.$\,$\ref{fig:AA_NNN_tac1_tac2o3}{]}. For
$\tau_{c}=1$ there is no SD energy for which the FS is invariant
under the $\varphi\leftrightarrow\kappa$ interchange, as seen in
the top row panel of Fig.$\,$\ref{fig:AA_NNN_tac1_tac2o3}. For the
higher-order approximant shown in the bottom row panel, $\tau_{c}=2/3$,
the invariance seems to occur for $E_{c}\approx0.07$. In fact, a
more detailed analysis shows that the invariance is not yet perfect
for this approximant, but becomes increasingly better for higher-order
approximants (see Appendix$\,$\ref{subsec:estimate_VR_tR_ER}). Indeed,
for all the models studied, we observe that the FS always acquires
the simple shapes shown in Fig.$\,$\ref{fig:phi_k_curves}(b), close
to the critical point and for CA of high-enough order.

\paragraph{Parametrization of constant energy curves near SD points---}

For a higher enough order of the approximant, the functional form
of the FS as a function of $\varphi$ and $\kappa$ can be well captured
by sinusoidal shapes, close enough to quasiperiodicity-driven localization-delocalization
transitions. Considering an Hamiltonian with parameters $\bm{\lambda}$,
we have at fixed energy $E$ (see Appendix$\,$\ref{sec:duality_transf_correct_method}
for details):

\begin{equation}
E=V_{R}(\bm{\lambda},E)\cos(\varphi-\varphi_{0})+2t_{R}(\bm{\lambda},E)\cos(\kappa-\kappa_{0})+E_{R}(\bm{\lambda},E).\label{eq:renorm_model_main}
\end{equation}
This is a generalization of the renormalized model defined in Ref.$\,$\citep{Szabo2018}
for the AAM, encoded in the energy-dependence of the parameters. $V_{R}$,
$t_{R}$ and $E_{R}$ are renormalized couplings independent of $\varphi$
and $\kappa$.

The ansatz of Eq. (\ref{eq:renorm_model_main}) assumes that for large
enough unit cells only the fundamental harmonics in $\varphi$ and
$\kappa$ survive. In fact $V_{R}$ ($t_{R}$) also becomes irrelevant,
i.e. vanishes, in the extended (localized) phase as the CA's order
is increased. This assumption is motivated phenomenologically here
by the universality observed in the FS of different models and in
the next section by some special models where Eq. (\ref{eq:renorm_model_main})
is exact. A deeper insight on its validity will be provided elsewhere
\citep{prepar}.

\begin{figure}[h]
\begin{centering}
\includegraphics[width=1\columnwidth]{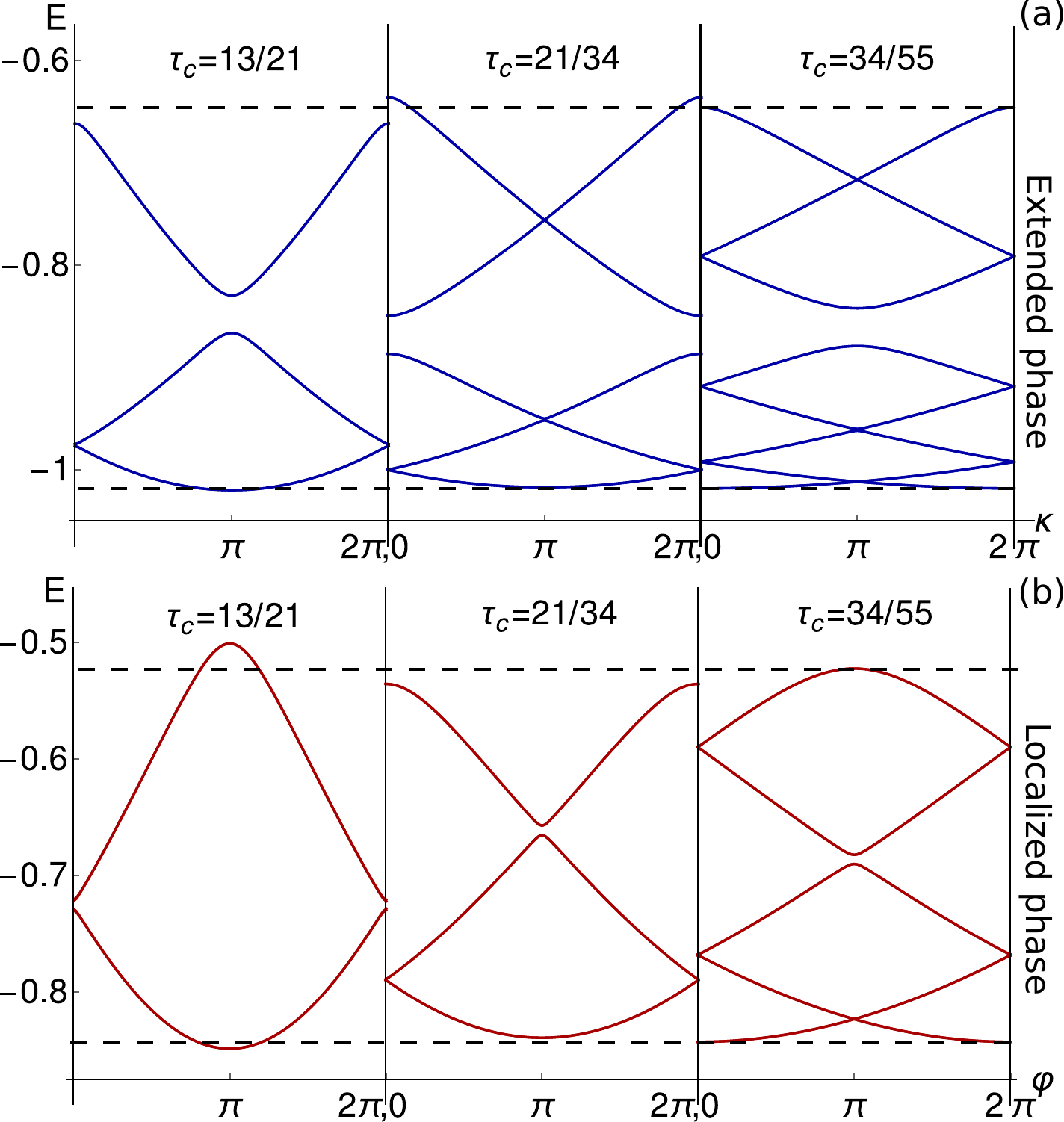}
\par\end{centering}
\caption{Examples of \textquotedblleft band-folding\textquotedblright{} for
some selected energy bands and different CA. These results are for
the AA-NNN, defined in Eq.$\,$\ref{eq:EAAM_H} and below it, with
$t_{2}=0.5$. (a) Energy dispersion with $\kappa$ for $V=1$ (extended
phase) and three different CA. The dispersion with $\varphi$ is smaller
than the line width. (b) Energy dispersion with $\varphi$ for $V=3.5$
(localized phase) and three different CA. The dispersion with $\kappa$
is smaller than the line width. The dashed lines are the band edges
of the lowest and highest energy bands shown, for $\tau_{c}=34/55$.
The energy-bands of higher-order CA are approximately band-foldings
of the energy bands of lower-order ones. \label{fig:band_folding_main}}
\end{figure}

According to our hypothesis, in the extended (localized) phase the
sub-bands of a CA of high-enough order can be described by a renormalized
$\kappa$-dependent ($\varphi$-dependent) single-band effective model.
In that case, further increasing the CA's order approximately folds
the model's energy band. Examples of the ``band-folding'' in the
extended and localized phases are given in Fig.$\,$\ref{fig:band_folding_main}.

At the critical point, both $V_{R}$ and $t_{R}$ are relevant: energy
gaps are opened irrespectively of the CA's order.

This hypothesis also explains why the regions of finite DOS for a
QPS are so well bounded by the energy bands of its CA in the extended
and localized phases: the band edges are not significantly changed
by the approximated band-foldings. Furthermore, the accuracy of the
commensurate approximation increases quadratically with the unit cell's
size. To understand why, consider a CA defined by $\tau_{c}^{m}=n_{2}^{m}/n_{1}^{m}$,
as the $m$-th order convergent of the irrational $\tau$ \citep{rockett1992continued}.
A well-known property of the convergents is that $|\tau_{c}^{m}-\tau|<(n_{1}^{m}n_{1}^{m+1})^{-1}$
\citep{rockett1992continued}. Therefore, for CA corresponding to
$\tau_{c}^{m}$ and the QPS characterized by $\tau$ only have significant
differences at length scales $L_{m}\sim n_{1}^{m}n_{1}^{m+1}>(n_{1}^{m})^{2}$.
On the other hand, in extended and localized phases, the correlation
length, $\xi$, is finite. Therefore, for $L_{m}>\xi$, the CA provides
a good approximation of the QPS: differences between both systems
only arise at length scales larger than $\xi$. The fast convergence
of the energy bands with the order of the approximate is a also consequence
of the fast increase in $L_{m}>(n_{1}^{m})^{2}$.

\paragraph{Models with CA-independent self-dualities.---}

Some special models exhibit a $\tau$-independent self-duality symmetry.
These include for instance the AAM, the MAAM and the model in \citep{Wang2020a}.
For these models, the $\varphi\leftrightarrow\kappa$ self-duality
is exact independently of the CA. The localization-delocalization
transition curve can then be easily computed analytically by analysing
the simplest possible CA.

For the AAM, the simplest approximant has one site per unit cell ($\tau_{c}=1$)
and the single band energy dispersion is given by

\begin{equation}
E_{{\rm AAM}}(V,\varphi,k)=V\cos\varphi+2\cos\kappa
\end{equation}

It is easy to see that for $V=2$, the energy dispersion is invariant
under $\varphi\leftrightarrow\kappa$. This interchanging operation
is identical to a $\pi/2$-rotation, $\mathcal{R}_{0}$, around $(\varphi_{0},\kappa_{0})=(0,0)$.
We can also check for a rotation symmetry around $(\varphi_{0},\kappa_{0})=(\pi,0)$.
To do so, we redefine $\varphi'=\varphi+\pi$, yielding $E_{{\rm AAM}}(V,\varphi',\kappa)=-V\cos\varphi'+2\cos\kappa$.
The model becomes now self-dual under the new rotation when $V=-2$.
Putting the two conditions together, we get the well known critical
potential (and self-dual condition) $|V|=2$.

For the MAAM, the energy dispersion for the CA with $\tau_{c}=1$
is

\begin{equation}
E_{{\rm MAAM}}(V,\alpha,\varphi,\kappa)=\frac{2V\cos\varphi}{1-\alpha\cos\varphi}+2\cos\kappa
\end{equation}
Rearranging, we get $E=(\alpha E+2V)\cos\varphi+2\cos\kappa-2\alpha\cos\varphi\cos\kappa$,
for $1-\alpha\cos\varphi\neq0$, which is self-dual under $\varphi\leftrightarrow\kappa$
if $E=2(1-V)/\alpha$. Defining $\varphi'=\varphi+\pi$, as for the
AAM, we get $E=-2(1+V)/\alpha$. Putting the two together, we get
the condition for the mobility edge found in Ref.~\citep{PhysRevLett.114.146601}.

\noindent Finally, consider some of the models in Ref.$\,$\citep{Wang2020a},
defined by

\begin{equation}
H=\sum_{j}(c_{j}^{\dagger}c_{j+1}+{\rm H.c.})+2\sum_{j}\lambda_{j}c_{j}^{\dagger}c_{j}
\end{equation}

\begin{equation}
\lambda_{j}=\begin{cases}
\lambda\cos(2\pi\tau j+\phi) & ,\mod(j,\nu)=0\\
0 & {\rm ,otherwise}
\end{cases}
\end{equation}

\noindent where $\nu$ is an integer. For $\nu=2$, the simplest possible
approximant has two sites per unit cell, and the following characteristic
polynomial:

\begin{equation}
\mathcal{P}(\kappa,\varphi)=C_{{\rm inv}}(\kappa,\varphi)-2\cos(\kappa)-2E\lambda\cos(\varphi)
\end{equation}

\noindent where $C_{{\rm inv}}(\kappa,\varphi)=E^{2}-2$ contains
the terms that are invariant under switching $\kappa$ and $\varphi$.
The phase invariance condition is therefore $E=\pm1/\lambda$, which
is the mobility edge expression in Eq.$\,$(5) of Ref.$\,$\citep{Wang2020a}.
For $\nu=3$, the simplest possible CA has three sites per unit cell
and a characteristic polynomial:

\begin{equation}
\mathcal{P}(\kappa,\varphi)=C_{{\rm inv}}(\kappa,\varphi)+2\cos(\kappa)+2(E^{2}-1)\lambda\cos(\varphi)
\end{equation}

\noindent where $C_{{\rm inv}}(\kappa,\varphi)=3E-E^{3}$, which gives
the following condition for the mobility edge, $E=\pm\sqrt{1\pm1/\lambda}$,
the same obtained in Ref.$\,$\citep{Wang2020a}.

\section{Generalized local duality transformation}

\label{sec:DualityTransformation} Having established a generalized
duality transformation in the spectrum, we now extend it to the eigenstates.

\subsection{Eigenstates and symmetries in $\left(\phi,k\right)$ space}

In this section we present the symmetries of the eigenstates of CA
in the $(\varphi,\kappa)$ space. Our goal is to obtain the transformation
properties of the real-space wave-function and its dual under translations
in $k$ and $\phi$. This will then be used as a starting point for
the definition of the generalized duality transformation in the follow-up
section.To simplify the discussion, we restrict ourselves to EAAM
in the following and latter show that our description can also be
applied to LM through an example. The Hamiltonian for this class of
models can be written as

\begin{equation}
\begin{aligned}H= & \sum_{n}\mathcal{V}(2\pi\tau_{c}n+\phi_{0}+\phi)c_{n}^{\dagger}c_{n}\\
 & +\sum_{n,n'=0}t(|n-n'|)e^{-ik_{0}(n-n')}c_{n}^{\dagger}c_{n'}
\end{aligned}
\label{eq:H_com2-1-1}
\end{equation}
where $\mathcal{V}(x)$ is a $2\pi$-periodic function, $t(|x|)$
is a generic function with $t(0)=0$ and we introduced a phase twist
$k_{0}$ in the hopping terms to further enlarge the class of models
described by this expression. After applying the transformation in
Eq.$\,$\eqref{eq:c_tilde}, this can be written as

\begin{equation}
\begin{aligned} & \tilde{H}(\phi,k)=\sum_{m=0}^{n_{1}-1}\mathcal{V}(2\pi\tau_{c}m+\phi_{0}+\phi)\tilde{c}_{m}^{\dagger}\left(k\right)\tilde{c}_{m}\left(k\right)\\
 & +\sum_{r\in\mathbb{Z}}\sum_{m,m'=0}^{n_{1}-1}t(|rn_{1}+m-m'|)e^{-i(k+k_{0})(m+rn_{1}-m')}\tilde{c}_{m}^{\dagger}\left(k\right)\tilde{c}_{m'}\left(k\right)
\end{aligned}
\label{eq:H_com2-1}
\end{equation}
The terms with $|r|\geq1$ correspond to hopping terms between different
unit cells.

\paragraph{Properties of the real-space wave functions.}

The eigenstates of $\tilde{H}(\phi,k)$ can be expanded as
\begin{align}
\left|\psi(k)\right\rangle  & =\sum_{m}\tilde{\psi}_{m}^{\text{r}}(\phi,k)\tilde{c}_{m}^{\dagger}(k)\left|0\right\rangle .\label{eq:psi_tilde_basis}
\end{align}
For the real-space wave-function $\tilde{\psi}_{m}^{\text{r}}(\phi,k)$,
shifts in $\phi$ of $\phi_{j}=2\pi j/n_{1}$ correspond to cyclical
translations in $\tilde{\psi}_{m}^{\text{r}}$, that is,

\begin{equation}
\tilde{\psi}_{m}^{\text{r}}(\phi+2\pi j/n_{1},k)=\tilde{\psi}_{\mod(m+\Delta m_{j},n_{1})}^{\text{r}}(\phi,k).\label{eq:psi_r_trans}
\end{equation}
This can be seen by absorbing $\phi_{j}$ in index $m$ of the potential
$\mathcal{V}(2\pi\tau_{c}m+\phi_{0}+\phi)$. In particular, we have
that $2\pi\tau_{c}m+\phi_{j}=2\pi\tau_{c}(m+\Delta m_{j})$, where
$\Delta m_{j}\equiv(j+ln_{1})/n_{2}$ and $l$ is an integer such
that $\Delta m_{j}$ is also an integer (the term with $l$ contributes
with an irrelevant phase $2\pi l$ in the $2\pi$-periodic potential
$\mathcal{V}(x)$). Furthermore, since $l$ satisfies the linear Diophantine
equation $n_{2}\Delta m_{j}=j+ln_{1}$, it always has a solution
since $n_{1}$ and $n_{2}$ are co-prime integers  \citep{goodstein1970},
see also \citep{PhysRevB.91.125411}. Finally, since $\mathcal{V}(2\pi\tau_{c}(m+n_{1})+\phi_{0}+\phi)=\mathcal{V}(2\pi\tau_{c}m+\phi_{0}+\phi)$,
a translation in the potential of $m\rightarrow m+\Delta m_{j}$ corresponds
to a cyclical translation of $\tilde{\psi}_{m}^{\text{r}}$ as written
in Eq.$\,$\ref{eq:psi_r_trans}. Therefore, as previously mentioned,
shifts of $\phi_{j}$ correspond to redefinitions of the unit cell.
There are $n_{1}$ possible such redefinitions that can be obtained
through the different shifts $\phi_{j}$, with $j=0,\cdots,n_{1}-1$.

On the other hand, under shifts in $k$ one obtains,
\begin{equation}
\tilde{\psi}_{m}^{\text{r}}(\phi,k+k_{j})=e^{-imk_{j}}\tilde{\psi}_{m}^{\text{r}}(\phi,k),\textrm{ }k_{j}=2\pi j/n_{1}.\label{eq:psi_tilde_twist}
\end{equation}
This can be seen by applying the gauge transformation $\tilde{c}_{m}^{\dagger}\left(k\right)=e^{im(k+k_{0})}c_{m}^{\dagger}\left(k\right)$,
such that

\begin{equation}
\begin{aligned} & \tilde{H}(\phi,k)=\sum_{m=0}^{n_{1}-1}\mathcal{V}(2\pi\tau_{c}m+\phi_{0}+\phi)c_{m}^{\dagger}\left(k\right)c_{m}\left(k\right)\\
 & +\sum_{r\in\mathbb{Z}}\sum_{m,m'=0}^{n_{1}-1}t(|rn_{1}+m-m'|)e^{-irn_{1}(k+k_{0})}c_{m}^{\dagger}\left(k\right)c_{m'}\left(k\right).
\end{aligned}
\end{equation}
In the new basis, $\tilde{H}$ is explicitly periodic under the transformations
$k\rightarrow k+2\pi j/n_{1}$ , with $j\in\mathbb{Z}$ and setting
$c_{m}^{\dagger}\left(k+2\pi j/n_{1}\right)\to c_{m}^{\dagger}\left(k\right)$.
Therefore, the new amplitudes $\psi_{m}^{\text{r}}(\phi,k)$ obeying

\begin{eqnarray}
\left|\psi(k)\right\rangle  & = & \sum_{m}\psi_{m}^{\text{r}}(\phi,k)c_{m}^{\dagger}(k)\left|0\right\rangle ,\label{eq:psi's-1-1}
\end{eqnarray}
are also periodic $\psi_{m}^{\text{r}}(\phi,k+2\pi j/n_{1})=\psi_{m}^{\text{r}}(\phi,k)$.
Using this equality and Eqs.$\,$\ref{eq:psi_tilde_basis},\ref{eq:psi's-1-1}
we can arrive at Eq. \eqref{eq:psi_tilde_twist}. 

\paragraph{Properties of the dual wave functions.}

The dual wave function, $\tilde{\psi}_{q}^{\t d}(\phi,k)$, can be
defined as in the AA case in Eq.$\,$\eqref{eq:AA_dual_eq}. The transformations
of $\tilde{\psi}_{q}^{\t d}(\phi,k)$ under shifts in $\phi$ of $k$
can also be obtained. Inverting relation Eq.$\,$\eqref{eq:AA_dual_eq}
and using the Eq.$\,$\eqref{eq:psi_tilde_twist} we obtain

\begin{align}
\tilde{\psi}_{q}^{\t d}(\phi,k+k_{j}) & =\frac{1}{\sqrt{n_{1}}}\sum_{m=0}^{n_{1}-1}e^{-i2\pi\tau_{c}qm}\tilde{\psi}_{m}^{\text{r}}(\phi,k+k_{j})\label{eq:duality}\\
 & =\frac{1}{\sqrt{n_{1}}}\sum_{m=0}^{n_{1}-1}e^{-im(2\pi\tau_{c}q+k_{j})}\tilde{\psi}_{m}^{\text{r}}(\phi,k).
\end{align}
From the right-hand side of this expression we can see that the term
$k_{j}$ might be absorbed into $q$, yielding 
\begin{align}
\tilde{\psi}_{q}^{\t d}(\phi,k+k_{j}) & =\frac{1}{\sqrt{n_{1}}}\sum_{m=0}^{n_{1}-1}e^{-im2\pi\tau_{c}\left(q+\frac{j}{n_{2}}+l\frac{n_{1}}{n_{2}}\right)}\tilde{\psi}_{m}^{\text{r}}(\phi,k)\\
 & =\tilde{\psi}_{\mod(q+\Delta q_{j},n_{1})}^{\t d}(\phi,k),\textrm{ }\label{eq:psi_k_trans}
\end{align}
where we again used that $k_{j}=2\pi j/n_{1}$ and $l$ is an integer
such that $\Delta q_{j}\equiv(j+ln_{1})/n_{2}$ is also an integer,
exactly as in the definition of $\Delta m_{j}$, below Eq.$\,$\ref{eq:psi_r_trans}.
Using the same arguments as for $\Delta m_{j}$, we can again always
find $l$ that makes $\Delta q_{j}$ an integer. Finally, since $\tilde{\psi}_{q+n_{1}}^{\t d}=\tilde{\psi}_{q}^{\t d}$,
we arrive at Eq.$\,$\ref{eq:psi_k_trans}.

For $k_{j}=2\pi j/n_{1}\in[0,2\pi[$, $\Delta q_{j}$ can take the
values $\Delta q_{j}=0,\cdots,n_{1}-1$. The different possible wave
functions $\tilde{\psi}_{q}^{\t d}(\phi,k+k_{j})$ are therefore the
$n_{1}$ possible cyclic translations of $\tilde{\psi}_{q}^{\t d}(\phi,k)$.

\paragraph{Cyclical translations under shifts in $k$ and $\phi$.}

Defining the $n_{1}$-component vectors $\tilde{\boldsymbol{\psi}}^{\t r}(\phi,k)=\left\{ \tilde{\psi}_{0}^{\t r}(\phi,k),\tilde{\psi}_{1}^{\t r}(\phi,k),...\right\} ^{T}$
and $\tilde{\boldsymbol{\psi}}^{\t d}(\phi,k)=\left\{ \tilde{\psi}_{0}^{\t d}(\phi,k),\tilde{\psi}_{1}^{\t d}(\phi,k),...\right\} ^{T},$
we can summarize the previous results as

\begin{equation}
\tilde{\boldsymbol{\psi}}^{\text{r}}(\phi+2\pi j/n_{1},k)=\mathcal{T}^{\Delta m_{j}}\tilde{\boldsymbol{\psi}}^{\text{r}}(\phi,k),\label{eq:psi_r_Tn}
\end{equation}

\begin{equation}
\tilde{\boldsymbol{\psi}}^{\t d}(\phi,k+2\pi j/n_{1})=\mathcal{T}^{\Delta q_{j}}\tilde{\boldsymbol{\psi}}^{\t d}(\phi,k),\label{eq:psi_d_Tn}
\end{equation}

\noindent where $\mathcal{T}_{ij}=\delta_{i,\mod\left(j+1,n_{1}\right)}$
denotes the cyclical translation matrix, and $\Delta m_{j}$ and $\Delta q_{j}$
are the integers defined previously.

At this point, a duality between the wave functions in Eqs.$\,$\ref{eq:psi_d_Tn},\ref{eq:psi_r_Tn},
switching the roles of $\phi$ and $k$, is already apparent. In the
following section, we will make use of this insight to define a duality
mapping between these wave functions.

\subsection{Definition of generalized duality transformation}

\paragraph{Motivation and definition.---}

The results in section \ref{sec:SpectralDuality} for the energy bands
of CA hinted at the existence of generalized dualities between the
energy bands in the extended and localized phases. Here we complete
our description by studying also the CA wave functions, and explicitly
constructing a generalized duality transformation.

\begin{figure}[h]
\centering{}\includegraphics[width=1\columnwidth]{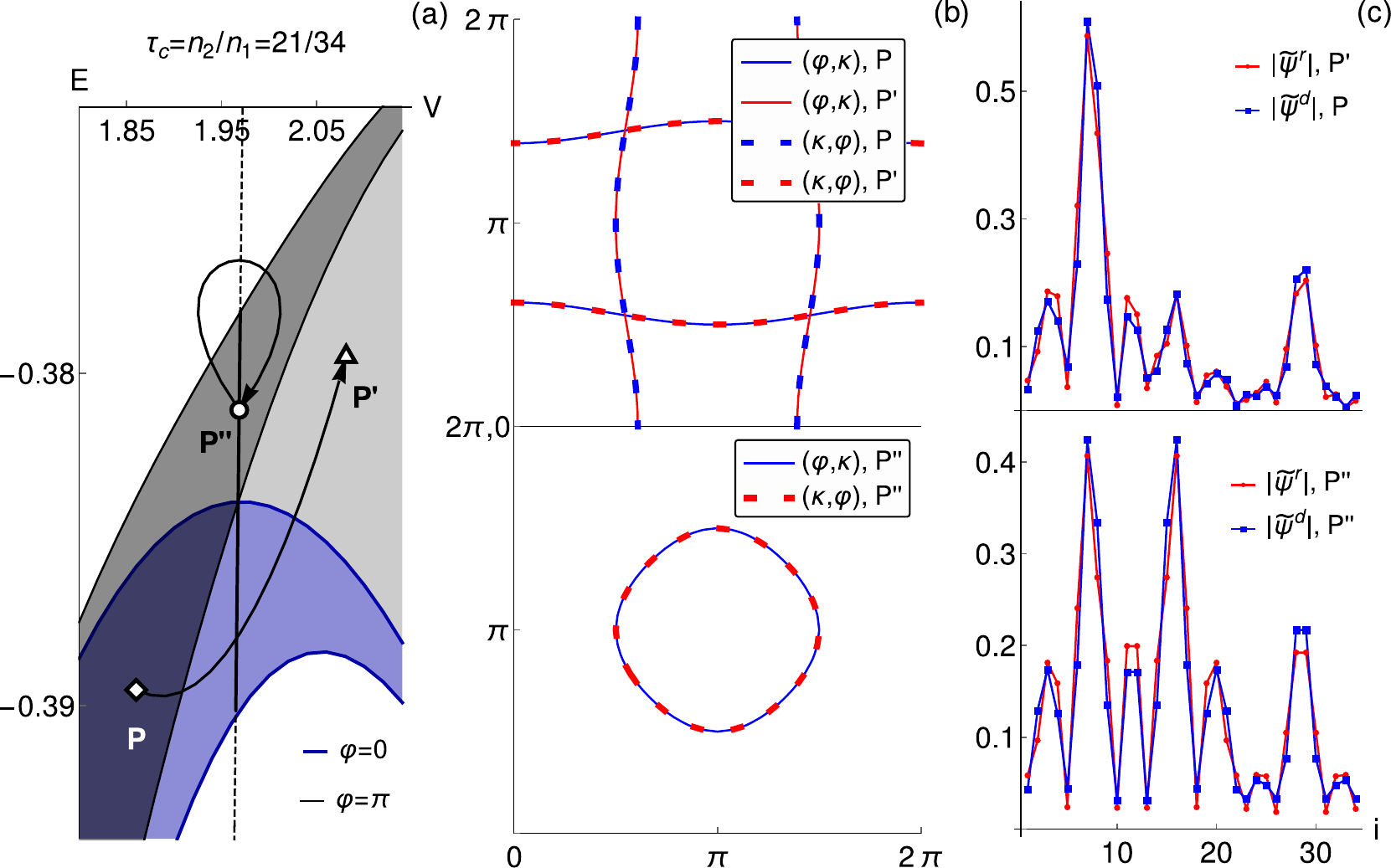}\caption{(a) Example of dual points for a randomly selected energy band of
the CA defined by $\tau_{c}=21/34$, for the AA-NNN, defined in Eq.$\,$\ref{eq:EAAM_H}
and below it, with next-nearest-neighbor hoppings $t_{2}=0.2$. (b)
Top: FS at P (blue) and P' (red) and their $\pi/2$-rotation around
$(\varphi_{0},\kappa_{0})=(\pi,\pi)$, respectively dashed blue and
dashed red. Bottom: FS for the SD point $P''$ (blue) and its rotation
(dashed red). (c) Top: Wave function $\tilde{\bm{\psi}}^{\text{r}}$
at $P'$ and its Aubry-André dual $\tilde{\bm{\psi}}^{d}$ at $P$.
Bottom: Wave function $\tilde{\bm{\psi}}^{\text{r}}$ and its Aubry-André
dual $\tilde{\bm{\psi}}^{d}$ at $P''$. \label{fig:DualTransf1}}
\end{figure}

For a Hamiltonian depending on a set of parameters $\bm{\lambda}$,
we can define dual points $P\equiv P\Big(\bm{\lambda},E(\bm{\lambda},\varphi,\kappa)\Big)$
and $P'\equiv P'\Big(\bm{\lambda}',E'(\bm{\lambda}',\mathcal{R}_{0}[\varphi,\kappa]^{T})\Big)$
for each energy band of a CA such that the FS at $P$ is the rotation
$\mathcal{R}_{0}$ of the FS at $P'$, see Fig.$\,$\ref{fig:DualTransf1}(a,b).
SD points are defined by $P=P'$. Recall that $\mathcal{R}_{0}$ is
a $\pi/2$-rotation in the $(\phi,k)$ plane, around point $(\phi_{0},k_{0})$.
These parameters were introduced in Eq.$\,$\eqref{eq:H_com2-1} and
encode the fixed point of the duality transformation in the $(\phi,k)$
plane {[}which does not have to be $(\phi_{0},k_{0})=(0,0)${]}. In
the same way, as stated before, the FS of dual points in the phase
diagram are identical under rotation $\mathcal{R}_{0}$ in the $(\varphi,\kappa)$
plane, around $(\varphi_{0},\kappa_{0})$. The latter are related
with $(\phi_{0},k_{0})$ through $(\varphi_{0},\kappa_{0})=n_{1}\big(\mod(\phi_{0},2\pi/n_{1}),\mod(k_{0},2\pi/n_{1})\big)$.

We can compare the real-space wave function vectors at point $P'$,
$\tilde{\bm{\psi}}^{\text{r}}(P';\mathcal{R}_{0}[\phi,k]^{T})$, \footnote{Note that we should specify both the point $P(\bm{\lambda},E)$ and
phases $\phi$ and $k$. This is because in general there is a subset
of infinitely many phases $(\phi,k)$ associated with point $P$ (the
FS is a line and not a point, in general). Even though two different
phases in this subset correspond to the same point $P$, they may
correspond to different wave functions.} with its Aubry-André dual at point $P$, $\tilde{\bm{\psi}}^{\text{d}}(P;\phi,k)$,
with entries are given by

\begin{equation}
\tilde{\psi}_{n}^{\t d}(P;\phi,k)=\frac{1}{\sqrt{n_{1}}}\sum_{m=0}^{n_{1}-1}e^{-i2\pi\tau_{c}mn}\tilde{\psi}_{m}^{\text{r}}(P;\phi,k),\label{eq:AA_dual_eq}
\end{equation}

\noindent where, besides the dependence on $\phi$ and $k$, we also
specified the points at which the wave functions are calculated. For
the AAM, $\tilde{\bm{\psi}}^{\text{r}}(P';\mathcal{R}_{0}[\phi,k]^{T})=\tilde{\bm{\psi}}^{\t d}(P;\phi,k)$
, as we have seen in Sec.$\,$\ref{sec:WarmUp}, with $k_{0}=0$ and
$\phi_{0}=0\vee\phi_{0}=\pi$. For generic models this is not the
case. However, interestingly, they can be very similar. An example
is shown in Fig.\ref{fig:DualTransf1}$\,$(c) for the AA-NNN.

We can define the exact generalized duality transformation that maps
all the wave functions at a point $P'$ to the Aubry-André dual wave
functions at $P$, as simply the matrix transformation 
\begin{align*}
\mathcal{O}_{c}\tilde{\bm{\psi}}^{\t d}(P;\phi,k) & =\tilde{\bm{\psi}}^{\text{r}}(P';\mathcal{R}_{0}[\phi,k]^{T}).
\end{align*}
 As this mapping respects cyclical translations, $\mathcal{O}_{c}$
obeys

\begin{equation}
\mathcal{O}_{c}\left[\mathcal{T}^{n}\tilde{\bm{\psi}}^{\t d}(P;\phi,k)\right]=\mathcal{T}^{n}\tilde{\bm{\psi}}^{\text{r}}(P';\mathcal{R}_{0}[\phi,k]^{T}),\hspace{1em}n=0,\cdots,n_{1}-1,\label{eq:definition_Oc}
\end{equation}

\noindent where $\mathcal{T}$ is the cyclic translation operator
previously defined \footnote{We assume the matrix $\mathcal{Q}_{d}$ containing $\mathcal{T}^{n}\tilde{\bm{\psi}}(P;\phi,k),\,n=0,\cdots,n_{1}-1$
in its columns is not rank-deficient, that is $\textrm{rank}(\mathcal{Q}_{d})=n_{1}$.
However, one should check whether this is true for the states being
used before computing $\mathcal{O}_{c}$, otherwise the latter is
not well-defined.}. This implies $\mathcal{O}_{c}$ is a circulant matrix, i.e. its
rows are just cyclic translations of the first row. Therefore, since
any circulant matrix is diagonalized by the discrete Fourier transform,
$\mathcal{O}_{c}$ is entirely defined by its eigenvalues. In other
words Eq. \eqref{eq:definition_Oc} determines the duality transformation
$\mathcal{O}_{c}$ uniquely for a given approximant.

To summarize, the full duality transformation can be obtained through
a two-step procedure:
\begin{enumerate}
\item Find dual points in the phase diagram $P\equiv P\Big(\bm{\lambda},E(\bm{\lambda},\varphi,\kappa)\Big)$
and $P'\equiv P'\Big(\bm{\lambda}',E'(\bm{\lambda}',\mathcal{R}_{0}[\varphi,\kappa]^{T})\Big)$,
associated with FS that are identical under rotation $\mathcal{R}_{0}$;
\item Find the duality matrix $\mathcal{O}_{c}$, defined in Eq.$\,$\eqref{eq:definition_Oc},
that maps the wave functions $\tilde{\bm{\psi}}^{\t d}(P;\phi,k)$
at $P$ and $\tilde{\bm{\psi}}^{\t r}(P';\mathcal{R}_{0}[\phi,k]^{T})$
at $P'$ and their cyclic translations.
\end{enumerate}
In \citep{SM}, we describe an efficient method to carry out step
1. Step 2 will be detailed with examples in the following sections.
Schematically, considering that the Hamiltonian depends on a set of
parameters $\bm{\lambda}$, the full duality transformation for a
given energy band reads

\begin{equation}
\begin{array}{ccc}
 & \textrm{Full duality}\\
P\Big(\bm{\lambda},E(\bm{\lambda},\varphi,\kappa)\Big) & \leftrightarrow & P'\Big(\bm{\lambda}',E'(\bm{\lambda}',\mathcal{R}_{0}[\varphi,\kappa]^{T})\Big)\\
\downarrow &  & \downarrow\\
 & \mathcal{O}_{c}\\
\mathcal{T}^{j}\tilde{\bm{\psi}}^{\t d}(P;\phi,k) & \leftrightarrow & \mathcal{T}^{j}\tilde{\bm{\psi}}^{\t r}(P';\mathcal{R}_{0}[\phi,k]^{T})\\
j=0,\cdots,n_{1}-1 &  & j=0,\cdots,n_{1}-1
\end{array}
\end{equation}

\begin{figure}[h]
\centering{}\includegraphics[width=1\columnwidth]{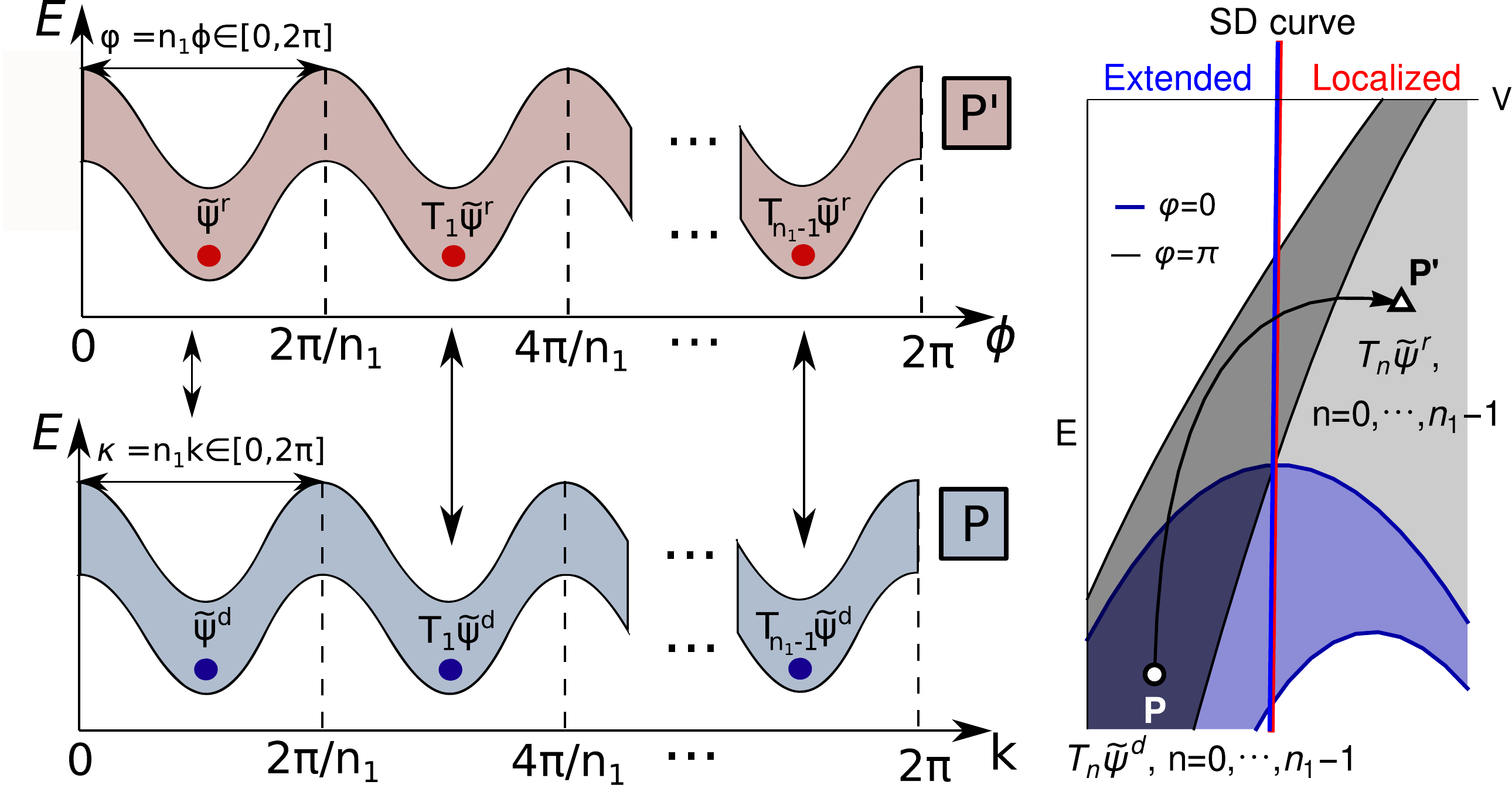}\caption{Pictorial description of the duality transformation. The energy bands
are periodic in $\phi$ and $k$ with period $2\pi/n_{1}$. However,
the wave function differs by cyclic translations in each of these
periods. The different $\phi$-periods ($k$-periods) contain all
the $n_{1}-1$ possible cyclic translations of the first-period real-space
wave function $\tilde{\bm{\psi}}^{\text{r}}$ (dual wave function
$\tilde{\bm{\psi}}^{\protect\t d}$). If $P$ and $P'$ are dual points
of a given CA, $\mathcal{O}_{c}$ is the transformation that transforms
$T_{n}\tilde{\bm{\psi}}^{\protect\t d}$ at $P$ into $T_{n}\tilde{\bm{\psi}}^{\text{r}}$
at $P'$, with $n=0,\cdots,n_{1}-1$. \label{fig:DualTransf2}}
\end{figure}

This duality is illustrated in Fig.$\,$\ref{fig:DualTransf2}. The
duality is only local, in the sense that it is defined for each energy
band of a CA, in the neighborhood of the critical (self-dual) point.
In general, these \textit{local dualities} may break-down sufficiently
away from SD points, i.e. there might be no pair $\bm{\lambda}$ and
$\text{\ensuremath{\bm{\lambda}}'}$ associated with FS identical
under any rotation $\mathcal{R}_{0}$. Nevertheless, for a the generic
families we studied, there is always a set of dual points in the vicinity
of a self-dual point. .\lyxdeleted{Miguel}{Sun Jun  5 14:14:45 2022}{ }

\paragraph{Testing the generalized duality.---}

As noted in the previous section, for the AAM, the duality matrix
$\mathcal{O}_{c}$ is always the identity because $\tilde{\bm{\psi}}^{\t d}(P;\phi,k)=\tilde{\bm{\psi}}^{\text{r}}(P';\mathcal{R}_{0}[\phi,k]^{T})$
for any CA. In order to test the validity of $\mathcal{O}_{c}$ in
a non-trivial example, we consider the MAAM, for which an analytical
expression for the duality transformation was found in Ref. \citep{PhysRevLett.114.146601}.
Note however that this transformation is not unique at the self-dual
points . In fact, we found that, defining the eigenstates for the
MAAM (particular case of Eq.$\,$\ref{eq:EAAM_H}) as $\ket{\psi}=\sum_{n}\psi_{n}c_{n}^{\dagger}\ket 0$
in the infinite-size incommensurate limit (for irrational $\tau$),
the wave function is also self-dual under the transformation (see
Appendix$\,$\ref{sec:derivation})

\begin{equation}
\psi'_{l}=\sum_{n}e^{-2\pi i\tau ln}\chi_{n}(\beta_{0},\tau)\psi_{n},\label{eq:dualTransf_MAAM_ThermoLimit}
\end{equation}

\noindent where $\psi'_{l}$ is the dual wave function (not to confuse
with the Aubry-André dual wave function $\psi_{l}^{d}=\sum_{n}e^{-2\pi i\tau ln}\psi_{n}$),
the parameters of the model are $\bs{\lambda}=\{t,V,\alpha\}$, $\chi_{n}(\beta_{0},\tau)=\sinh\beta_{0}[\cosh\beta_{0}-\cos(2\pi\tau n)]^{-1}$,
with $\beta_{0}$ defined as $2t\cosh\beta_{0}=E+2V\alpha^{-1}$.
For a given CA defined by $\tau_{c}$, we define

\begin{equation}
\tilde{\psi}'_{p}=\sum_{m=0}^{n_{1}-1}e^{-2\pi i\tau_{c}pm}\chi_{c}^{m}\tilde{\psi}_{m}^{\text{r}},
\end{equation}

\begin{figure}[h]
\centering{}\includegraphics[width=1\columnwidth]{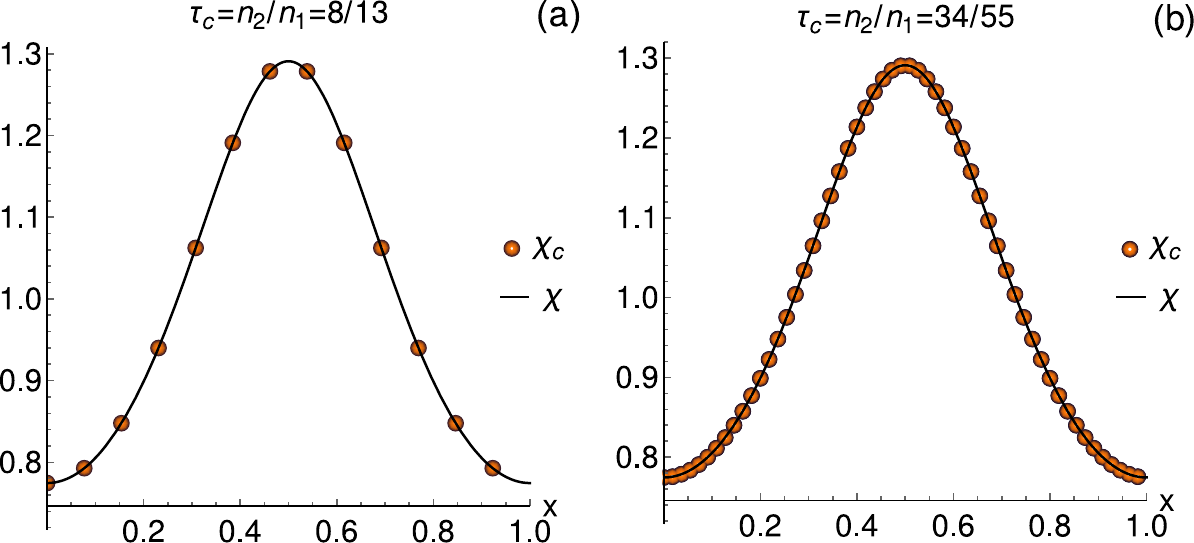}\caption{Results for the MAAM, defined in Eq.$\,$\ref{eq:EAAM_H} and below
it. Eigenvalues of matrices $\mathcal{O}_{c}$ (Eq.$\,$\ref{eq:chi_c_Oc})
defined for different energy bands of CA with $\tau_{c}=8/13$ (a)
and $\tau_{c}=34/55$, for $\cosh\beta=4$ and $\phi=k=0$.(b). A
total of $7$ and $28$ energy bands (points) were used respectively
in (a) and (b). Different points have different sizes and colors,
perfectly overlapping. The eigenvalues are compared with the function
$\chi(\beta_{0},x)$ defined in the text below Eq.$\,$\ref{eq:chi_c_Oc}.
They were rescaled so that the eigenvalue corresponding to $x_{p}=0$
matches the value $\chi(\beta_{0},0)$. \label{fig:DualTransf3}}
\end{figure}

\noindent where $\chi_{c}^{m}\equiv\chi_{m}(\beta_{0},\tau_{c})$.
This transformation reduces to Eq.$\,$\eqref{eq:dualTransf_MAAM_ThermoLimit}
when $n_{1}\rightarrow\infty$ ($\tau_{c}\rightarrow\tau$). Note
that we used the prime notation in $\tilde{\psi}'_{p}$ to not confuse
it with $\tilde{\psi}_{p}^{\textrm{d}}$, the Aubry-André dual wave
function defined in Eq.$\,$\ref{eq:AA_dual_eq}. Using the vector
notation, it can also be written as

\begin{equation}
\tilde{\boldsymbol{\psi}}'=U^{\dagger}\bs{\chi}_{c}\tilde{\bm{\psi}}^{\text{r}}=(U^{\dagger}\bs{\chi}_{c}U)U^{\dagger}\tilde{\bm{\psi}}^{\text{r}}=(U^{\dagger}\bs{\chi}_{c}U)\tilde{\bm{\psi}}^{\text{d}},\label{eq:dualTransfDasSarma}
\end{equation}

\noindent where $\tilde{\boldsymbol{\psi}}'=\left\{ \tilde{\psi}'_{0},\cdots,\tilde{\psi}'_{n_{1}-1}\right\} ^{T}$,
$U$ and $\bs{\chi}_{c}$ are matrices with entries, respectively,
$U_{np}=e^{2\pi i\frac{n_{2}}{n_{1}}np}$ and $\bs{\chi}_{c}^{np}=\chi_{c}^{p}\delta_{np}$,
with $n,p=0,\cdots,n_{1}-1$, and we used that $U^{\dagger}\tilde{\bm{\psi}}^{\text{r}}=\tilde{\bm{\psi}}^{\text{d}}$.
At SD points, the transformation in Eq.$\,$\eqref{eq:dualTransfDasSarma}
maps the wave function into itself, up to a normalization, and therefore
$\tilde{\boldsymbol{\psi}}'\propto\tilde{\bm{\psi}}^{\text{r}}$.
Assuming that $\tilde{\bm{\psi}}^{\text{r}}$ and $\tilde{\boldsymbol{\psi}}'$
are normalized , we have that, at SD points:

\begin{equation}
\mathcal{O}_{c}=U^{\dagger}\bs{\chi}_{c}U.\label{eq:chi_c_Oc}
\end{equation}

We may now check if our definition for $\mathcal{O}_{c}$ in Eq.$\,$\eqref{eq:definition_Oc}
matches the one obtained above. Since $\mathcal{O}_{c}$ is a circulant
matrix, it is diagonalized by the unitary transformation $U$. Therefore,
we just need to compute the eigenvalues of $\mathcal{O}_{c}$ and
check if they match the values $\chi_{c}^{p}$. Note that if we define
the function $\chi(\beta_{0},x)=\frac{\sinh\beta_{0}}{\cosh\beta_{0}-\cos(2\pi x)}$,
$\chi_{c}^{p}$ are just evaluations of this function at points $x_{p}=\mod(n_{2}p/n_{1},1)$.
For a CA with $n_{1}$ sites in the unit cell, we sample $n_{1}$
points of function $\chi(x,\beta_{0})$. The results are shown in
Fig.$\,$\ref{fig:DualTransf3} for the model in Eq.$\,$6 of \citep{PhysRevLett.114.146601}.
We can see that the eigenvalues of the computed matrices $\mathcal{O}_{c}$
perfectly fall on top of the $\chi(\beta_{0},x)$ curve, after a global
rescaling. We have computed $\mathcal{O}_{c}$ for fixed $\cosh\beta=4$,
using $\tilde{\bm{\psi}}^{\text{r}}$ and $\tilde{\bm{\psi}}^{\text{d}}$
at multiple SD points of a given CA, with $k=\phi=0$ (note that at
SD points, $\beta_{0}=\beta$). The obtained eigenvalues shown in
Fig.$\,$\ref{fig:DualTransf3} were always the same, up to normalization.
This was expected as for fixed $\beta$, $\chi_{c}^{p}$ does not
depend on energy nor on the rest of the Hamiltonian's parameters.

\begin{figure*}[t]
\centering{}\includegraphics[width=0.7\paperwidth]{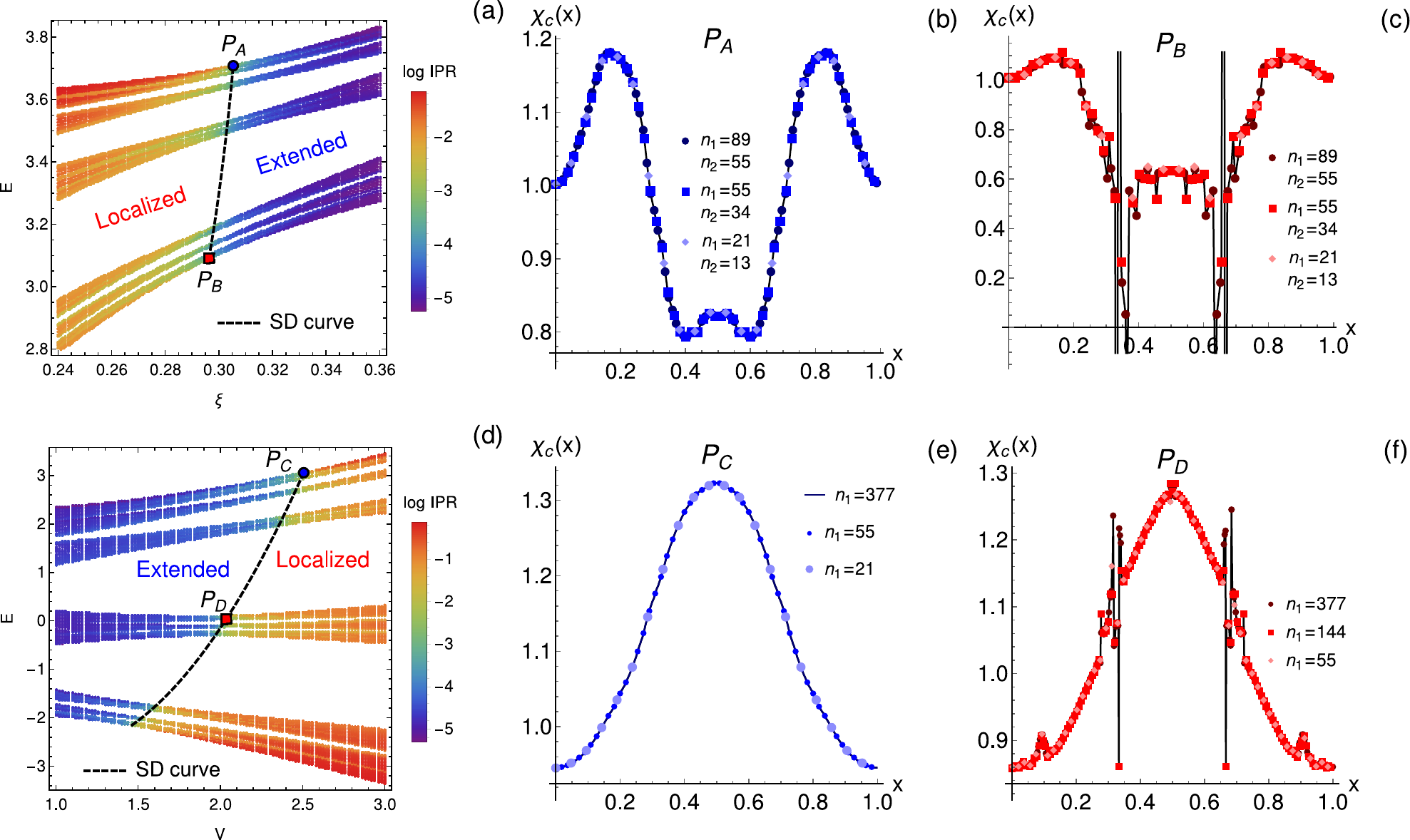}\caption{(a) IPR for the LM, defined in Eq.$\,$\ref{eq:ILM_H}, with $V=3,\Lambda=2.5a$,
for $\tau_{i}=144/233$ {[}highest energy states in Fig.$\,$\ref{fig:main}(c){]}.
The dashed black line corresponds to interpolated SD curves, computed
for $\tau_{c}=34/55$. (b,c) Eigenvalues of matrices $\mathcal{O}_{c}$
defined in Eq.$\,$\ref{eq:chi_c_Oc}, labeled as $\chi_{c}(x_{p})\equiv\chi_{c}^{p}$,
computed for different CA, for points in the SD curve marked in (a).
(d) IPR for the AA-NNN, defined in Eq.$\,$\ref{eq:EAAM_H}, with
$t_{2}=0.1$, for $\tau_{i}=144/233$ (system with $233$ sites).
The SD curves were computed for $\tau_{c}=34/55$. (b,c) $\chi_{c}(x_{p})$
computed for the points marked in (d). In (c,f), all the points were
connected with a black line to guide the eye. Function $\chi_{c}$
was computed for states with $\phi=k=0$.\label{fig:fig10}}
\end{figure*}

\paragraph{Application to generic models.---}

The definition of $\bs{\chi}_{c}$ can be easily generalized to other
models using that the duality transformation, $\mathcal{O}_{c}$,
defined in Eq.$\,$\eqref{eq:definition_Oc}, is a circulant matrix
and thus can always be written in the form of Eq.$\,$\eqref{eq:chi_c_Oc}.
Two dual points $P$ and $P'$ of a given approximant, $\tau_{c}=n_{2}/n_{1}$,
of a quasiperiodic model define a set of eigenvalues $\chi_{c}^{p}$,
which can be parametrized as the function $\chi_{c}(x)$ evaluated
at $x=x_{p}=\mod(n_{2}p/n_{1},1)$. In the following, we provide numerical
evidence that the function $\chi_{c}$ can converge very fast with
the order of the approximant, i.e. $\chi_{c}(x)\simeq\chi\left(x\right)$.
Thus, the method we describe above to explicitly compute $\mathcal{O}_{c}$
provides a way of effectively approximating $\chi$ for generic models.

\begin{figure}[h]
\centering{}\includegraphics[width=1\columnwidth]{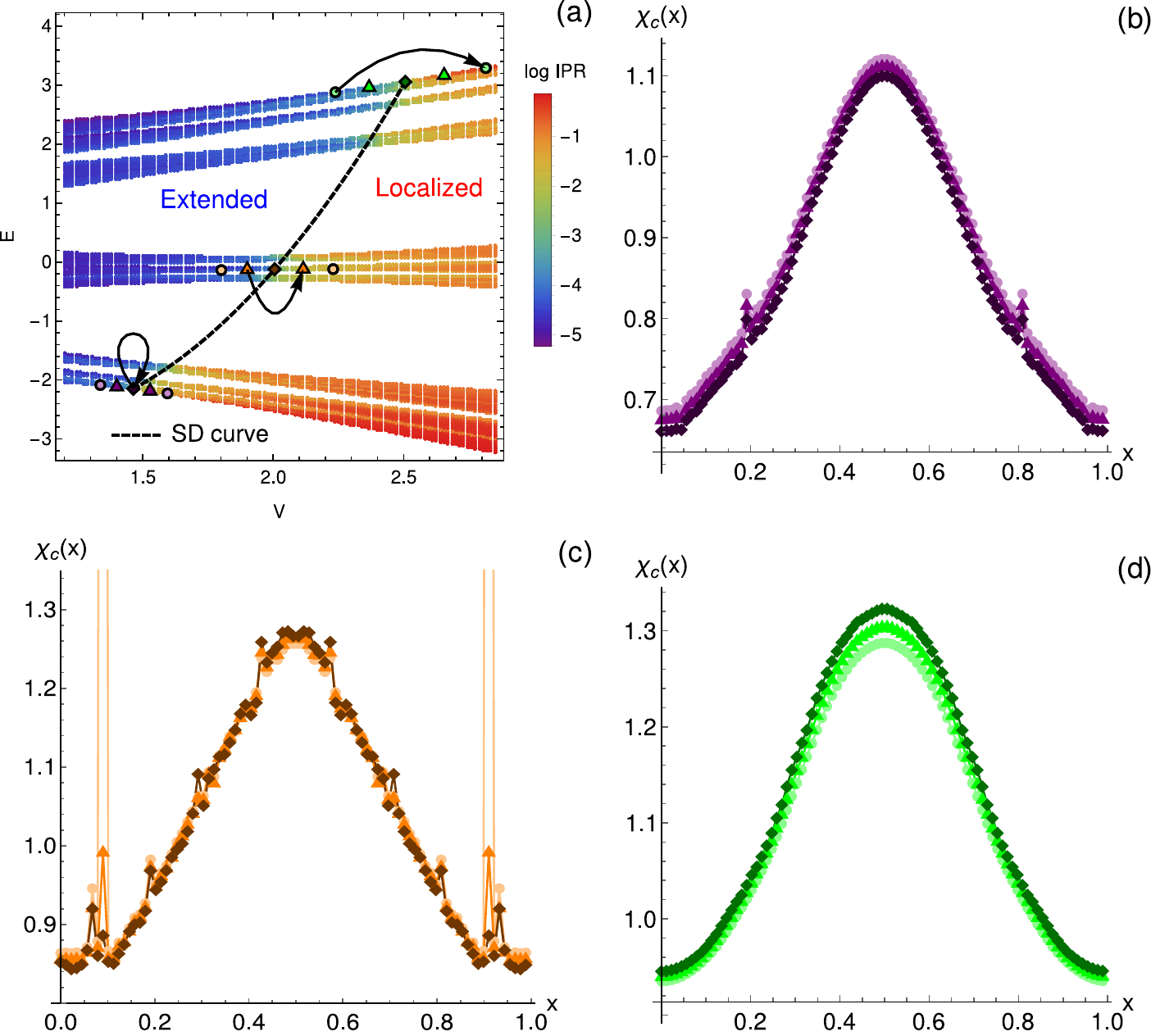}\caption{(a) IPR for the AA-NNN, defined in Eq.$\,$\ref{eq:EAAM_H}, with
$t_{2}=0.1$ and selected dual points. These include points in the
SD curve, that are self-dual, and non-self-dual dual points in the
extended and localized phases represented by the same marker. (b-d)
Eigenvalues of matrices $\mathcal{O}_{c}$ defined in Eq.$\,$\ref{eq:chi_c_Oc},
labeled as $\chi_{c}(x_{p})$, computed for the dual points in (a),
for the CA with $\tau_{c}=55/89$. The points $\chi_{c}(x_{p})$ in
(b-d) are plotted with the markers of the corresponding dual points
in (a).\label{fig:fig11}}
\end{figure}

To illustrate our findings, we show in Fig.$\,$\ref{fig:fig10} the
function $\chi_{c}(x)$ obtained for the AA-NNN and the LM. The SD
points are indicated in the phase diagrams in Figs.$\,$\ref{fig:fig10}(a,d).
The CA's states used to define $\mathcal{O}_{c}$ were selected so
that their energies were the closest possible to the selected points
in the phase diagram. For the LM, $\mathcal{O}_{c}$ was defined from
the real-space and dual wave functions and their translations within
the upper layer (containing $n_{1}$ sites by definition). A similar
duality transformation could be defined for the lower layer, from
the real-space and dual wave functions within this layer.

Figs.$\,$\ref{fig:fig10}(b,e) show that $\chi_{c}(x)$ is well-behaved
in the sense that interpolations of the points $\chi_{c}(x_{p})$
are essentially independent of the approximant. This means that the
thermodynamic-limit function $\chi(x)$ can be obtained even for low-order
CA. However, such clean behaviour only occurs for fine-tuned states,
such as the highest-energy states marked with points $P_{A}$ and
$P_{C}$ in Figs.$\,$\ref{fig:fig10}(a,d). For other critical states,
$\chi_{c}(x)$ expresses a more irregular behaviour, as shown in the
examples of Figs.$\,$\ref{fig:fig10}(c,f). In this case, $\chi_{c}(x)$
has singularity-like features that may only be resolved for large-order
CA. The non-triviality of $\chi_{c}(x)$ in these cases shows that
finding analytical descriptions of duality transformations in certain
models can be challenging.

It is worth reinforcing that the duality transformation defined above
is not restricted to SD points, and can be defined for any dual points
$P$ and $P'$. An example of $\chi_{c}(x)$ computed at both SD and
non-SD points is given in Fig.$\,$\ref{fig:fig11}. Note that $\chi_{c}(x)$
changes smoothly as we move away from the critical point.

As a final remark, we note that we used here examples of duality transformations
for which $(\phi_{0},k_{0})=(0,0)$. In general, $(\phi_{0},k_{0})$
can take other values and that is observed for other regions in the
phase diagram, as for the AA-NNN and the LM. Typically, branches of
the phase diagram that share continuously connected mobility edges
(SD curves) share the same $(\phi_{0},k_{0})$. For instance, in the
example of Fig.$\,$\ref{fig:main}(d), the energy bands that cross
the upper branch with positive concavity have $(\phi_{0},k_{0})=(0,0)$,
while for the lower branch with negative concavity we have $(\phi_{0},k_{0})=(\pi(n_{1}+n_{2})/n_{1},0)$,
see Appendix.$\,$\ref{sec:wave_duality_remarks}.

\section{Discussion}

\label{sec:Discussion}

We have shown several examples of 1D QPS for which transitions between
extended and localized phases are associated with local hidden dualities.
We could not find any instance where such hidden duality was not present,
at least sufficiently close to localization-delocalization critical
point. Therefore, we conjecture that such dualities exist generically
in 1D QPS and provide further arguments in the following.\lyxdeleted{Miguel}{Sun Jun  5 14:14:45 2022}{ }

The existence of dualities is deeply connected with the invariance
of QPS under $\varphi$ and $\kappa$-shifts that have a period inversely
proportional to the unit cell's size. In generic QPS, this invariance
may be more complicated than in the previous examples. For instance,
for EAAM, besides the on-site quasiperiodic energies, we could have
other inhomogeneities between different sites. For concreteness, we
consider $N$ consecutive sites that have different on-site energies,
in addition to the quasiperiodic potential ($N$ should be bounded,
otherwise the system would be disordered). In that case, the $\phi$-periodicity
for a CA would change from $2\pi/n_{1}$ to $2\pi N/n_{1}$. For the
latter, we should define $\varphi=n_{1}\phi/N$, so that $\Delta\varphi=2\pi$
contains a period of the energy dispersion. Obviously, the unit cell
of the lowest-order CA is constrained to have, at least, $n_{1}=N$
sites.

We tested this scenario in two different ways. The first was a simple
example of the AAM with an additional staggered potential between
any two consecutive sites ($\eta$-AAM, see Appendix$\,$\ref{sec:Emergence-of-duality}).
In this case, after defining $\varphi=n_{1}\phi/2$, we observe that
even though no signs of duality exist for $\tau_{c}=1/2$ (the lowest-order
possible CA), $\tau_{c}=21/34$ already hosts SD points for which
the FS is invariant under interchanging the new $\varphi$ with $\kappa$.

We studied a second example of an AAM with additional different on-site
energies for any three consecutive sites (3ICS-AAM, see Appendix$\,$\ref{sec:Emergence-of-duality}).
In this case, $\varphi=n_{1}\phi/3$ and there is an important qualitative
difference with respect to the other examples that we studied. In
particular, the center of rotation $(\varphi_{0},\kappa_{0})$ depends
smoothly on energy and on the Hamiltonian's parameters. Remarkably,
besides this difference, the FS for a CA of high-enough order around
criticality also has the universal behavior shown in Fig.$\,$\ref{fig:phi_k_curves}(b),
being perfectly described locally by the model in Eq.$\,$\eqref{eq:renorm_model_main}.
The SD points again perfectly matched the mobility edge of the limiting
QPS.

All the results shown so far are in favour of the two main ideas of
this work: (i) CA of generic QPS share the Aubry-André FS universality
around criticality; (ii) Transitions between extended and localized
phases in 1D QPS are associated with hidden dualities that manifest
around criticality.

Regarding (i), we conjecture that the FS universality of CA around
criticality is connected to an existing universality in localization-delocalization
transitions in 1D QPS. For a given CA, the ``critical'' FS of Fig.$\,$\ref{fig:phi_k_curves}(b)
(dispersive both in $\kappa$ and $\varphi$) occur only within a
narrow region of parameters around criticality. This region shrinks
as the unit cell is increased, eventually collapsing to the QPS's
critical points.

The critical universal FS that we observe for large unit cells are
compatible with the vanishing of all the harmonics in $\varphi$ and
$\kappa$ other than the fundamental in this limit (so that Eq.$\,$\eqref{eq:renorm_model_main}
is valid). The implications of the vanishing of non-fundamental harmonics
near criticality for generic 1D QPS will be discussed elsewhere \citep{prepar}.

The hidden dualities mentioned in (ii) are deeply connected with the
symmetry of CA to displacements encoded in phase $\phi$. This is
apparent in our definition of $\mathcal{O}_{c}$ in Eq.$\,$\eqref{eq:definition_Oc}.
Such symmetry exists for CA of generic QPS, therefore the presence
of hidden dualities is expected to arise for a much larger set of
systems than the ones studied in this work. Our findings open a new
route for the study and understanding 1D QPS. In particular, they
provide a working criterion for the existence of mobility edges and
a way of generating models of QPS with analytical phase diagrams by
explicitly creating dualities in their simplest CA \citep{prepar}.

The scope of this work was to characterize quasiperiodicity-driven
localization-delocalization transitions for which the FS for different
CA can be characterized by the effective model in Eq.$\,$\ref{eq:renorm_model_main},
at least close enough to the transition and for a large enough unit
cell. Such transitions are associated with hidden dualities manifested
in $\varphi$ and $\kappa$, that can be clearly seen through this
effective model. However, in some cases, the effective model can be
more general. This is the case for models that have phases with critical
states over a finite range of parameters (and not only at fine-tuned
points as at the critical points of localization-delocalization transitions).
In this case, hidden dualities can also arise and may still be captured
analytically in some models. We will cover this case in detail elsewhere
\citep{prepar}. On the other hand, there may be some models for which
no localization-delocalization transition exists for any finite quasiperiodicity
and therefore no hidden duality can be defined. An example is the
Maryland model that we discuss in Appendix$\,$\ref{sec:maryland},
for which all the eigenstates are localized for any strength of the
quasiperiodic potential. In this case, the effective model is given
in terms of a $\tan(\varphi/2)$ term instead of $\cos(\varphi)$.
Since there is no $\kappa$-dependent term of the type $\tan(\kappa/2)$,
it is clear that no hidden duality exists in this case. Nonetheless,
we can still see that the renormalized coupling that multiplies the
$\kappa$-dependent term becomes irrelevant with respect to the $\varphi$-dependent
term as the unit cell is increased for any finite $V$. This is in
accordance with our view of the localized regime, where only the latter
coupling should survive in the limit of large unit cell.\lyxdeleted{Miguel}{Sun Jun  5 14:14:45 2022}{ }

A natural next step is to understand if our ideas extend to more complex
1D systems and to higher dimensions where more exotic localization
phenomena arise \citep{Devakul2017,PhysRevB.100.144202,Szabo2020}.
More complex 1D systems may include (i) models with internal degrees
of freedom, for which the unit cell contains more than one site even
in the absence of the quasiperiodic term in the Hamiltonian; and (ii)
models with multiple quasiperiodic potentials \footnote{A simple example is a model with two quasiperiodic potentials characterized
by different irrationals $\tau_{1}\neq\tau_{2}$ (BCM). Interestingly,
CA of the $\eta$-AAM are also CA of the BCM, with $\tau_{1}=\tau$
and $\tau_{2}$ an irrational close to $1/2$. The same holds for
the 3ICS-AAM, but in that case, $\tau_{2}$ can be an irrational close
to $1/3$ or $2/3$. Therefore, the existence of hidden dualities
in the $\eta$-AAM and 3ICS-AAM suggests that they are also present
in CA of the BCM.}. Our results for the $\eta$-AAM and the 3ICS-AAM, examples of type-(i)
models, suggest that hidden dualities are also present for models
of this type close to localization-delocalization transitions. Another
interesting open question is the influence of interactions on generalized
dualities. Mobility edges for interacting systems were found to depart
from the single-particle description in some regimes \citep{PhysRevLett.122.170403,PhysRevLett.126.040603}.
Thus a natural question is whether these many-body mobility edges
are also associated with hidden dualities that depart from CA.
\begin{acknowledgments}
The authors acknowledge partial support from Fundação para a Ciência
e Tecnologia (Portugal) through Grant and UID/CTM/04540/2019. BA and
EVC acknowledge partial support from FCT-Portugal through Grant No.
UIDB/04650/2020. MG acknowledges further support through the Grant
SFRH/BD/145152/2019. BA acknowledges further support from FCT-Portugal
through Grant No. CEECIND/02936/2017. We finally acknowledge the Tianhe-2JK
cluster at the Beijing Computational Science Research Center (CSRC),
the Baltasar-Sete-Sóis cluster, supported by V. Cardoso's H2020 ERC
Consolidator Grant no. MaGRaTh-646597, and the OBLIVION supercomputer
(based at the High Performance Computing Center - University of Évora)
funded by the ENGAGE SKA Research Infrastructure (reference POCI-01-0145-FEDER-022217
- COMPETE 2020 and the Foundation for Science and Technology, Portugal)
and by the BigData@UE project (reference ALT20-03-0246-FEDER-000033
- FEDER and the Alentejo 2020 Regional Operational Program. Computer
assistance was provided by CSRC, CENTRA/IST and the OBLIVION support
team.
\end{acknowledgments}

\bibliographystyle{apsrev4-1}
\bibliography{1D_Hidden_SD_Paper}




\clearpage\onecolumngrid

\beginsupplement
\begin{center}
\textbf{\large{}APPENDICES\vspace{0.1cm}
}{\large\par}
\par\end{center}

\vspace{0.3cm}

\twocolumngrid

\appendix

\section{Scaling analysis}

\begin{figure*}[t]
\centering{}\includegraphics[width=0.8\paperwidth]{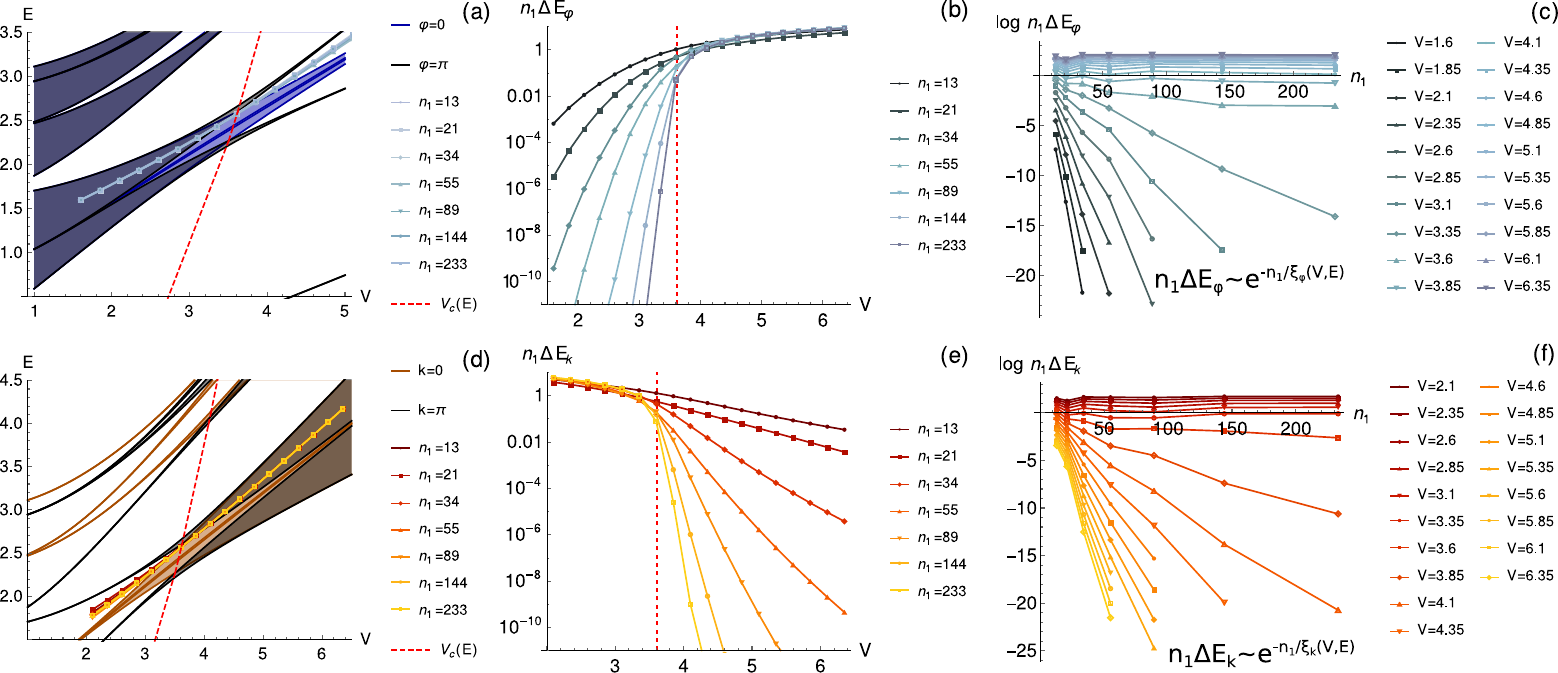}\caption{Examples of $\Delta E_{\varphi}\equiv\Delta E_{\varphi}(\kappa^{*}=0)$
and $\Delta E_{\kappa}\equiv\Delta E_{\kappa}(\varphi^{*}=0)$ for
the AA-NNN with $t_{2}=0.5$. These quantities were computed for the
points marked in (a,b). In (b,e) we plot fixed-$n_{1}$ curves for
variable $V$ and in (c,f) we do the opposite. The energy bands shown
in (a,b) are for $\tau_{c}=8/13$. See text for a more detailed explanation.\label{fig:scaling_dek_dephi_procedure}}
\end{figure*}

\label{sec:scaling_anal}

In this section, we show that it is possible to carry out a scaling
analysis in terms of the $\varphi$ and $\kappa$-dependent energy
dispersions, $\Delta E_{\varphi}(\kappa^{*})=|E_{\varphi=\pi}(\kappa^{*})-E_{\varphi=0}(\kappa^{*})|$
and $\Delta E_{\kappa}(\varphi^{*})=|E_{\kappa=\pi}(\varphi^{*})-E_{\kappa=0}(\varphi^{*})|$.
The aim is to inspect how these dispersions change upon increasing
the order of the approximant.

To carry out the scaling analysis, it is important to study different
CA at the same point in the phase diagram, which may be challenging.
A possible way to do it is to recall that in the extended phase of
the limiting QPS, for a CA of high-enough order, $E(\kappa,\varphi)\approx E(\kappa)$,
while in the localized phase, $E(\kappa,\varphi)\approx E(\varphi)$.
As seen in the main text, using larger approximants in such cases
is similar to a band folding, in the $\kappa$ or $\varphi$ direction.
We can fix a point $(\bm{\lambda},E)$ starting deep in the extended
phase and, considering that $E(\kappa,\varphi)\approx E(\kappa)$,
compute the wave vectors corresponding to such point for different
approximants, $\kappa^{*}(\bm{\lambda},E,\tau_{c})$. We can then
see how states with fixed $\kappa^{*}$ evolve with $\bm{\lambda}$
for different approximants. As long as we are sufficiently away from
criticality or using a CA of high-enough order, if there is a state
at point $(\bm{\lambda},E)$, there will also be a state at the same
point for a higher-oder CA. In that case, a finite-size scaling analysis
can be carried out at point $(\bm{\lambda},E)$, by increasing the
size of the unit cell (the CA's order).

\begin{figure}[h]
\centering{}\includegraphics[width=1\columnwidth]{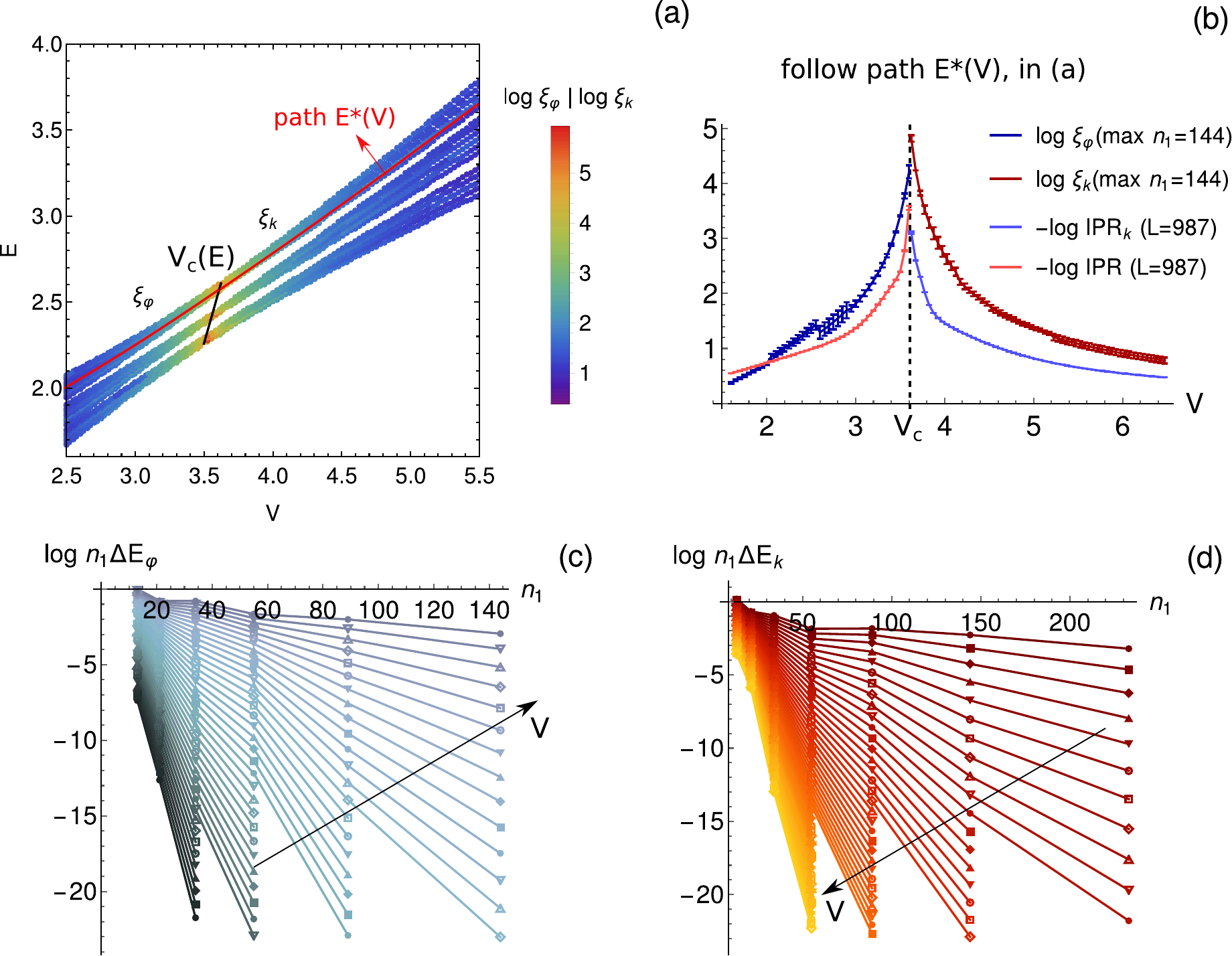}\caption{(a) Density plot of $\log\xi_{\varphi}$ in the localized phase, and
$\log\xi_{\kappa}$ in the extended phase. This example is for the
the AA-NNN with $t_{2}=0.5$. The full black line indicates the mobility
edge. (b) $\log\xi_{\varphi}$ and $\log\xi_{\kappa}$ computed for
the path shown in (a) and comparison with the ${\rm IPR}$ and ${\rm IPR}_{\kappa}$
along the same path, for a system with $L=987$ sites. (c,d) $\log n_{1}\Delta E_{\varphi}$
and $\log n_{1}\Delta E_{\kappa}$ as a function of the unit cell's
size, $n_{1}$. $V$ increases in the direction of the arrow. Note
that the apparent discontinuities in (b) for $\xi_{\kappa}$ and $\xi_{\varphi}$
can be understood in (c/d): for a CA of high-enough order, $\Delta E_{\varphi}$
and $\Delta E_{\kappa}$ fall below machine precision and such data
points cannot be considered in the fit to extract $\xi_{\kappa}$
and $\xi_{\varphi}$. Therefore, the fitting procedure involves different
approximants, depending on the coupling $V$. For the fit, we considered
the 3 approximants corresponding to the smallest $\Delta E_{\varphi}$
or $\Delta E_{\kappa}$. As we are not in the thermodynamic limit
and there is still some dependence on the size of the unit cell, a
change in the group of fitted approximants can translate in apparent
slight discontinuities in $\xi_{\kappa}$ and $\xi_{\varphi}$. \label{fig:scaling_dek_dephi_IPR_IPRk}}
\end{figure}

As an example of application, we use the AA-NNN with $t_{2}=0.5$.
In Fig.\ref{fig:scaling_dek_dephi_procedure}(a), we can see that
the points of fixed $\kappa^{*}(V,E,\tau_{c})$ in the figure change
with $V$, starting at the smallest-$V$ point. These points fall
on top of each other for different approximants as a consequence of
the arguments given above. The same can be done in the localized phase,
but there one must compute the phase $\varphi^{*}(V,E,\tau_{c})$
for a point $(V,E)$ deep in the localized phase, and then follow
this state upon varying $V$. An example of application is in Fig.\ref{fig:scaling_dek_dephi_procedure}(d).

We can finally compute $\Delta E_{\varphi}\equiv\Delta E_{\varphi}(\kappa^{*}=0)$
and $\Delta E_{\kappa}\equiv\Delta E_{\kappa}(\varphi^{*}=0)$ for
different $(V,E)$ points, which are exponentially small respectively
in the extended and localized phases. $\Delta E_{\varphi}$ is shown
in Figs.$\,$\ref{fig:scaling_dek_dephi_procedure}(b,c) for the points
represented in Fig.$\,$\ref{fig:scaling_dek_dephi_procedure}(a)
and $\Delta E_{\kappa}$ is in Figs.$\,$\ref{fig:scaling_dek_dephi_procedure}(e,f)
for the points represented in Fig.$\,$\ref{fig:scaling_dek_dephi_procedure}(d).
In the extended phase, we have $\Delta E_{\kappa}\sim n_{1}^{-1}$,
while in the localized phase, $\Delta E_{\varphi}\sim n_{1}^{-1}$.
This is because the bands of higher-order CA are, approximately, just
being folded. Therefore, in Figs.$\,$\ref{fig:scaling_dek_dephi_procedure}(b,c,e,f)
we plot the quantities $n_{1}\Delta E_{\kappa}$ and $n_{1}\Delta E_{\varphi}$.
In Figs.$\,$\ref{fig:scaling_dek_dephi_procedure}(c,f), we see that
in the localized phase, $n_{1}\Delta E_{\kappa}\sim e^{-n_{1}/\xi_{\kappa}(V,E)}$
and in the extended phase, $n_{1}\Delta E_{\varphi}\sim e^{-n_{1}/\xi_{\varphi}(V,E)}$.
These exponential decays define the correlation lengths $\xi_{\kappa}$
and $\xi_{\varphi}$, respectively in the localized and extended phases.
Such length scales can be extracted by fitting the data in Figs.$\,$\ref{fig:scaling_dek_dephi_procedure}(c,f).

By considering other paths similar to the ones in Figs.$\,$\ref{fig:scaling_dek_dephi_procedure}(a,d),
we can compute the correlation lengths at different points in the
phase diagram. An example of such computation is shown Fig.$\,$\ref{fig:scaling_dek_dephi_IPR_IPRk}.
There, we see that $\xi_{\kappa}$ and $\xi_{\varphi}$ diverge at
the critical point. In Fig.$\,$\ref{fig:scaling_dek_dephi_IPR_IPRk}(b)
we compare the correlation lengths with the inverse participation
ratio (IPR) and the momentum-space inverse participation ratio (${\rm IPR}_{\kappa}$).
The former is defined in the main text, while the latter is defined
as ${\rm IPR}_{\kappa}(E)=(\sum_{n}|\psi_{n}^{\kappa}(E)|^{2})^{-2}\sum_{n}|\psi_{n}^{\kappa}(E)|^{4}$,
where $\psi_{n}^{\kappa}(E)$ is the Fourier-transform of the real-space
wave function. The IPR and ${\rm IPR}_{\kappa}$ also diverge at the
critical point and can define correlation lengths respectively in
the localized and extended phase. $\xi_{\kappa}$ and $\xi_{\varphi}$
provide alternative definitions for the correlation lengths without
the explicit knowledge of the wave function.

\section{Finding SD points and Mapping dual points in the phase diagram}

\label{sec:duality_transf_correct_method}

We have seen in the main text that the knowledge of the CA's FS in
the $(\varphi,\kappa)$ plane is a powerful way to identify dual points
in the phase diagram, and, in particular, SD points. This can be done
analytically for low-order approximants, but may become challenging
for higher-order ones. In this section we provide methods to compute
dual points for generic models and CA.

\subsection{Renormalized single-band model}

\begin{figure}[h]
\centering{}\includegraphics[width=1\columnwidth]{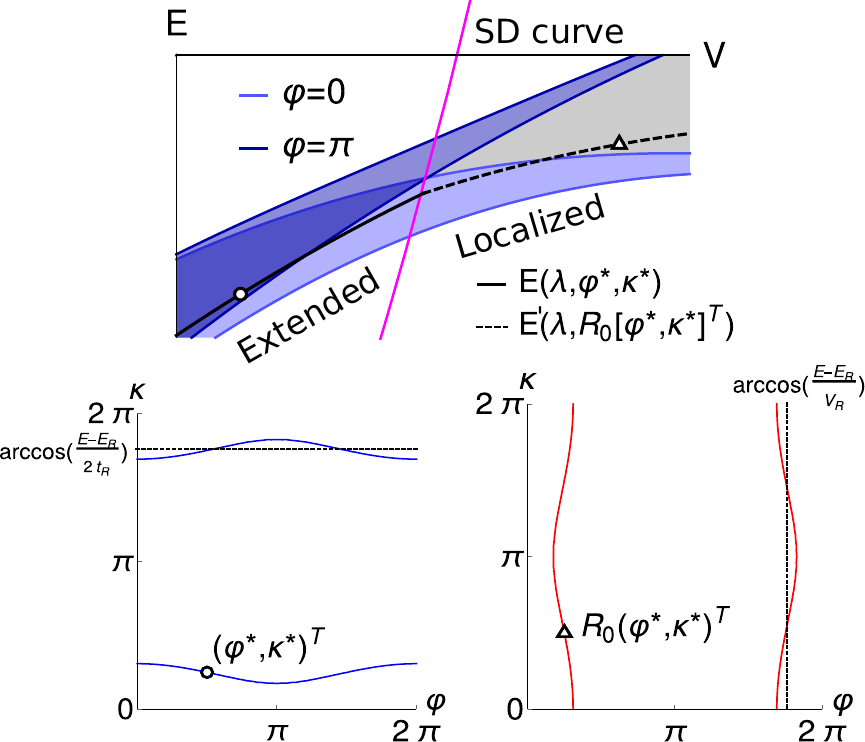}\caption{Sketch to illustrate how dual points can be found. For a given energy
band, fixing the phases $\varphi^{*}$ and $\kappa^{*}$, we can define
the curve $E(\bm{\lambda},\varphi^{*},\kappa^{*})$ in the extended
phase (full black line) and the curve $E'(\bm{\lambda},\mathcal{R}_{0}[\varphi^{*},\kappa^{*}]^{T})$
in the localized phase (dashed line). These curves cross exactly at
the SD curve (magenta). The FS associated with arbitrarily chosen
dual points within these curves (circle and triangle) are shown in
the bottom panel. Note that these dual points automatically satisfy
Eqs.$\,$\ref{eq:condDual1},\ref{eq:condDual2}, with the latter
being automatically fulfilled due to the choice of curves $E$ and
$E'$.\label{fig:dual_points_explain}}
\end{figure}

For a CA of high-enough order and close enough to quasiperiodicity-driven
localization-delocalization transitions, we have observed that the
FS in the $(\varphi,\kappa)$ plane reduces to the sinusoids shown
in Figs.$\,$\ref{fig:Ecuts_ext_loc} and~\ref{fig:phi_k_curves},
irrespectively of the model that we studied. Such FS can be simply
described through the following expression for the energy dispersion
around energy $E_{R}$:

\begin{equation}
\begin{aligned}E= & E_{R}(\bm{\lambda},E)+V_{R}(\bm{\lambda},E)\cos[\varphi-\varphi_{0}(\bm{\lambda},E)]\\
 & +2t_{R}(\bm{\lambda},E)\cos[\kappa-\kappa_{0}(\bm{\lambda},E)]
\end{aligned}
\label{eq:renormalized_model}
\end{equation}

\noindent where $V_{R}$ and $t_{R}$ are renormalized couplings,
$(\varphi_{0},\kappa_{0})$ is the point around which the FS is invariant
under a suitable $\varphi\leftrightarrow\kappa$ interchange at SD
points, and $\bm{\lambda}$ are parameters of the Hamiltonian. This
is the energy dispersion of a renormalized single-band AAM, being
the generalization of the renormalized model defined in Ref.$\,$\citep{Szabo2018}
(due to the energy-dependence of the renormalized couplings). It is
important to let the renormalized couplings depend both on $\bm{\lambda}$
and $E$. If the sub-bands were perfect Aubry-André single-bands (from
AAM with $\tau_{c}=1$), there should be no energy dependence. However,
in that case, there would be no mobility edges. To see this, just
notice that, for fixed $E$:

\begin{equation}
\cos(\varphi-\varphi_{0})=\frac{E-E_{R}}{V_{R}}-\frac{2t_{R}}{V_{R}}\cos(\kappa-\kappa_{0})
\end{equation}

\begin{equation}
\cos(\kappa-\kappa_{0})=\frac{E-E_{R}}{2t_{R}}-\frac{V_{R}}{2t_{R}}\cos(\varphi-\varphi_{0}).
\end{equation}
The generalized duality conditions that map points $(\bm{\lambda},E)$
to dual points $(\bm{\lambda}',E')$ are

\begin{equation}
\frac{2t_{R}(\bm{\lambda},E)}{V_{R}(\bm{\lambda},E)}=\frac{V_{R}(\bm{\lambda}',E')}{2t_{R}(\bm{\lambda}',E')}\label{eq:condDual1}
\end{equation}

\begin{equation}
\frac{E-E_{R}(\bm{\lambda},E)}{V_{R}(\bm{\lambda},E)}=\frac{E'-E_{R}(\bm{\lambda}',E')}{2t_{R}(\bm{\lambda}',E')}.\label{eq:condDual2}
\end{equation}
If $t_{R}$ and $V_{R}$ only depend on $\bm{\lambda}$, then the
first equation fixes the transformation $\bm{\lambda}'(\bm{\lambda})$
which becomes energy independent. This is not what is observed in
generic models.

Equations$\,$\eqref{eq:condDual1} and~\eqref{eq:condDual2} allow
us to find dual points in the phase diagram, including SD points $(\bm{\lambda}_{c},E_{c})$
that satisfy

\begin{equation}
V_{R}(\bm{\lambda}_{c},E_{c})=2t_{R}(\bm{\lambda}_{c},E_{c}).\label{eq:condDual1-1}
\end{equation}
In practice, if we compare curves $E(\bm{\lambda}(t),\varphi^{*},\kappa^{*})$
with curves $E'(\bm{\lambda}'(t),\mathcal{R}_{0}[\varphi^{*},\kappa^{*}]^{T})$,
where $\bm{\lambda}(t)$ and $\bm{\lambda}'(t)$ are parametric curves
in the space of parameters $\bm{\lambda}$ constructed by fixing the
phases $\varphi^{*},\kappa^{*}$, the condition in Eq.$\,$\eqref{eq:condDual2}
is immediately satisfied for all points in those curves. Then we just
need to identify dual points within these curves that satisfy Eq.$\,$\eqref{eq:condDual1},
or in the case of SD points, Eq.$\,$\eqref{eq:condDual1-1}. An illustration
of the curves $E(\bm{\lambda})$ and $E'(\bm{\lambda}')$, along with
the FS on both sides of the transition, is shown in Fig.$\,$\ref{fig:dual_points_explain}.
A detailed discussion on how to determine the renormalized couplings
is given below.

\begin{figure}[h]
\centering{}\includegraphics[width=1\columnwidth]{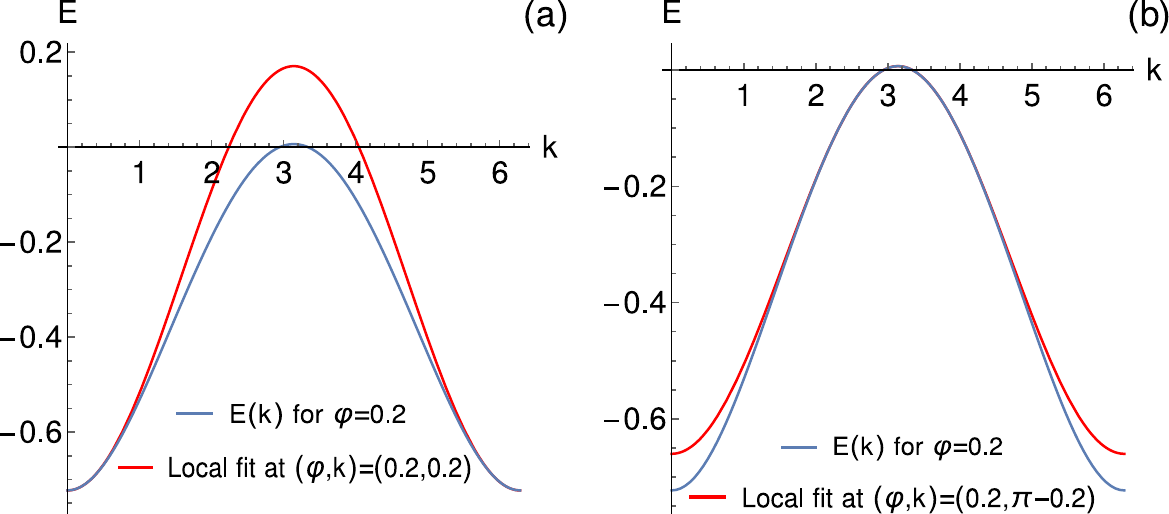}\caption{Examples of computation of the renormalized parameters $V_{R},t_{R}$
and $E_{R}$. We locally estimate these parameters for a given sub-band,
with the procedure introduced in Sec.$\,$\ref{subsec:estimate_VR_tR_ER}.
In this figure we show examples of such estimation for $(\kappa,\varphi)=(0.2,0.2)$
(a) and $(\kappa,\varphi)=(\pi-0.2,0.2)$, for $\delta\kappa=\delta\varphi=0.15$.
The obtained model (in red) is energy-dependent. These examples are
for the AAM with $\tau_{c}=2/3$, for the sub-band of intermediate
energy. \label{fig:cossine_fitting}}
\end{figure}

\subsection{Estimation of the renormalized couplings}

\label{subsec:estimate_VR_tR_ER}

\begin{figure}[h]
\centering{}\includegraphics[width=1\columnwidth]{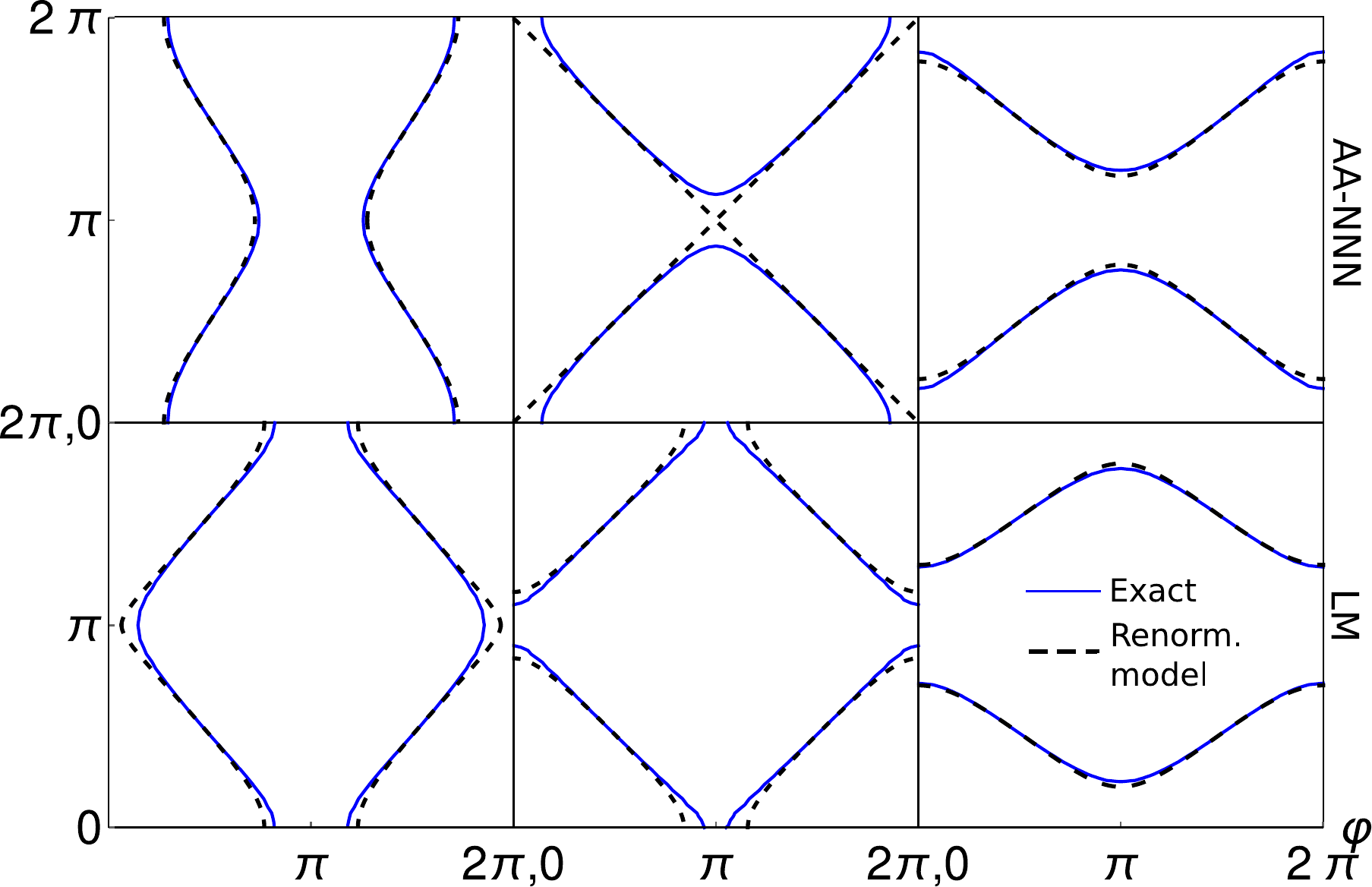}\caption{Examples of FS obtained for the CA defined by $\tau_{c}=2/3$, for
some selected points in the phase diagram. The exact result (full
blue) is compared with the FS obtained with the single-band model
in Eq.$\,$\eqref{eq:renormalized_model}. The results in the upper
row are for the AA-NNN, with $t_{2}=0.5$, lowest energy band and
(from left to right) $V=3.5;2.76;2.4$, $(\varphi^{*},\kappa^{*})=(\pi/2,\pi/2)$.
The results in the lower row are for the LM, with $V=3$, $\Lambda=2.5$
and (from left to right) $\xi=0.35;0.365;0.4$, $(\varphi^{*},\kappa^{*})=(\pi/2,\pi/2)$.
We used $\delta\kappa=\delta\varphi=\pi/100$.\label{fig:FS_exact_vs_RenormModel_BAD}}
\end{figure}

\begin{figure}[h]
\centering{}\includegraphics[width=1\columnwidth]{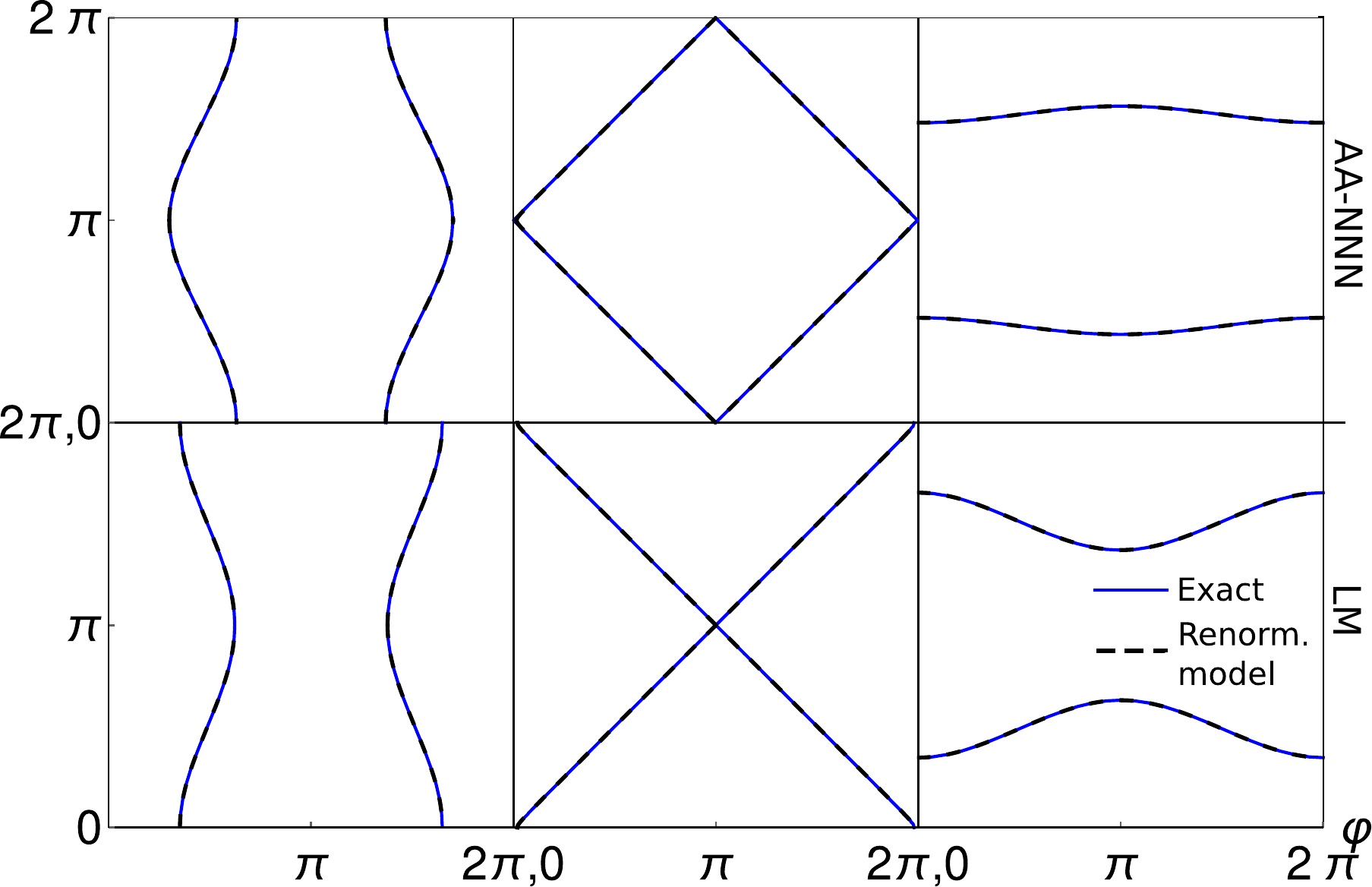}\caption{Examples of FS obtained for the CA defined by $\tau_{c}=13/21$, for
some selected points in the phase diagram. The exact result (full
blue) is compared with the FS obtained with the single-band model
in Eq.$\,$\eqref{eq:renormalized_model}. The results in the upper
row are for the AA-NNN, with $t_{2}=0.5$, $2^{\textrm{nd}}$ lowest
energy band and (from left to right) $V=2.2;2.0915;1.8$, $(\varphi^{*},\kappa^{*})=(\pi/2,\pi/2);(\pi,0);(\pi/2,\pi/2)$.
The results in the lower row are for the LM, with $V=3$, $\Lambda=2.5$
and (from left to right) $\xi=0.35;0.3655;0.38$, $(\varphi^{*},\kappa^{*})=(\pi/2,\pi/2);(\pi,\pi);(\pi/2,\pi/2)$.
We used $\delta\kappa=\delta\varphi=\pi/100$.\label{fig:FS_exact_vs_RenormModel}}
\end{figure}

We here assume that point $(\varphi_{0},\kappa_{0})$ is known and
use new coordinates $(\varphi,\kappa)\rightarrow(\varphi-\varphi_{0},\kappa-\kappa_{0})$
such that $(\varphi_{0},\kappa_{0})$ is shifted into the origin.
For fixed $\bm{\lambda}=\bm{\lambda}^{*}$ and a given sub-band, we
choose a grid of points $(\varphi,\kappa)$. Let $(\varphi^{*},\kappa^{*})$
be one of such points. It fixes an energy $E^{*}=E(\bm{\lambda}^{*},\varphi^{*},\kappa^{*})$.
For the phase diagram point $(\bm{\lambda}^{*},E^{*})$, we estimate
the couplings as follows:

\begin{equation}
V_{R}=\frac{E(\varphi^{*}+\delta\varphi/2,\kappa^{*})-E(\varphi^{*}-\delta\varphi/2,\kappa^{*})}{\cos(\varphi^{*}+\delta\varphi/2)-\cos(\varphi^{*}-\delta\varphi/2)}\label{eq:VR_estimate}
\end{equation}

\begin{equation}
t_{R}=\frac{1}{2}\frac{E(\varphi^{*},\kappa^{*}+\delta\kappa/2)-E(\varphi^{*},k^{*}-\delta\kappa/2)}{\cos(\kappa^{*}+\delta\kappa/2)-\cos(\kappa^{*}-\delta\kappa/2)}\label{eq:tR_estimate}
\end{equation}

\begin{equation}
E_{R}=E(\varphi^{*},\kappa^{*})-2t_{R}\cos(\kappa^{*})-V_{R}\cos(\varphi^{*})\label{eq:ER_estimate}
\end{equation}

\noindent where $\delta\kappa,\delta\varphi\ll2\pi$. Note that for
a perfect sinusoidal sub-band, these couplings are energy-independent
{[}and therefore do not depend on the point $(\varphi^{*},\kappa^{*})${]}.
In general however, we will not have perfect sinusoidal sub-bands,
and the couplings become energy dependent as in the example shown
in Fig.$\,$\ref{fig:cossine_fitting} \footnote{It is important to be careful with points $P_{\varphi\kappa}$ around
which $E(\kappa)$ is even {[}such as $(\varphi^{*},\kappa^{*})=(0,0)${]}
otherwise the renormalized couplings as defined in Eqs.$\,$\eqref{eq:VR_estimate}-\eqref{eq:ER_estimate}
will diverge. In such cases, we can just choose points $P_{\varphi\kappa}+(\epsilon,\epsilon)$,
with small $\epsilon$.}.

\begin{figure*}[t]
\centering{}\includegraphics[width=0.75\textwidth]{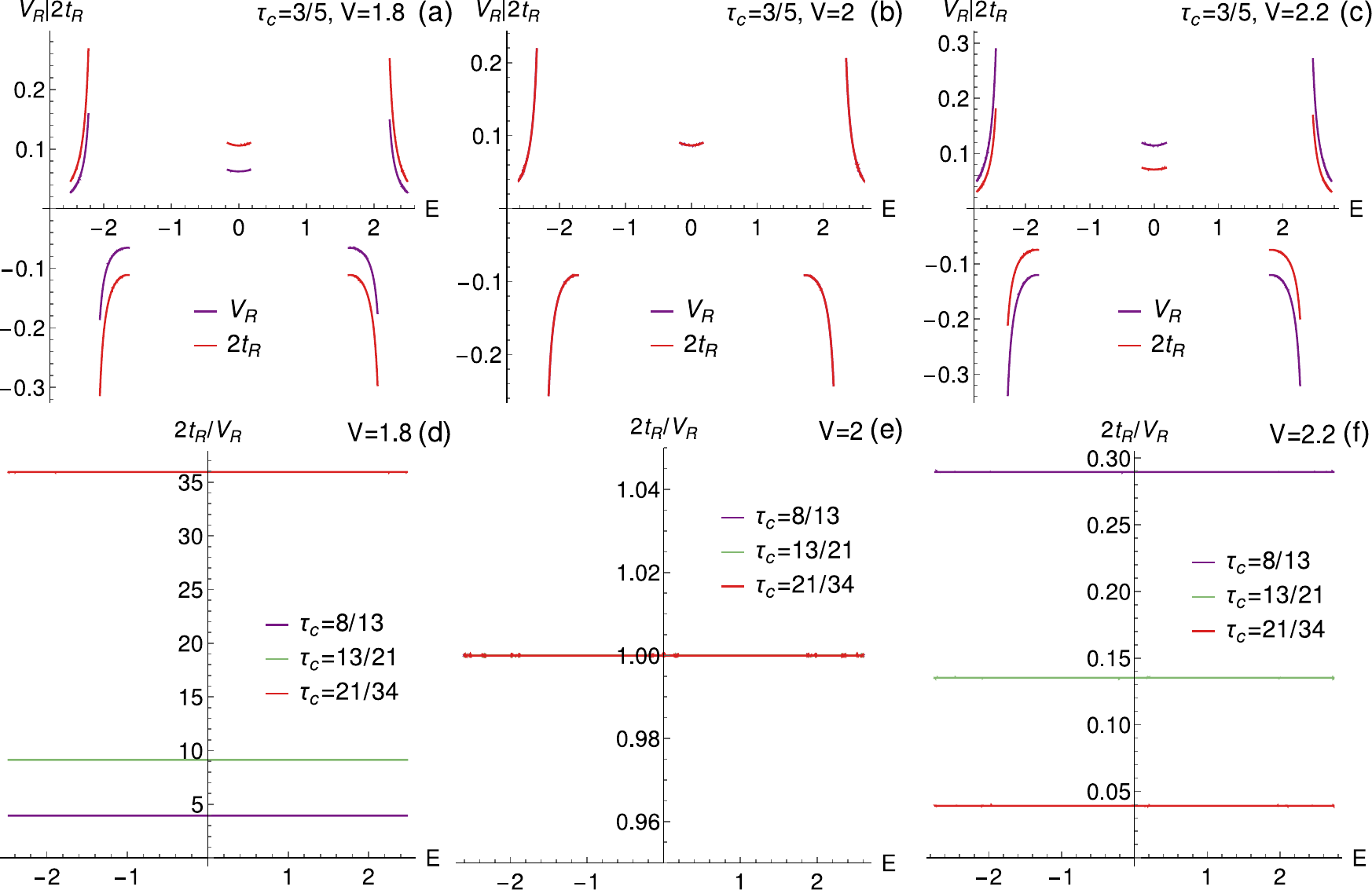}\caption{(a-c) Renormalized couplings $V_{R}$ and $2t_{R}$ for all the energy
bands of the CA with $\tau_{c}=3/5$, for $V=1.8$ (extended phase),
$V=2$ (critical point) and $V=2.2$ (localized phase). Note that
the couplings are energy dependent which means that the sub-bands
are not simple cosines. This is also the case for larger approximants.
(d-f) Ratio $2t_{R}/V_{R}$ as a function of energy, for different
approximants and $V$. This ratio is energy-independent as a consequence
of the lack of a mobility edge and the energy-independent correlation
lengths in the AAM \citep{AubryAndre}. For $V=1.8$, the ratio increases
with the CA's order, meaning that $V_{R}$ becomes irrelevant with
respect to $t_{R}$, as expected in the extended phase. For $V=2.2$,
$2t_{R}/V_{R}$ decreases with the approximant - $t_{R}$ becomes
irrelevant as expected in the localized phase. At the critical point,
$V=2$, the ratio is scale-invariant. \label{fig:renormalized_couplings_AAM}}
\end{figure*}

In Figs.$\,$\ref{fig:FS_exact_vs_RenormModel_BAD} and \ref{fig:FS_exact_vs_RenormModel}
we selected some example points in the phase diagram for the AA-NNN
and the LM, and compared the exact FS with the one obtained through
the renormalized single-band model in Eq.$\,$\eqref{eq:renormalized_model}.
For $\tau_{c}=2/3$, there are some deviations: the approximation
is good close to the chosen point $(\varphi^{*},\kappa^{*})=(\pi/2,\pi/2)$,
but fails away from it. For the higher-order approximant, $\tau_{c}=13/21$,
the agreement is perfect for the whole FS.

\subsection{Behaviour of renormalized couplings}

Here we exemplify the behaviour of the renormalized couplings for
the AAM. The idea of coupling renormalization has been previously
discussed in Ref.$\,$\citep{Szabo2018}.

We start by analyzing the examples provided in Fig.$\,$\ref{fig:renormalized_couplings_AAM}.
We see that in the extended phase, we have $|V_{R}|<|2t_{R}|$ {[}Fig.$\,$\ref{fig:renormalized_couplings_AAM}(a){]},
while in the localized phase, we have $|V_{R}|>|2t_{R}|$ {[}Fig.$\,$\ref{fig:renormalized_couplings_AAM}(c){]}.
At the critical point, $V_{R}=2t_{R}$ {[}Fig.$\,$\ref{fig:renormalized_couplings_AAM}(b){]}.
As the CA's order is increased, the ratio $2t_{R}/V_{R}$ increases
in the extended phase {[}Fig.$\,$\ref{fig:renormalized_couplings_AAM}(d){]}
and decreases in the localized phase {[}Fig.$\,$\ref{fig:renormalized_couplings_AAM}(f){]}.
At the critical point, it remains unity {[}Fig.$\,$\ref{fig:renormalized_couplings_AAM}(e){]}.
Even though we used the AAM as an example, this is what occurs in
generic models:
\begin{itemize}
\item The coupling $V_{R}$ ($t_{R}$) becomes irrelevant in the extended
(localized) phase as $\tau_{c}\rightarrow\tau$;
\item The couplings $V_{R}$ and $t_{R}$ are always relevant at the critical
point.
\end{itemize}
We have seen in Sec.$\,$\ref{sec:scaling_anal} that $\Delta E_{\varphi}\sim e^{-n_{1}/\xi_{\varphi}(V,E)}$
in the ballistic phase and $\Delta E_{\kappa}\sim e^{-n_{1}/\xi_{\kappa}(V,E)}$
in the localized phase. Therefore, we can also conclude that $V_{R}$
and $t_{R}$, which measure energy dispersion with $\varphi$ and
$\kappa$, respectively, also behave as $V_{R}\sim e^{-n_{1}/\xi_{\varphi}(V,E)}$
in the extended phase and $t_{R}\sim e^{-n_{1}/\xi_{\kappa}(V,E)}$
in the localized phase.

\subsection{Mapping dual points}

\label{subsec:duality_transform_through_renormalized_couplings}

We have already seen that the conditions in Eqs.$\,$\eqref{eq:condDual1}
and~\eqref{eq:condDual2} allow us to map dual points. Here we present
examples of application for the AAM and the AA-NNN.

\begin{figure}[h]
\centering{}\includegraphics[width=1\columnwidth]{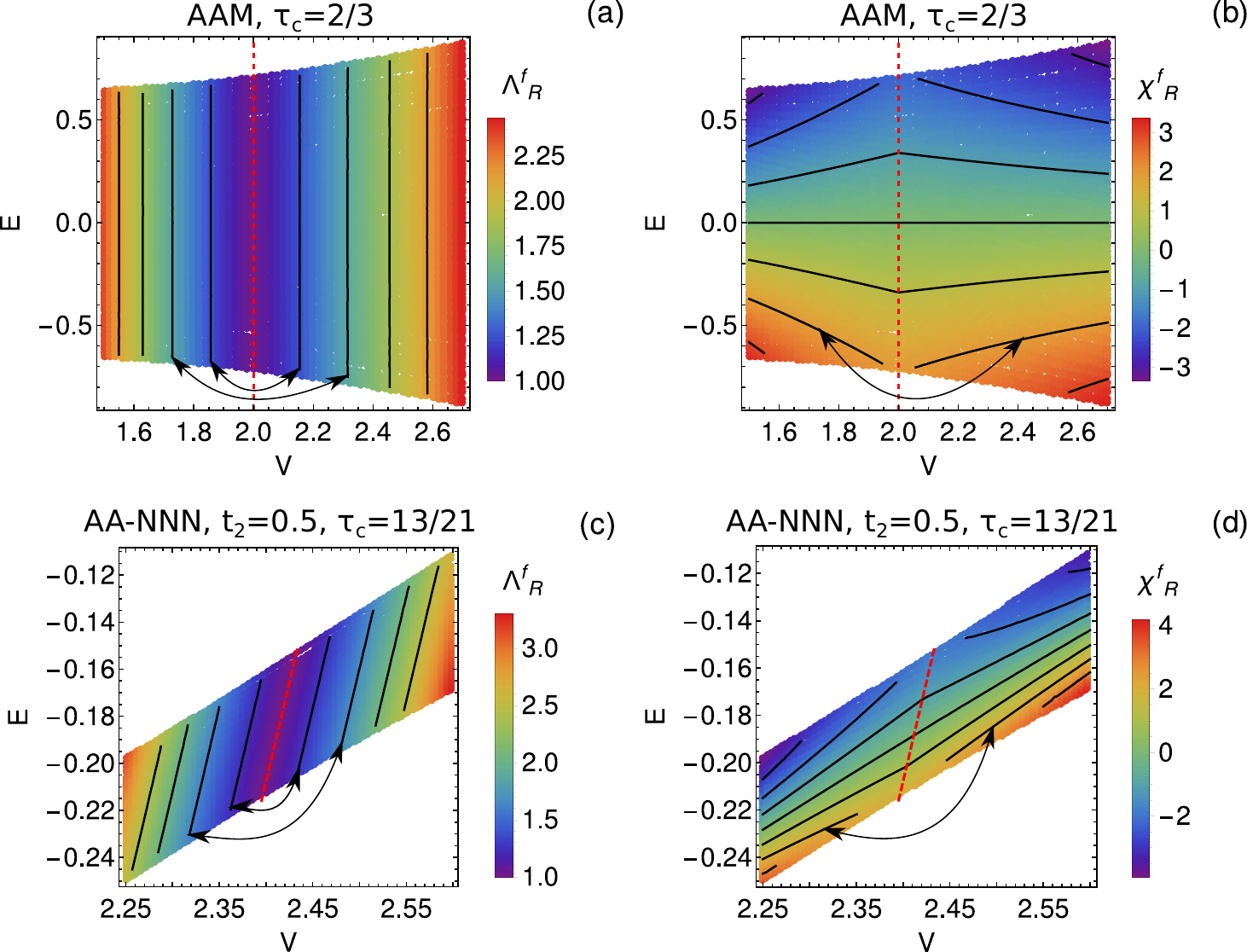}\caption{Quantities $\chi_{R}^{f}$ and $\Lambda_{R}^{f}$ defined in Eqs.$\,$\eqref{eq:chi_f_R}
and~\eqref{eq:lambda_f_R} for the AAM with $\tau_{c}=2/3$ (a,b)
and the AA-NNN with $t_{2}=0.5$ and $\tau_{c}=13/21$ (c,d), for
selected energy bands. The black lines depict contours over which
these quantities are constant. The red dashed lines depict the critical
line, for which $\Lambda_{R}^{f}=1$. The arrows relate examples of
contours below and above the critical point with the same $\chi_{R}^{f}$
or $\Lambda_{R}^{f}$. Dual points share both the same $\chi_{R}^{f}$
and $\Lambda_{R}^{f}$.\label{fig:dualPoints_densityPlot}}
\end{figure}

Dual points can be mapped with a two-step procedure:
\begin{enumerate}
\item Find curves $E(V)$ of constant $\chi_{R}=\frac{E-E_{R}}{V_{R}}$.
Map them into curves $E'(V')$ of constant $\chi'_{R}=\frac{E'-E'_{R}}{2t'_{R}}=\chi_{R}$.
This can be done by mapping curves $E(V,\varphi,\kappa)$ into curves
$E'(V',\mathcal{R}_{0}[\varphi,\kappa]^{T})$;
\item Map dual points within the curves above by considering $\Lambda_{R}\equiv2t_{R}/V_{R}=V'_{R}/(2t'_{R})\equiv\Lambda'_{R}$
.
\end{enumerate}
Defining the following quantities,

\begin{equation}
\chi_{R}^{f}=\begin{cases}
\chi_{R}\equiv\frac{E-E_{R}}{V_{R}} & \chi_{R}\leq\chi'_{R}\\
\chi'_{R}\equiv\frac{E'-E'_{R}}{2t'_{R}} & \chi_{R}>\chi'_{R}
\end{cases}\label{eq:chi_f_R}
\end{equation}

\begin{equation}
\Lambda_{R}^{f}=\begin{cases}
\Lambda_{R}\equiv2t_{R}/V_{R} & \Lambda_{R}\leq\Lambda'_{R}\\
\Lambda'_{R}\equiv V'_{R}/(2t'_{R}) & \Lambda_{R}>\Lambda'_{R}
\end{cases}\label{eq:lambda_f_R}
\end{equation}

\noindent we can visually observe the duality between the extended
and localized phases, given some energy band. Examples are in Fig.$\,$\ref{fig:dualPoints_densityPlot}.

\subsection{Absence of duality for low-order CA}

As seen in the main text, for generic models a (almost) perfect FS
duality only emerges for CA of high-enough order. For these models,
the description of the sub-bands of a low-order CA in terms of Eq.$\,$\eqref{eq:renormalized_model}
should fail, as already hinted in Fig.$\,$\ref{fig:FS_exact_vs_RenormModel_BAD}.
For CAs of high-enough order, this simple description of the FS ensures
the existence of dual points as long as the conditions in Eqs.$\,$\eqref{eq:condDual1}
and~\eqref{eq:condDual2} are satisfied.

\begin{figure*}[t]
\centering{}\includegraphics[width=0.8\textwidth]{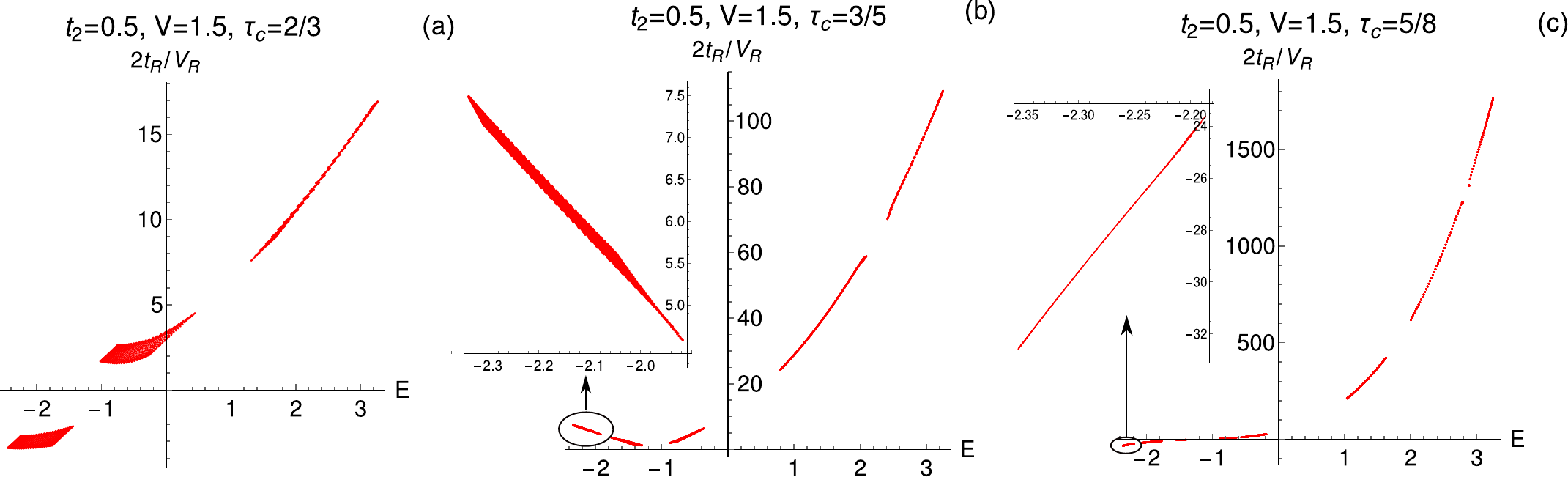}\caption{Ratio of renormalized couplings, $2t_{R}/V_{R}$, for fixed $t_{2}=0.5$
and $V=1.5$ (AA-NNN model), for the whole spectrum of each CA. The
CA's order is increased from (a) to (c). The renormalized parameters
were computed for a grid of points $(\varphi,k)$, as explained in
Sec.$\,$\ref{subsec:estimate_VR_tR_ER}. The insets show a close-up
of the results for the lowest energy band to show that the values
of $2t_{R}/V_{R}$ collapse to a well-defined curve as the CA's order
is increased.\label{fig:renormalized_couplings_AA-NNN}}
\end{figure*}

Here we exemplify the breakdown of the renormalized single-band description
for low-order approximants, in the AA-NNN model. In Fig.$\,$\ref{fig:renormalized_couplings_AA-NNN}
we compute the ratio $2t_{R}/V_{R}$ for fixed $t_{2}=0.5$ and $V=1.5$,
for the whole spectrum and different approximants. There are two important
comments:
\begin{itemize}
\item For smaller approximants, there is no well defined curve $\Lambda_{R}(E)\equiv2t_{R}(E)/V_{R}(E)$.
For a given energy, $\Lambda_{R}$ can take a finite range of values,
which means that the description of the FS in terms of Eq.$\,$\eqref{eq:renormalized_model}
is not suitable. Under the conjecture that the duality emerges when
the energy dispersion becomes sinusoidal, no true duality exists for
small approximants. It only emerges when we consider higher-order
approximants for which the curve $\Lambda_{R}(E)$ becomes well-defined;
\item The ratio $\Lambda_{R}(E)$ is energy-dependent for a fixed approximant,
in contrast with the AAM. This follows from the existence of a mobility
edge and energy-dependent correlation lengths.
\end{itemize}

\section{Emergence of (almost) perfect duality for higher-oder commensurate
approximants}

\label{sec:Emergence-of-duality}

\begin{figure}[h]
\centering{}\includegraphics[width=0.7\columnwidth]{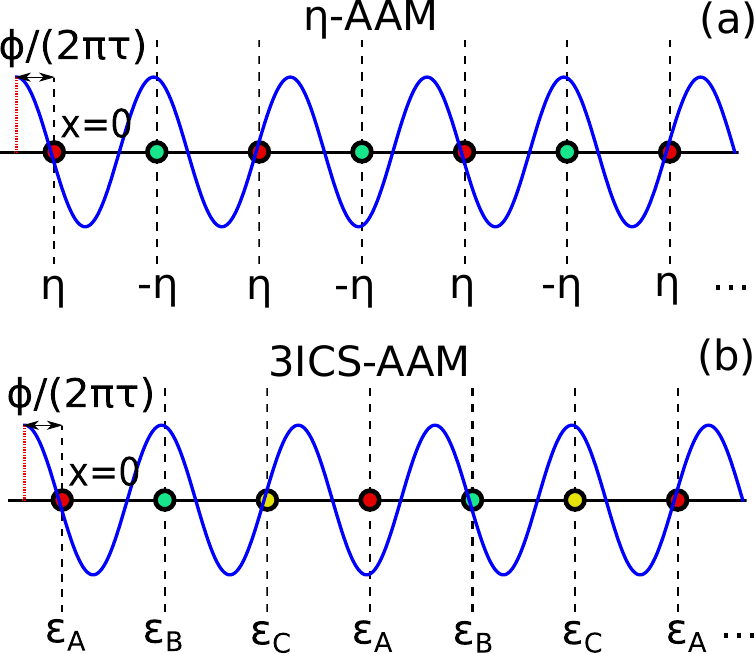}\caption{(a) AAM with additional staggered potential of strength $\pm\eta$
for every two consecutive sites ($\eta$-model). (b) AAM with additional
on-site energies $\epsilon_{A}\protect\neq\epsilon_{B}\protect\neq\epsilon_{C}$
for every three consecutive sites (3ICS-AAM). We fix the on-site energies
to $\epsilon_{A}=1,\epsilon_{B}=-1,\epsilon_{C}=0.5$, in units of
$t$. \label{fig:new_models_sketch}}
\end{figure}

We have seen that for some special models, including the AAM, there
is a $\tau$-independent duality symmetry. In these cases, the expression
for the mobility edge may be obtained by using the simplest possible
CA. However, this cannot be done for more generic models. Remarkably,
even in such models, new duality symmetries emerge when the size of
the unit cell is increased and the incommensurate limit is approached,
as stated in the main text. To see that, we consider three variations
of the AAM, already mentioned in the main text:
\begin{enumerate}
\item AA-NNN: AAM with next-nearest-neighbors;
\item $\eta$-AAM: AAM with additional staggered potential $+\eta$ in even
sites and $-\eta$ in odd sites {[}see Fig.$\,$\ref{fig:new_models_sketch}(a){]};
\item 3ICS-AAM: AAM with additional different on-site energies for any three
consecutive sites {[}see Fig.$\,$\ref{fig:new_models_sketch}(b){]}.
\end{enumerate}

\begin{figure}[h]
\centering{}\includegraphics[width=1\columnwidth]{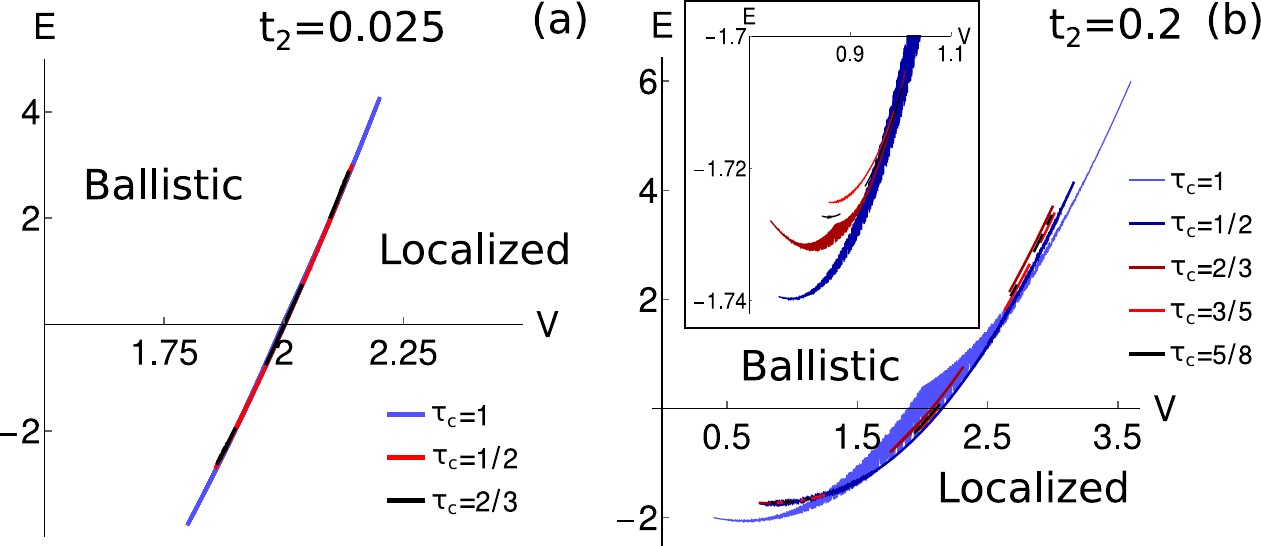}\caption{Set of intersection points $\{(E_{i},V_{i})\}$ that satisfy the condition
$E_{i}(V_{i},\varphi,\kappa)=E_{i}(V_{i},\mathcal{R}_{0}[\varphi,\kappa]^{T})$,
for different approximants, in the AA-NNN. (a) For $t_{2}=0.025$,
the duality symmetry of the AAM is only weakly broken and all these
points approximately fall into a well-defined curve for any approximant.
(b) For $t_{2}=0.2$ the duality symmetry of the AAM is strongly broken
and the points $\{(E_{i},V_{i})\}$ fall into a region for approximants
characterized by smaller unit cells. For such approximants, there
is no true SD curve: there is no well-defined curve $E_{c}(V)$ for
which the FS is invariant under $\mathcal{R}_{0}$, for all $\varphi,\kappa$.
However, when the order of the approximant is increased, the region
of intersection points converges into a well-defined SD curve. \label{fig:intersection_points}}
\end{figure}

We first analyze the AA-NNN model for $\tau_{c}=1$. We have

\begin{equation}
E=V\cos\varphi+2\cos\kappa+2t_{2}\cos2\kappa
\end{equation}
Note that for $t_{2}\neq0$ the energy dispersion is no longer invariant
under a suitable interchange $\varphi\leftrightarrow k$ at $V=2$
(and at any point).

For the $\eta$-AAM, the smallest possible unit cell corresponds to
$\tau_{c}=1/2$. The energy bands can be computed by solving:

\begin{equation}
E^{2}=2-\frac{V^{2}}{2}-\eta^{2}-2\cos k-\frac{V}{2}\Big(4\eta\cos\varphi+V\cos2\varphi\Big)\label{eq:eta_AAM_1o2}
\end{equation}
Note that here we have identified $\varphi=\phi n_{1}/2$ and not
$\varphi=\phi n_{1}$, as usual. This is due to the staggered potential,
which increases the periodicity of $\phi$ by a factor of 2. For $\eta=0$,
we recover the $2\pi/n_{1}$-periodicity for $\phi$ and the correct
definition for $\varphi$ is recovered by making $\varphi\rightarrow\varphi/2$.
As implied by Eq.$\,$\eqref{eq:eta_AAM_1o2}, for $\eta\neq0$ the
energy bands are never invariant under a suitable interchange $\varphi\leftrightarrow k$,
for $\tau_{c}=1/2$.

We start by focusing on the AA-NNN in Fig.$\,$\ref{fig:intersection_points}.
For different approximants of $\tau^{-1}=(\sqrt{5}-1)/2$, we plot
the set of points $\{(E_{i},V_{i})\}$ that satisfy the condition
$E_{i}(V_{i},\varphi,\kappa)=E_{i}(V_{i},\mathcal{R}_{0}[\varphi,\kappa])$.
In Fig.$\,$\ref{fig:intersection_points}(a) we use $t_{2}=0.025$.
In this case, the AAM's duality is only weakly broken and all the
points fall into a seemingly well-defined line, to a very good approximation,
even for $\tau_{c}=1$. Very similar curves are obtained for different
approximants, showing that the commensurate SD points quickly converge
to the incommensurate SD points, even for small unit cells. This is
what is expected if there are well-defined SD points for any CA: for
each energy $E$, the unrotated and rotated FS fully intersect at
a single coupling strength $V^{*}(E)$, to a very good approximation.For
$t_{2}=0.2$ {[}Fig.$\,$\ref{fig:intersection_points}(b){]}, however,
the intersection points clearly fall into a region (and not a curve)
for the lowest-order approximants. In this case, there is no well-defined
SD curve for which the FS is invariant under $\mathcal{R}_{0}$, for
all $\varphi,\kappa$. Interestingly, as we increase the order of
the approximant the region shrinks, eventually looking like a curve.
A perfectly well-defined SD curve is only obtained in the limit $\tau_{c}\rightarrow\tau$.
This is also the case for $t_{2}=0.025$, in Fig.$\,$\ref{fig:intersection_points}(a),
although in this case the curve seems essentially perfect within numerical
accuracy, even for very small unit cells. This is because in practice,
CA of high-enough order already provide an essentially perfect description
of the incommensurate thermodynamic limit SD curve. In Fig.$\,$\ref{fig:intersection_points}(b),
for instance, the intersection points already fall into an almost
perfect curve for $\tau_{c}=5/8$.

\begin{figure*}[t]
\centering{}\includegraphics[width=0.8\paperwidth]{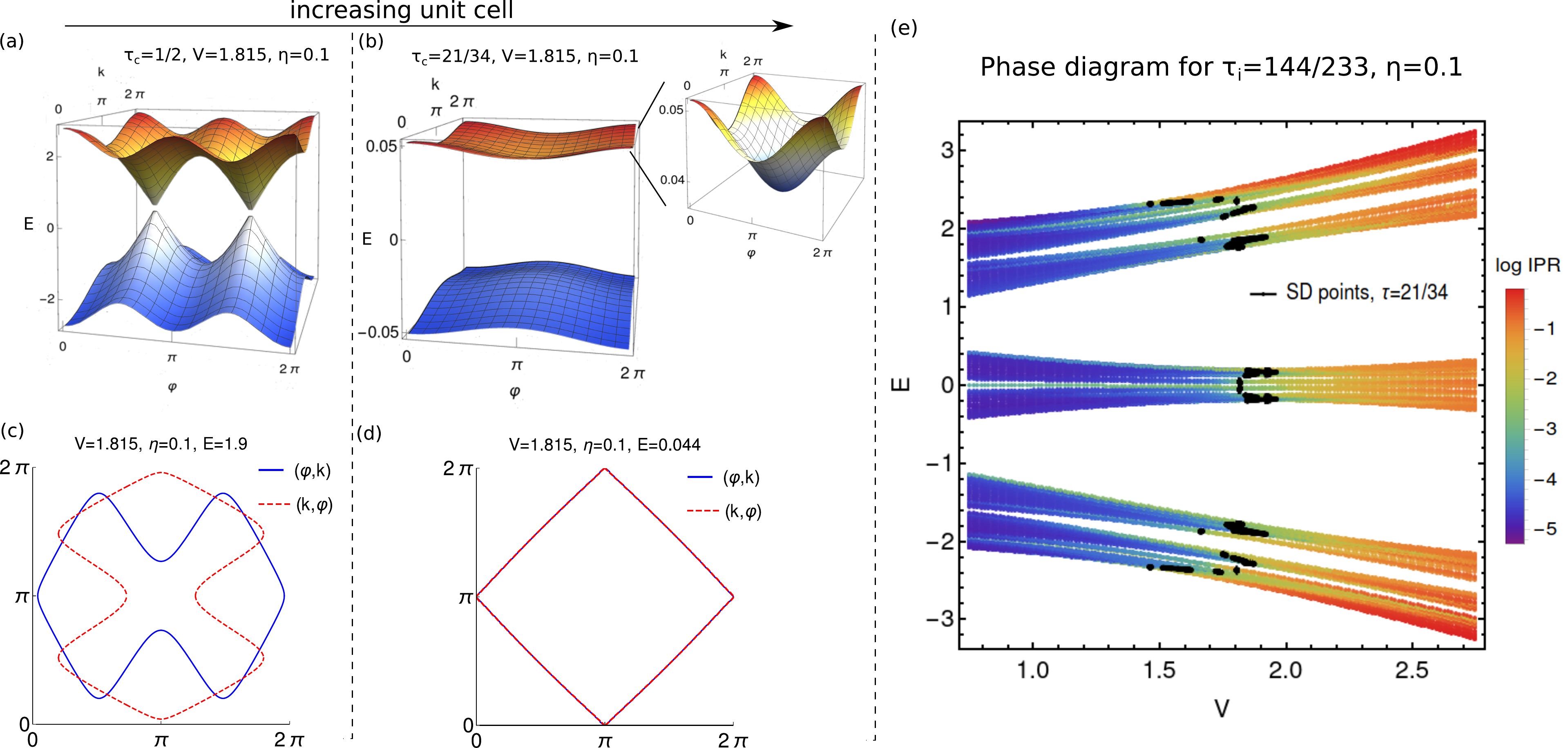}\caption{Results for the $\eta$-model, with $\eta=0.1$. (a,b) $E(\varphi,\kappa)$
energy bands for $\tau_{c}=1/2$ and $\tau_{c}=21/34$. For the latter,
the two bands closer to $E=0$ were plotted. The inset shows a close-up
of one of the bands. (c,d) Examples of constant energy cuts of the
$E(\varphi,\kappa)$ bands. For $\tau_{c}=1/2$, there is no cut for
which the SD symmetry is observed. For $\tau_{c}=21/34$ such cuts
exist for most of the bands. (e) Spectrum and IPR for $\tau_{i}=144/233$
along with predicted SD points for $\tau_{c}=21/34$. \label{fig:eta_AAM_panel}}
\end{figure*}

For the $\eta$-AAM, the main results are explained in Fig.$\,$\ref{fig:eta_AAM_panel}.
For $\tau_{c}=1/2$, we can see that the energy bands are almost (but
not exactly) $\varphi$-periodic with period $\Delta\varphi=\pi$
{[}Fig.$\,$\ref{fig:eta_AAM_panel}(a){]}. This is expected because
we have used a small $\eta=0.1$ and therefore the model is similar
to the AAM (where $\varphi\rightarrow2\varphi$). In this case it
is obvious that no SD symmetry can be found {[}Fig.$\,$\ref{fig:eta_AAM_panel}(c){]}.
For higher-order approximants, however, the almost $\pi$-periodicity
is completely lost {[}Fig.$\,$\ref{fig:eta_AAM_panel}(b){]} and
a SD symmetry with the correct $2\pi$-periodicity emerges {[}Fig.$\,$\ref{fig:eta_AAM_panel}(d){]}.

\begin{figure}[h]
\begin{centering}
\includegraphics[width=1\columnwidth]{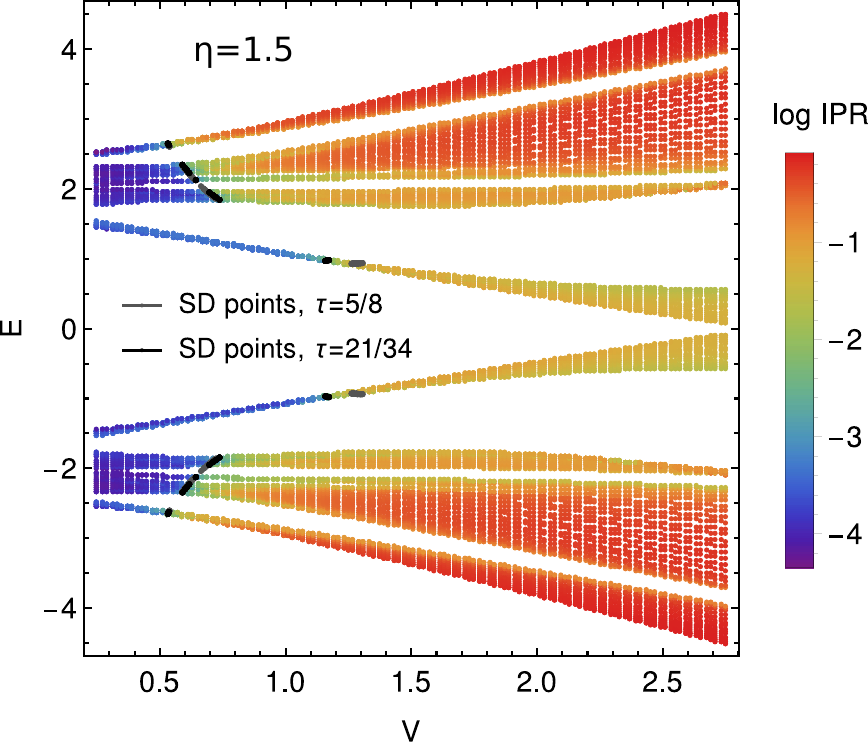}
\par\end{centering}
\caption{Spectrum and IPR for $\tau_{i}=144/233$ along with predicted SD points
for $\tau_{c}=5/8$ (gray) and $\tau_{c}=21/34$ (black).\label{fig:eta_AAM_panel_mob_edge_eta_0.5}}
\end{figure}

Note that for small $\eta$ the mobility edges seem to vary in a chaotic
manner, as illustrated in Fig.$\,$\ref{fig:eta_AAM_panel}(e). The
reason is that the $\eta$-perturbation defines a new periodicity
for $\phi$, different from the one in the parent model. The consequence
is that the duality symmetry also becomes completely different from
the one in the parent model. With the correct definition of $\varphi=\phi n_{1}/2$,
only the fundamental harmonic {[}$\cos(\varphi)${]} survives for
a CA of high-enough order. However, for low-order CA and small $\eta$,
the system is still very similar to the AAM: the universal dualities
will only appear for high-order CA and may give rise to significantly
different SD curves (mobility edges), even for bands that are close
in energy. If we instead consider a large $\eta$, the mobility edge
is almost entirely well captured for lower-order CA. In this case,
the $\phi$-periodicity of the AAM is completely broken even for the
lower-order CA: the almost perfect SD symmetry with the correct periodicity
emerges even for these. An example is in Fig.$\,$\ref{fig:eta_AAM_panel_mob_edge_eta_0.5},
for $\eta=1.5$.

We finish this section by analyzing the 3ICS-AAM. Differently from
the $\eta$-model, this model does not host any obvious $\phi^{*}$
around which the system's properties are symmetric. For the $\eta$-model,
$\phi^{*}=0$ is obviously one of such points. In that case, the FS
of CA in the $(\varphi,\kappa)$ plane can be described by simple
cossines, with $\varphi_{0}=0$ (FS is even around $\varphi_{0}=0$,
due to the symmetry around $\phi_{0}=0$). In the new model, shifting
the potential to negative $\phi$ (towards $\epsilon_{C}$) is not
equivalent to shifting it to positive $\phi$ (towards $\epsilon_{B}$).
This results in an important qualitative change with respect to previous
models that we studied: $(\varphi_{0},\kappa_{0})$ acquires a dependence
on energy and on the Hamiltonian's parameters.

We start by noticing that the energy bands of the 3ICS-AAM are periodic
with respect to $\phi$-translations, with period $3\times2\pi/n_{1}$.
Notice the factor of $3$ that comes from the existence of three consecutive
inequivalent sites. This motivates the definition of $\varphi$ as
$\varphi=n_{1}\phi/3$. For the following results, we consider $\epsilon_{A}=1,\epsilon_{B}=-1,\epsilon_{C}=0.5$.

\begin{figure}[h]
\begin{centering}
\includegraphics[width=1\columnwidth]{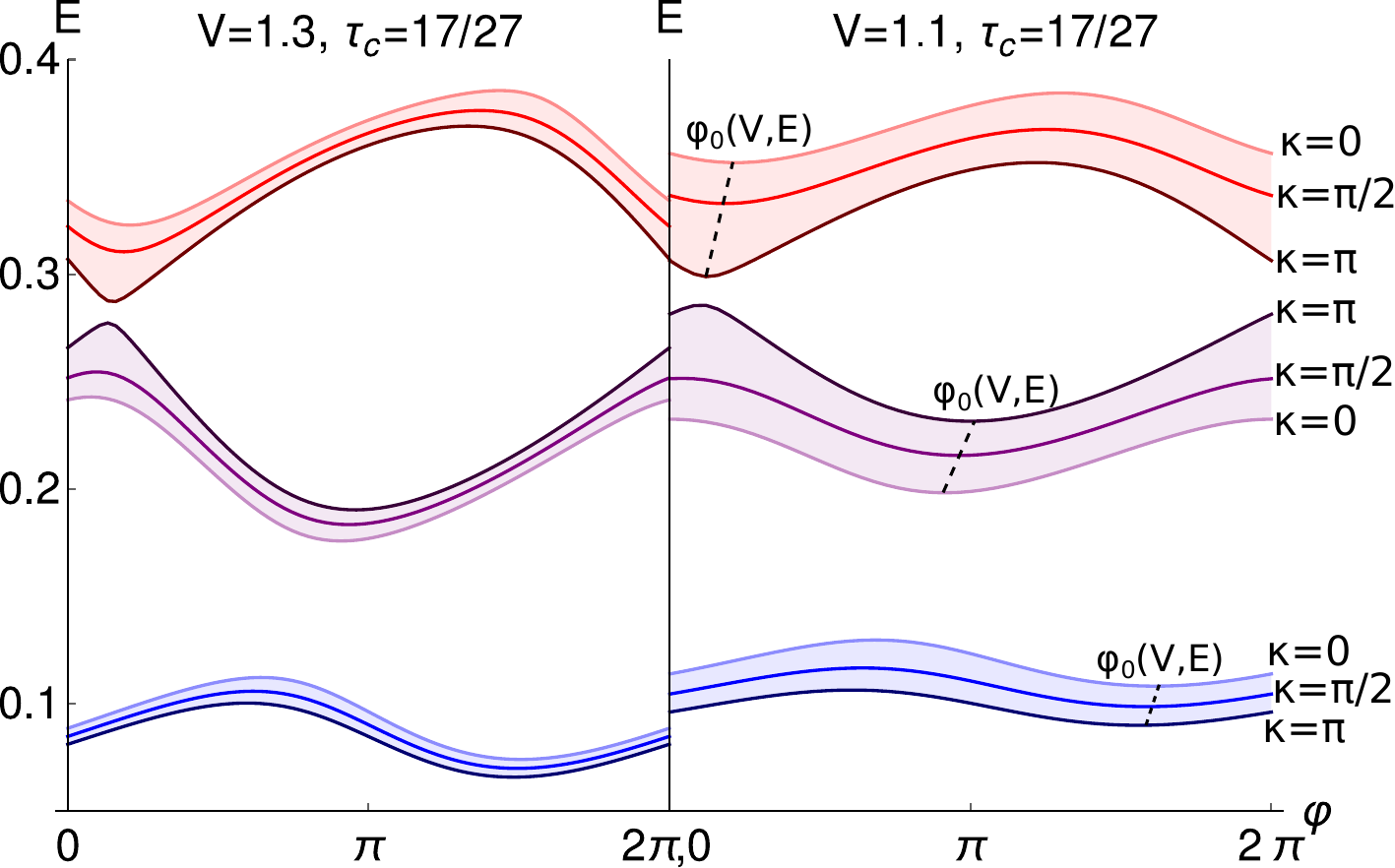}
\par\end{centering}
\caption{Energy bands of the CA defined by $\tau_{c}=17/27$ for the 3ICS-AAM
, as a function of $\varphi$ and for different $V$. For each band,
we plot contours corresponding to $\kappa=0,\pi/2,\pi$.\label{fig:bands_phi_3ICS-AAM }}
\end{figure}

In Fig.$\,$\ref{fig:bands_phi_3ICS-AAM } we show some energy bands
as a function of $\varphi$. $\varphi_{0}$ can be seen as the phase
$\varphi$ corresponding to the minimum of a given energy band, for
a fixed $\kappa$. As previously stated, we see that $\varphi_{0}$
depends on the Hamiltonian's parameters and on energy. For previous
models, we always obtained $\varphi_{0}=0$ or $\varphi_{0}=\pi$.
This was a consequence of the models' symmetry under shifts around
these points. In the present case, there are no obvious points for
which the system is symmetric under shifts.

\begin{figure*}[t]
\centering{}\includegraphics[width=0.8\paperwidth]{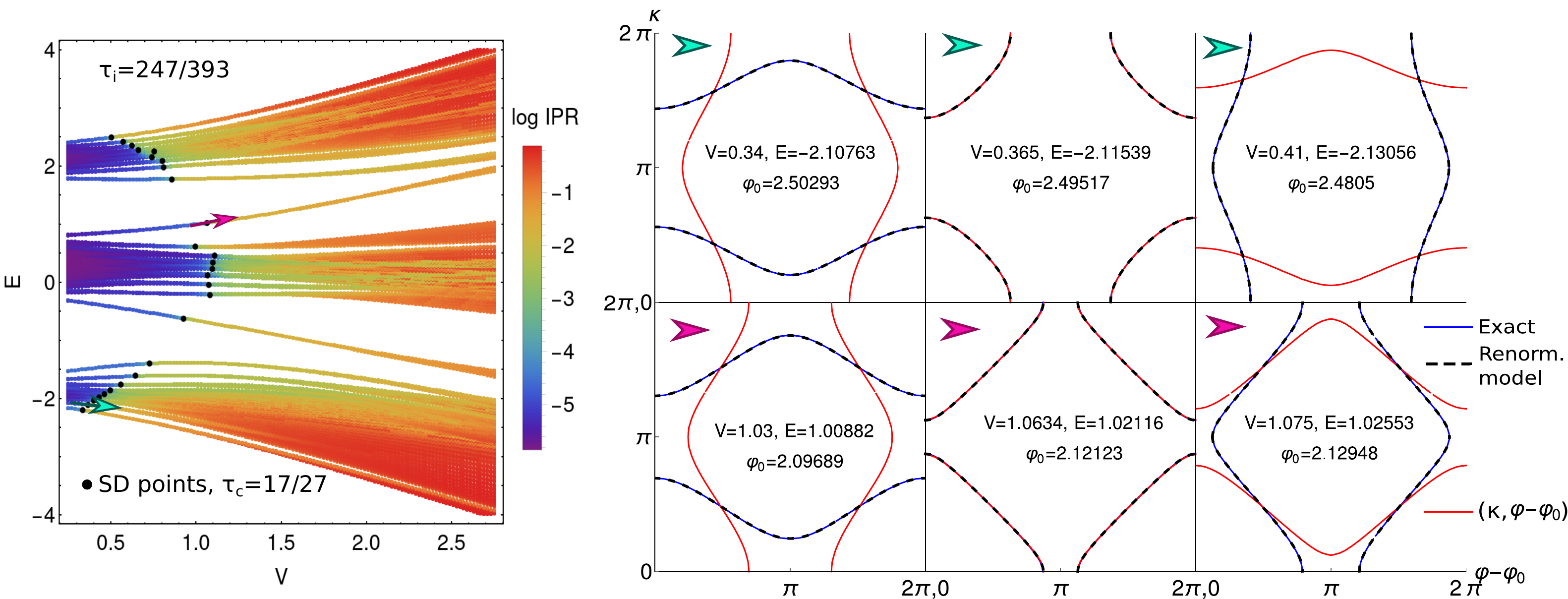}\caption{Left: IPR for the 3ICS-AAM, for a QPS with $\tau_{i}=247/393$ (system
with $393$ sites), together with SD points computed for the CA with
$\tau_{c}=17/27$. Right: Examples of FS around the critical point.
The blue and red lines correspond to the exact FS, respectively with
$\varphi-\varphi_{0}$ and $\kappa$ in the x-axis. The dashed black
lines correspond to the results of the renormalized model in Eq.$\,$\ref{eq:renormalized_model}.
Note that in the middle panel, the red and blue lines are superimposed
because these FS were computed at SD points.\label{fig:PD_and_FS_3ICS-AAM}}
\end{figure*}

Remarkably, we see that the FS again become perfectly sinusoidal around
the critical point, but now with a shift $\varphi_{0}$ that varies
smoothly with the model's parameters. This can be seen in Fig.$\,$\ref{fig:PD_and_FS_3ICS-AAM}.
There, we start by showing the IPR for a QPS with $\tau_{i}=247/393$,
that shows a clear transition from an extended to a localized phase.
In black, we plot SD points for a CA with $\tau_{c}=17/27$. These
were computed by shifting the FS by $-\varphi_{0}(V,E)$ in the $\varphi$
direction and then computing the renormalized parameters as described
in Sec.$\,$\ref{subsec:estimate_VR_tR_ER}. Some examples of FS are
shown inthe right panel of Fig.$\,$\ref{fig:PD_and_FS_3ICS-AAM},
where we can see the perfect description of the FS by the model in
Eq.$\,$\ref{eq:renormalized_model} and the existence of SD points
that match the localization-delocalization transition.

\begin{figure}[h]
\centering{}\includegraphics[width=1\columnwidth]{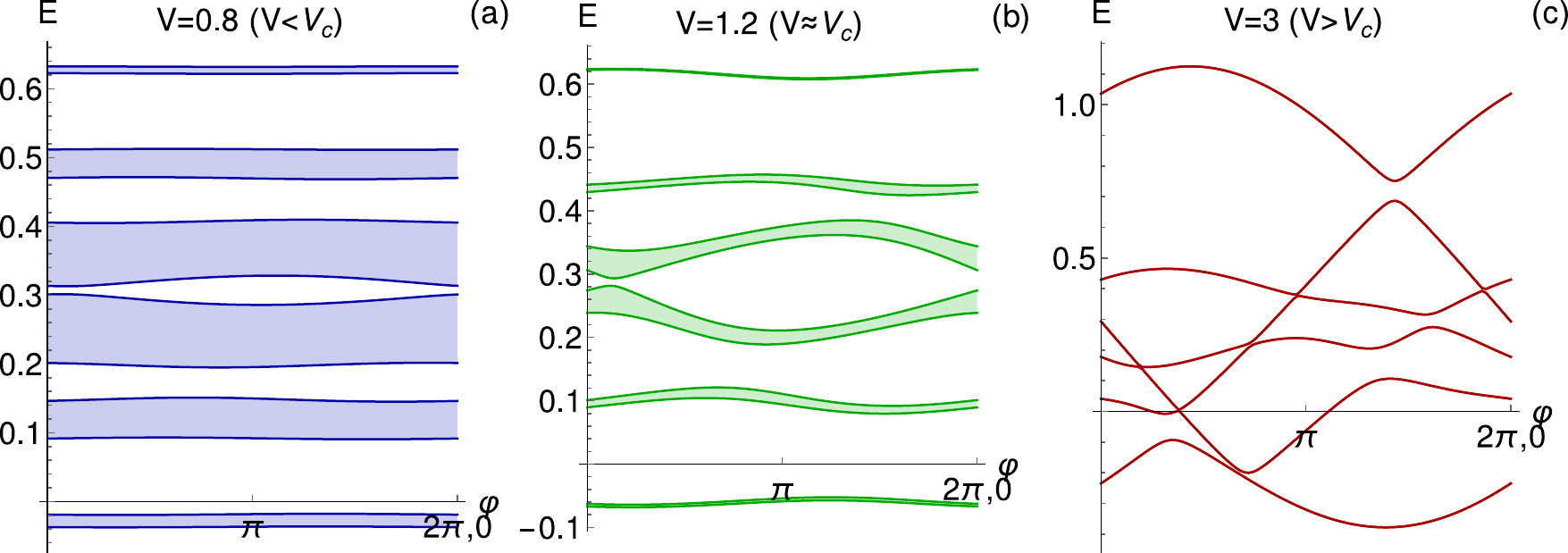}\caption{Examples of energy bands of the 3ICS-AAM. These bands are integrated
over $\kappa$ and plotted for variable $\varphi$ and different $V$
below (a), close-to (b) and above (c) criticality.\label{fig:Ebands2_3ICS-AAM}}
\end{figure}

Finally, we emphasize that in general the simple sinusoidal description
of the FS is only valid around the critical point. The current model
is an example of that: if we look at the energy bands away from criticality,
in particular for large-$V$, we see a lot of non-trivial $\varphi$-driven
band-crossings {[}see Fig.$\,$\ref{fig:Ebands2_3ICS-AAM}(c){]}.
In such cases, the FS is more complicated (even though being non-dispersive
in $\kappa$). Nonetheless, sufficiently close to the critical point,
no band crossings should occur: at criticality, band-gaps are opened
at any CA's order \footnote{This follows from the infinite correlation length at the critical
point. Increasing the order of the approximant can reflect in changes
that are only observed at very large length scales. But if the correlation
length is infinite, these changes will always be relevant and gaps
will be opened at any order.}.

\section{Derivation of Eq.$\,$\ref{eq:dualTransf_MAAM_ThermoLimit}}

\label{sec:derivation}

In the main text we claimed that the Hamiltonian in Ref.$\,$\citep{PhysRevLett.114.146601}
is self-dual under the following duality transformation,

\begin{equation}
f_{k}=\sum_{p}e^{-2\pi i\tau kp}\chi_{p}(\beta_{0})u_{p}
\end{equation}

\noindent where $\chi_{p}(\beta)=\frac{\sinh\beta}{\cosh\beta-\cos(2\pi\tau p)}$,
$\cosh\beta=\alpha^{-1}$ and $\beta_{0}$ is defined as $2t\cosh\beta_{0}=E+2\lambda\cosh\beta$.
Here we prove that statement. The Schrodinger equation for this model
can be written as (Eq.$\,$(7) of Ref.$\,$\citep{PhysRevLett.114.146601}):

\begin{equation}
t(u_{p-1}+u_{p+1})+g\chi_{p}(\beta)u_{p}=(E+2\lambda\cosh\beta)u_{p}\label{eq:Schrodinger_dasSarma}
\end{equation}
We know from Ref.$\,$\citep{PhysRevLett.114.146601} that the self-duality
condition implies that $\beta=\beta_{0}$. The derivation is made
below term by term:
\begin{center}
\textbf{Term $t(u_{p-1}+u_{p+1})$}
\par\end{center}

Starting with $tu_{p-1}$, we multiply by $e^{-2\pi i\tau kp}$ and
sum over $p$,

\begin{equation}
\begin{aligned}tu_{p-1}\rightarrow & \sum_{p}tu_{p-1}e^{-2\pi i\tau kp}\chi_{p-1}^{-1}(\beta_{0})\chi_{p-1}(\beta_{0})\\
 & =\frac{1}{\sinh\beta_{0}}\sum_{p}tu_{p-1}[\cosh\beta_{0}-\cos[2\pi\tau(p-1)]]\\
 & \times e^{-2\pi i\tau k(p-1)}\chi_{p-1}(\beta_{0})e^{2\pi i\tau k}\\
\textrm{ } & =\frac{1}{\sinh\beta_{0}}\sum_{p}tu_{p}[\cosh\beta_{0}-\cos(2\pi\tau p)]\\
 & \times e^{-2\pi i\tau kp}\chi_{p}(\beta_{0})e^{2\pi i\tau k}.
\end{aligned}
\end{equation}
In the expression obtained above, the first term becomes $\frac{t\cosh\beta_{0}}{\sinh\beta_{0}}f_{k}e^{2\pi i\tau k}$,
while the second one is $-\frac{t}{2\sinh\beta_{0}}(f_{k+1}+f_{k-1})e^{2\pi i\tau k}$.
In a similar way, for $tu_{p+1}$,

\begin{equation}
\begin{aligned}tu_{p+1}\rightarrow & \sum_{p}tu_{p+1}e^{-2\pi i\tau kp}\chi_{p+1}^{-1}(\beta_{0})\chi_{p+1}(\beta_{0})\\
 & =\frac{t\cosh\beta_{0}}{\sinh\beta_{0}}f_{k}e^{-2\pi i\tau k}-\frac{t(f_{k+1}+f_{k-1})}{2\sinh\beta_{0}}e^{-2\pi i\tau k}.
\end{aligned}
\end{equation}
Combining the two terms,

\begin{equation}
\begin{aligned}t(u_{p-1}+u_{p+1})\rightarrow-\frac{t\cos(2\pi\tau k)}{\sinh\beta_{0}}(f_{k+1}+f_{k-1})\\
+\frac{2t\cosh\beta_{0}}{\sinh\beta_{0}}\cos(2\pi\tau k)f_{k}.
\end{aligned}
\end{equation}

\begin{center}
\textbf{Term $g\chi_{p}(\beta)u_{p}$}
\par\end{center}

Multiplying by $e^{-2\pi i\tau kp}$ and summing over $p$,
\begin{equation}
\begin{aligned}g\chi_{p}(\beta)u_{p}\rightarrow g\sum_{p}\chi_{p}(\beta)\chi_{p}^{-1}(\beta_{0})u_{p}e^{-2\pi i\tau kp}\chi_{p}(\beta_{0})\end{aligned}
\end{equation}

We can simplify the result here by using \textit{a priori} the self-duality
condition $\beta=\beta_{0}$. In that case, this term simply becomes
$gf_{k}$.
\begin{center}
\textbf{Term $(E+2\lambda\cosh\beta)u_{p}$}
\par\end{center}

Again, we multiply by $e^{-2\pi i\tau kp}$ and sum over $p$,
\begin{equation}
\begin{aligned}(E+2\lambda\cosh\beta)u_{p}\rightarrow & \sum_{p}(E+2\lambda\cosh\beta)\chi_{p}^{-1}(\beta_{0})\chi_{p}(\beta_{0})\\
 & \times u_{p}e^{-2\pi i\tau kp}.
\end{aligned}
\end{equation}
Recalling that $\chi_{p}^{-1}(\beta_{0})=\frac{\cosh\beta_{0}-\cos(2\pi\tau p)}{\sinh\beta_{0}}$,
the first term becomes

\begin{equation}
\begin{aligned} & \frac{\cosh\beta_{0}}{\sinh\beta_{0}}\sum_{p}(E+2\lambda\cosh\beta)\chi_{p}(\beta_{0})u_{p}e^{-2\pi i\tau kp}\\
 & =\frac{\cosh\beta_{0}(E+2\lambda\cosh\beta)}{\sinh\beta_{0}}f_{k},
\end{aligned}
\end{equation}

and the second term,

\begin{equation}
\begin{aligned} & -\frac{1}{\sinh\beta_{0}}\sum_{p}(E+2\lambda\cosh\beta)\cos(2\pi\tau p)u_{p}e^{-2\pi i\tau kp}\\
 & =-\frac{E+2\lambda\cosh\beta}{2\sinh\beta_{0}}(f_{k+1}+f_{k-1}).
\end{aligned}
\end{equation}

\begin{center}
\textbf{Combine everything}
\par\end{center}

Combining everything, we have

\begin{equation}
\begin{aligned}\Big[-\frac{t\cos(2\pi\tau k)}{\sinh\beta_{0}}+\frac{E+2\lambda\cosh\beta}{2\sinh\beta_{0}}\Big](f_{k+1}+f_{k-1})+gf_{k}\\
=\frac{\cosh\beta_{0}}{\sinh\beta_{0}}[E+2\lambda\cosh\beta-2t\cos(2\pi\tau k)]f_{k}.
\end{aligned}
\end{equation}

We now verify that for the duality condition $\beta=\beta_{0}$, the
model above is self-dual of the model in Eq.$\,$\eqref{eq:Schrodinger_dasSarma}.
This condition implies that $E+2\lambda\cosh\beta=2t\cosh\beta$.
The term with square brackets on the left hand side of Eq.$\,$\eqref{eq:Schrodinger_dasSarma}
becomes

\begin{equation}
\begin{aligned} & \frac{t}{\sinh\beta}\Big[-\cos(2\pi\tau k)+\cosh\beta\Big](f_{k+1}+f_{k-1})\\
= & t\chi_{k}^{-1}(\beta)(f_{k+1}+f_{k-1}),
\end{aligned}
\end{equation}

\noindent while the right hand side can be written as

\begin{equation}
\begin{aligned} & \frac{\cosh\beta}{\sinh\beta}[E+2\lambda\cosh\beta-2t\cos(2\pi\tau k)]f_{k}\\
= & 2t\cosh\beta\chi_{k}^{-1}(\beta)f_{k}.
\end{aligned}
\end{equation}

Combining everything again:

\begin{equation}
\begin{aligned}t\chi_{k}^{-1}(\beta)(f_{k+1}+f_{k-1})+gf_{k} & =2t\cosh\beta\chi_{k}^{-1}(\beta)f_{k}\\
\Leftrightarrow t(f_{k+1}+f_{k-1})+g\chi_{k}(\beta)f_{k} & =(E+2\lambda\cosh\beta)f_{k},
\end{aligned}
\end{equation}

\noindent which is clearly dual of Eq.$\,$\eqref{eq:Schrodinger_dasSarma}.

\newpage

\section{Additional remarks on wave function duality}

\label{sec:wave_duality_remarks}

We have seen in the main text that an existing duality symmetry maps
the real-space wave function $\psi^{r}(\phi,k)$ and its dual $\psi^{d}(\mathcal{R}_{0}[\phi,k])$,
where $\mathcal{R}_{0}$ is a $\pi/2$-rotation in the $(\phi,k)$
plane around point $(\phi_{0},k_{0})$. In the text, we mostly dealt
with examples for which $\phi_{0}=0$ or $\phi_{0}=\pi$. In this
section we show that this 'center of rotation' can be less trivial.

\begin{figure}[h]
\begin{centering}
\includegraphics[width=1\columnwidth]{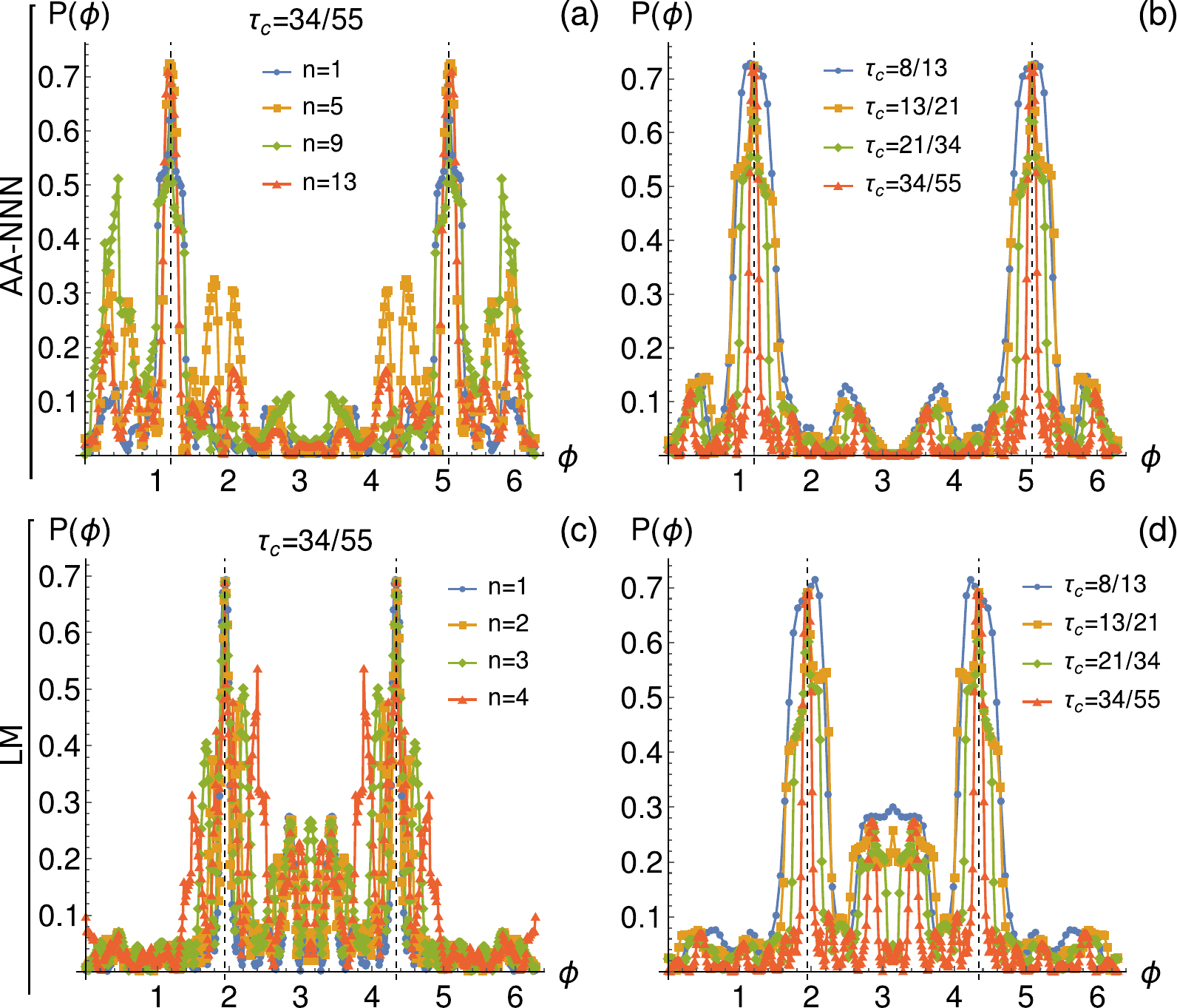}
\par\end{centering}
\caption{Overlap $P(\phi)=|\protect\braket{\tilde{\bm{\psi}}^{\textrm{r}}(\phi,k=0)}{\tilde{\bm{\psi}}^{\textrm{d}}(\phi,k=0)}|$,
computed at SD points. (a,b) Results for the AA-NNN, with $t_{2}=0.5$.
In (a) $\tau_{c}=34/55$ is fixed and $P(\phi)$ is computed for some
SD points within the interpolated SD curve with negative concavity
in Fig.$\,$\ref{fig:main}(d). In (b), we vary $\tau_{c}$ and compute
$P(\phi)$ for the lowest-energy SD point. The dashed lines correspond
to $\phi=\pi(n_{1}-n_{2})/n_{1}\vee\phi=2\pi-\pi(n_{1}-n_{2})/n_{1}$.
(c,d) Results for the LM, with $V=3$,$\Lambda=2.5$. In (c) we fix
$\tau_{c}=34/55$ and use SD points within the lowest-energy interpolated
SD curve with negative concavity in Fig.$\,$\ref{fig:main}(c). In
(d), we vary $\tau_{c}$ and compute $P(\phi)$ for the lowest-energy
SD point. The dashed lines correspond to $\phi=\pi n_{2}/n_{1}\vee\phi=2\pi-\pi n_{2}/n_{1}$.
\label{fig:overlap_phi0_AANNN_LM}}
\end{figure}

In general it may be challenging to find the correct 'center of rotation'
$(\phi_{0},k_{0})$. The knowledge of such point is essential to correctly
define the duality transformation for the wave function. A possible
quantity that can hint this point is the overlap between $\tilde{\bm{\psi}}^{\textrm{r}}(\phi,k)$
and $\tilde{\bm{\psi}}^{\textrm{d}}(\phi,k)$ at dual points, for
variable $(\phi,k)$. A large overlap may be found at point $(\phi_{0},k_{0})$
which is, by definition, invariant upon application of the duality
transformation. In fact, for the examples that we tested, the real-space
wave function and its Aubry-André dual were always similar, which
was reflected in a relatively large overlap. With this in mind, we
define the quantity,

\begin{equation}
P(\phi)=|\braket{\tilde{\bm{\psi}}^{\textrm{r}}(\phi,k=0)}{\tilde{\bm{\psi}}^{\textrm{d}}(\phi,k=0)}|,
\end{equation}

\noindent as the overlap between $\tilde{\bm{\psi}}^{\textrm{r}}(\phi,k)$
and $\tilde{\bm{\psi}}^{\textrm{d}}(\phi,k)$, at SD points, for $k=0$
(we have $k_{0}=0$ in the examples that follow and only need to search
for $\phi_{0}$). For fixed $k_{0}=0$, we define $\phi_{0}$ as the
phase that maximizes $P(\phi)$, checking that it remains stable as
the CA's order is increased. In the cases shown in the main text (Figs.$\,$\ref{fig:DualTransf3}-\ref{fig:fig11}),
$\phi_{0}=0$ or $\phi_{0}=\pi$. In Fig.$\,$\ref{fig:overlap_phi0_AANNN_LM},
we show examples for the AA-NNN and LM, for which we identified, respectively
$\phi_{0}=\pi(n_{1}-n_{2})/n_{1}\vee\phi_{0}=2\pi-\pi(n_{1}-n_{2})/n_{1}$
and $\phi_{0}=\pi n_{2}/n_{1}\vee\phi_{0}=2\pi-\pi n_{2}/n_{1}$.
For the AA-NNN these phases are for critical states within the SD
curve with negative concavity in Fig.$\,$\ref{fig:main}(d). For
the LM, we presented the center of rotation for the lowest-energy
interpolated SD curve with negative concavity, see Fig.$\,$\ref{fig:main}(c).
For the mentioned $\phi_{0}$ phases, $P(\phi_{0})$ is a stable maximum
in the sense that it remains a maximum when the CA's order is increased.

We finish by alerting that for some cases, the wave function and its
Aubry-André dual may be distinct - the generalized duality may be
very different from the Aubry-André duality. In such cases, we have
no systematic way of defining the duality transformation for the wave
function. The mobility edge, on the other hand, may still be predicted.
For the latter, we work with variables $\varphi$ and $\kappa$ which
allow us to find SD points.

\section{Possible generalizations: example of Maryland model}

\label{sec:maryland}

In the models that we have seen the FS in the $(\varphi,\kappa)$
space can be well described by Eq.$\,$\ref{fig:FS_exact_vs_RenormModel}
at each energy. However, the effective model describing FS can be
different, even though the treatment in terms of renormalized couplings
that measure the dependence on $\varphi$ and $\kappa$ still holds.
Example that we will treat in detail in \citep{prepar} are models
that also contain a coupling $C_{R}\cos(\varphi)\cos(\kappa)$. These,
as we will show, are responsible for the existence of phases with
critical eigenstates that exist over a region of parameters (and not
only at fine-tuned points such as at the critical point between a
delocalized and a localized phase). Here we show yet another example
in which the effective model is different than the one in Eq.$\,$\ref{fig:FS_exact_vs_RenormModel}:
the Maryland model \citep{PhysRevLett.49.833}. This model has an
unbounded quasiperiodic potential with Hamiltonian is given by\lyxdeleted{Miguel}{Sun Jun  5 14:14:45 2022}{ }

\begin{equation}
H=V\sum_{n}\mathcal{\tan}[(2\pi\tau n+\phi)/2]c_{n}^{\dagger}c_{n}+\sum_{n}\Big(c_{n}^{\dagger}c_{n+1}+\textrm{h.c.}\Big)\label{eq:maryland}
\end{equation}

\begin{figure}[h]
\begin{centering}
\includegraphics[width=1\columnwidth]{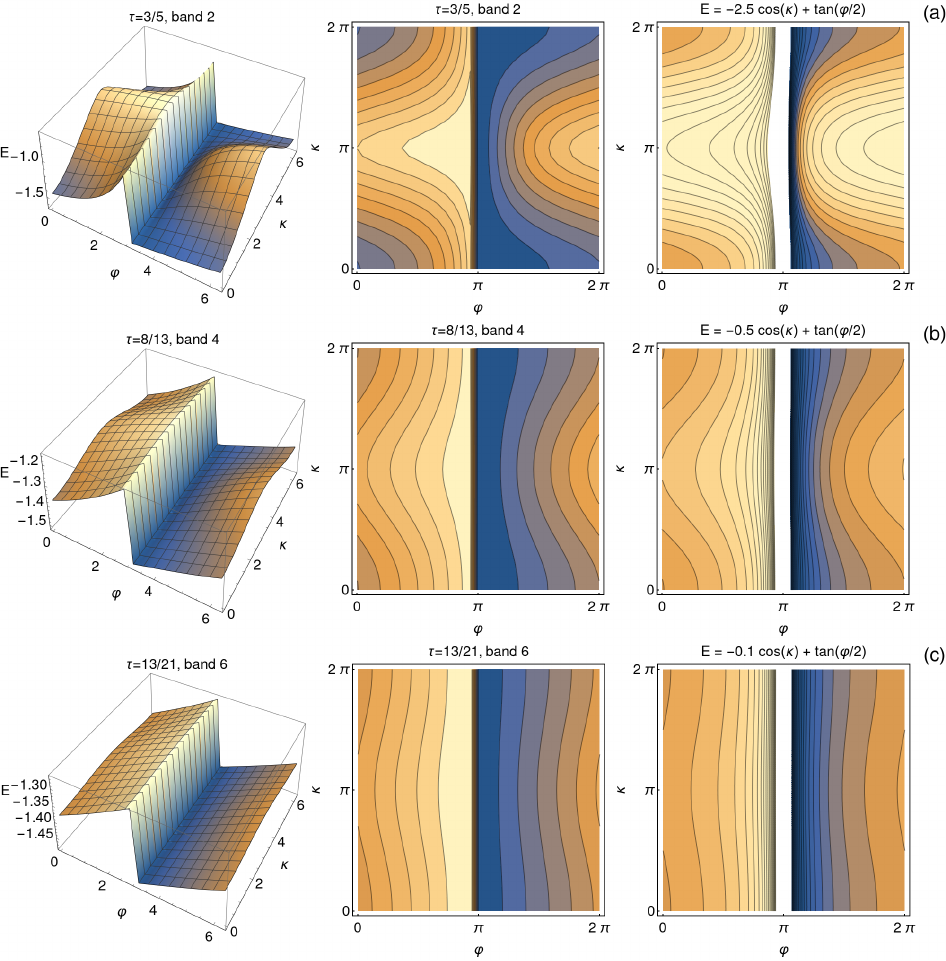}\caption{Energy bands and FS as a function of $\varphi=n_{1}\phi$ and $\kappa$
for selected energy bands of CA of the model in Eq.$\,$\ref{eq:maryland}.
(a) $\tau_{c}=3/5$, $2^{{\rm nd}}$ band (ordered from lowest to
higher energy); (b) $\tau_{c}=8/13$, $4^{{\rm th}}$ band; (c) $\tau_{c}=13/21$,
$6^{{\rm th}}$ band. The rightmost plot of each sub-figure shows
how the effective model in Eq.$\,$\ref{eq:effective_maryland} can
describe the FS obtained numerically for the different CA.\label{fig:Maryland}}
\par\end{centering}
.\lyxdeleted{Miguel}{Sun Jun  5 14:14:45 2022}{ }
\end{figure}

In this case, the FS for different CA can be captured through\lyxdeleted{Miguel}{Sun Jun  5 14:14:45 2022}{ }

\begin{equation}
E=V_{R}(V,E){\rm tan}(\varphi)+2t_{R}(V,E)\cos(\kappa)+E_{R}(V,E),\label{eq:effective_maryland}
\end{equation}
as exemplified in Fig.$\,$\ref{fig:Maryland}. Indeed, in the Maryland
model, the system becomes localized for any finite $V$. Therefore,
for a large enough unit cell the coupling $t_{R}$ becomes irrelevant
with respect to the coupling $V_{R}$. We show in Fig.$\,$\ref{fig:Maryland}
that this is the case using an example for $V=0.2$.

\end{document}